\documentclass[final,5p,times,twocolumn]{elsarticle}
\usepackage{graphics,epsfig}
\usepackage{color,epsfig,bm}
\usepackage{fancybox,amsmath,amssymb,rotating,pifont}
\usepackage{epic,eepic,calc,array,shadow}
\usepackage{npsfont}
\usepackage{graphicx}
\usepackage{dcolumn}  
\usepackage{subfigure}
\usepackage[section]{placeins}
\usepackage{multirow}
\usepackage[section]{placeins}
\usepackage{here}

\makeatletter
\newlength{\earraycolsep}
\def\eqnarray{\stepcounter{equation}\let\@currentlabel%
\theequation \global\@eqnswtrue\m@th
\global\@eqcnt\z@\tabskip\@centering\let\\\@eqncr
$$\halign to\displaywidth\bgroup\@eqnsel\hskip\@centering
$\displaystyle\tabskip\z@{##}$&\global\@eqcnt\@ne \hskip
2\earraycolsep \hfil$\displaystyle{##}$\hfil &\global\@eqcnt\tw@
\hskip 2\earraycolsep $\displaystyle\tabskip\z@{##}$\hfil
\tabskip\@centering&\llap{##}\tabskip\z@\cr} \makeatother

\journal{Physics Reports}

\begin{document}
\bibliographystyle{try}
\def\beq{\begin{equation}}
\def\enq{\end{equation}}
\def\beqa{\begin{eqnarray}}
\def\enqa{\end{eqnarray}}
\def\nnb{\nonumber}
\def\rar{\rightarrow}
\def\MeV{\nobreak\,\mbox{MeV}}
\def\GeV{\nobreak\,\mbox{GeV}}
\def\keV{\nobreak\,\mbox{keV}}
\def\fm{\nobreak\,\mbox{fm}}
\def\Tr{\mbox{ Tr }}
\def\qq{\lag\bar{q}q\rag}
\def\uu{\lag\bar{u}u\rag}
\def\dd{\lag\bar{d}d\rag}
\def\qqs{\lag\bar{s}s\rag}
\def\mix{\lag\bar{q}g\si.Gq\rag}
\def\mixs{\lag\bar{s}g\si.Gs\rag}
\def\Gd{\lag g^2G^2\rag}
\def\G3{\lag g^3G^3\rag}
\def\pli{p^\prime}
\def\ka{\kappa}
\def\la{\lambda}
\def\La{\Lambda}
\def\ga{\gamma}
\def\Ga{\Gamma}
\def\om{\omega}
\def\rh{\rho}
\def\si{\sigma}
\def\ps{\psi}
\def\ph{\phi}
\def\de{\delta}
\def\al{\alpha}
\def\be{\beta}
\def\alma{\alpha_{max}}
\def\almi{\alpha_{min}}
\def\bemi{\beta_{min}}
\def\lb{\label}
\def\nn{\nonumber}
\def\kab{\left[(\al+\be)m_c^2-\al\be s\right]}
\def\xsla{x\kern-.5em\slash}
\def\psla{p\kern-.5em\slash}
\newcommand{\rag}{\rangle}
\newcommand{\lag}{\langle}
\newcommand{\bph}{\mbox{\bf $\phi$}}
\newcommand{\rf}{\ref}
\newcommand{\ct}{\cite}

\title{New Charmonium States in  QCD Sum Rules: a Concise Review }
\author[label1]{Marina Nielsen}
\address[label1]{Instituto de F\'{\i}sica, Universidade de S\~{a}o Paulo,
C.P. 66318, 05389-970 S\~{a}o Paulo, SP, Brazil}
\cortext[cor1]{Corresponding author}
\ead{mnielsen@if.usp.br}
\author[label2]{Fernando S. Navarra}
\ead{navarra@if.usp.br}
\address[label2]{Instituto de F\'{\i}sica, Universidade de S\~{a}o Paulo,
C.P. 66318, 05389-970 S\~{a}o Paulo, SP, Brazil}
\author[label3]{Su Houng Lee}
\ead{suhoung@phya.yonsei.ac.kr}
\address[label3]{Institute of Physics and Applied Physics, Yonsei
University, Seoul 120-749, Korea}


\begin{abstract}
In the past years there has been a revival of  hadron spectroscopy. 
Many interesting new hadron states were discovered experimentally, some 
of which do not fit easily into the quark model. This situation motivated 
a vigorous theoretical activity. 
This is a rapidly evolving field with enormous amount 
of new experimental information.
In the present report we include and discuss data which were 
released very recently. The present review is the first one 
written from the perspective of QCD sum rules (QCDSR), where we present the 
main 
steps of concrete calculations and compare the results with other approaches 
and with experimental data.   

\end{abstract}

\date{\today}
\maketitle

\clearpage\setcounter{equation}{0} \markboth{\sl Interpretation of the New
Charmonium States} {\sl Table of contents} \tableofcontents

\markboth{\sl Interpretation of the New Charmonium States }{\sl Introduction}
\setcounter{equation}{0} \section{\label{Introduction}Introduction}
We are approaching the end of a decade which will be remembered as the 
``decade of the revival of hadron spectroscopy''. During these years 
several  $e^+e^-$ colliders started to operate and produce a large body 
of experimental information. At the same time new data came from the existing 
$p - \bar{p}$ colliders and also from the $e - p $ accelerators. In 
Table~\ref{tabnew} we give a list of the new charmonium states observed
in these accelerators.

\begin{table}[h]

  \begin{center}
    \caption{Charmonium states observed in the last years.}
  \label{tabnew}
    \begin{tabular}{|c|c|c|c|} \hline
      state    & production mode& decay mode  & ref.  \\ \hline
  $X(3872)$ & $B\to KX(3872)$ & $J/\psi\pi\pi¯$  &  \cite{belle1}  \\
  $X(3915)$ & $\gamma\gamma\to X(3915)$ & $J/\psi\omega$ & \cite{bellegg}\\
  $Z(3930)$ & $\gamma\gamma\to Z(3930)$  & $D\bar{D}$ & \cite{bellez3930}\\
  $Y(3930)$ & $B\to KY(3930)$ & $J/\psi\omega$ & \cite{belley3}\\
  $X(3940)$ & $e^+e^-\to J/\psi X(3940)$ & $D\bar{D}^*$ & \cite{bellecc}\\
  $Y(4008)$ & $e^+e^-\to\gamma_{ISR}Y(4008)$ & $J/\psi\pi\pi¯$ & \cite{yexp}  \\
  $Z_1^+(4050)$ & $B^0\to K^-Z_1^+(4050)$ & $\chi_{c1}\pi^+$  & \cite{belle3} \\
  $Y(4140)$ & $B\to KY(4140)$ & $J/\psi\phi$  & \cite{cdfy} \\
  $X(4160)$ & $e^+e^-\to J/\psi X(4160)$ & $D^*\bar{D}^*$ & \cite{Abe:2007sya}\\
  $Z_2^+(4250)$ & $B^0\to K^-Z_1^+(4250)$ & $\chi_{c1}\pi^+$  & \cite{belle3} \\
  $Y(4260)$ & $e^+e^-\to\gamma_{ISR}Y(4260)$ & $J/\psi\pi\pi¯$  & \cite{babar1}\\
  $X(4350)$ & $\gamma\gamma\to X(4350)$ &$J/\psi\phi$ & \cite{belleggs} \\
  $Y(4360)$ & $e^+e^-\to\gamma_{ISR}Y(4260)$ & $\psi^\prime\pi\pi¯$ &
\cite{belle4} \\
  $Z^+(4430)$ & $B^0\to K^-Z^+(4430)$ & $\psi^\prime\pi^+$  & \cite{bellez}\\
  $X(4630)$ & $e^+e^-\to\gamma_{ISR}X(4630)$ & $\Lambda^+\Lambda^-$ & 
\cite{belle4630}\\
  $Y(4660)$ & $e^+e^-\to\gamma_{ISR}Y(4660)$ & $\psi^\prime\pi\pi¯$ & 
\cite{belle4} \\
\hline
     \end{tabular}
  \end{center}
\end{table}

In what follows we will review and comment all this information 
\cite{belle1,bellegg,bellez3930,belley3,bellecc,yexp,belle3,cdfy,Abe:2007sya,
babar1,belleggs,belle4,bellez,belle4630,
cdf,d0,babarx,babaree,cleox,cdfx2,
pdg,belleE,babar2,cdf2,cdf3,babar09,belleD,babar3,cleo,belle0810,
babary2,babary,babar4,belle5,babar5,babar6,bellez2,babarz,
belleB0,babarB0,babary3}.
 
The study of spectroscopy and the decay properties of the heavy flavor 
mesonic states provides us with useful information about the dynamics of 
quarks and gluons at the hadronic scale. The remarkable progress on  the 
experimental side, with various high energy machines has opened up new 
challenges in the theoretical understanding of heavy flavor  hadrons.

\subsection{New experiments}

The $B$-factories,  the PEPII at SLAC in the U.S.A., and the KEKB at KEK in
Japan, were constructed to test the Standard Model mechanism for 
CP violation. However, their most interesting achievement was to contribute 
to the field of hadron spectroscopy, in particular in the area of charmonium 
spectroscopy. They are $e^+e^-$ colliders operating at a CM energy near
10,580 MeV. The $B\bar{B}$ pairs produced are measured by the BaBar (SLAC)
and Belle (KEK) collaborations. Charmonium states are copiously produced at
the $B$-factories in a variety of processes. At the quark level, the $b$ 
quark decays weakly to a $c$ quark accompanied by the emission of a virtual 
$W^-$ boson. Approximately half of the time, the $W^-$ boson materializes as a 
$s\bar{c}$ pair. Therefore, half of the $B$ meson decays result in a final state that
contains a $c\bar{c}$ pair. When these $c\bar{c}$ pairs are produced close to
each other in phase space, they can coalesce to form a $c\bar{c}$
charmonium meson.

\begin{figure}[h]
\scalebox{0.9}{\includegraphics[angle=0]{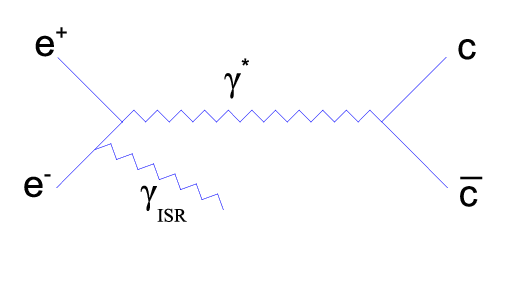}}
\vspace{-.6cm}
\caption{Illustration for the initial state radiation (ISR) process.}
\label{irs}
\end{figure}

The simplest charmonium producing $B$ meson decay is: $B\to K(c\bar{c})$.
Another interesting  form to produce charmonium in $B$-factories is
directly from the $e^+e^-$ collision, when the initial state $e^+$ or $e^-$
occasionally radiates a high energy $\gamma$-ray, and the  $e^+e^-$
subsequently annihilate at a corresponding reduced CM energy, as ilustrated 
in Fig.~\ref{irs}. 

When the
energy of the radiated $\gamma$-ray $(\gamma_{ISR})$ is between 4000
and 5000 MeV, the $e^+e^-$ annhilation occurs at CM energies that correspond
to the range of mass of the charmonium mesons. Thus, the initial state
radiation (ISR) process can directly produce charmonium states with
$J^{PC}=1^{--}$.

\subsection{New states}

Many states observed  by BaBar and Belle collaborations, like
the $X(3872)$, $Y(3930)$, $Z(3930)$, $X(3940)$, $Y(4008)$, $Z_1^+(4050)$, 
$Y(4140)$, $X(4160)$, $Z_2^+(4250)$, $Y(4260)$, $Y(4360)$, $Z^+(4430)$ and 
$Y(4660)$, remain controversial. A common feature of these states is that
they are seen to decay to final states that contain charmed and anticharmed 
quarks. Since their masses and decay modes are not in agreement with 
the predictions from potential models, they are considered as candidates 
for exotic states. By exotic we mean a more complex structure than the 
simple quark-antiquark state, like hybrid, molecular or tetraquark states. 
The idea of unconventional quark structures is quite old and despite decades 
of progress, no exotic meson has been conclusively identified. In particular,
those with $q\bar{q}$ quantum numbers should mix with ordinary mesons and
are thus hard to understand. Therefore, the observation of these new states
is  a challenge to our understanding of QCD.

In 2003 a great deal of attention was given to  pentaquark states, which may 
be exotic  objects \cite{theory_penta,mnnrl04}. After many 
negative results  the study of these states was abandoned. However, in that same 
year 
the discovery by BELLE of the unusual charmonium state named X prompted a lively 
and deep 
discussion about the existence of exotic states in the charmonium sector  
(i.e. non pure $c - \bar{c}$ states),  exactly where, up to that moment,  the  
calculations 
based on potential models had worked so well.  

Establishing the existence of these states means already a remarkable progress 
in hadron 
physics. Moreover it poses the new question: what is the structure of these new 
states?  
The debate on exotic hadron structure was strongly revived during the  short 
``pentaquark  era''. Unfortunately, because of the uncertainties on the 
experimental 
side, it was very soon aborted. However, some interesting ideas were passed along  
to the subsequent discussion about the nature of the $X$, $Y$ and $Z$ states, 
which is
still in progress.

\subsection{New quark configurations}
  
Concerning the structure we can say that there are still attempts to interprete 
the 
new states as  $c - \bar{c}$. One can pursue this approach introducing 
corrections in 
the potential, such as quark pair creation. This ``screened potential'' changes 
the 
previous results, obtained with the unscreened potential and allows to 
understand some of
the new data in the  $c - \bar{c}$ approach \cite{lichao}. Departing 
from the   
$c - \bar{c}$ assignment, the next option  is a system composed by four quarks, 
which can 
be uncorrelated, forming a kind of bag, or can be grouped  in diquarks which 
then form a 
bound system \cite{maiani,maiani2,maiani3}. These configurations are called 
tetraquarks. 
Alternatively,  these four quarks can  form two mesons which then interact 
and form a bound 
state. If the mesons contain only one charm quark or antiquark, 
this configuration is referred to as a molecule \cite{torn,close,swan1}. 
If one of the 
mesons is a charmonium, then the configuration is called hadro-charmonium 
\cite{volz12}. 
Another possible configuration is a hybrid charmonium \cite{manke,luoliu}. 
In this case, apart from the 
$c - \bar{c}$ pair, the state contains excitations of the gluon field.  
In some 
implementations of the hybrid, the excited gluon field is represented by a 
``string'' or 
flux tube, which can oscillate in normal modes \cite{swan1}. 

The configurations mentioned in the previous paragraph are quite different 
and are governed 
by different dynamics. In quarkonia states the quarks have a short range 
interaction 
dominated by one gluon exchange and a long range non-perturbative confining 
interaction, 
which is often parametrized by a linear attractive potential. In tetraquarks 
besides these 
two types of interations, we may have a diquark-antidiquark interaction, 
which is not very well 
known. In molecules and hadro-charmonium the interaction occurs through  
meson exchange. 
Finally, in some models inspired by lattice QCD results, there is a flux 
tube formation 
between color charges and also string junctions. With these building blocks 
one can construct 
very complicated ``stringy'' combinations of quarks and antiquarks and their 
interactions 
follow the rules of string fusion and/or recombination 
\cite{suganuma,rich_steiner}. 
In principle, the knowledge of the interaction should be enough to determine 
the spatial 
configuration of the system. In practice, this is only feasible in simple 
cases, such as the 
charmonium in the non-relativistic approach, where having the potential one 
can solve the 
Schr\"odinger equation and determine the wave function. In other approaches 
the spatial 
configuration must be guessed and it may play a crucial role in the production 
and decay of 
these states. 

\subsection{New mechanisms of decay and production}

The theoretical study of production and decays of the new quarkonium states 
is still in the beginning. Some of the states are quite narrow and this is 
difficult to 
understand  in some of the theoretical treatments, as for example, in the  
molecular approach.

The narrowness of a  state  might be, among other possibilities,  a 
consequence of its 
spatial configuration. 
Assuming, for example, that the  $X(3872)$ is a set of $c \, 
\overline{c} \, u \, \overline{u}$ quarks confined 
in a bag but with sufficient (or nearly sufficient) energy to decay into the 
two mesons $D$ and $\overline{D}^{*}$, 
why is it so difficult for them to do so?  What mechanism hinders this 
seemingly simple coalescence process, which is 
observed in other hadronic reactions?   In a not so distant past the same 
questions was asked in the case of the 
$\Theta^{+}$ pentaquark. Now we know that the existence of this particle is, 
to say the least, not very likely.  Nevertheless, as 
mentioned in a previous subsection, some of the ideas 
advanced in the pentaquark years might be retaken now in a different context 
to answer the question raised above. In fact,  
the pentaquark decay $\Theta^{+} (u u d d \overline{s})  \rightarrow n 
(u d d) \, + \, K^{+} (u \overline{s})$  was 
energetically ``superallowed'', since there was enough phase space available 
and no process of quark pair creation nor 
annihilation.  Why was this quark rearrangement and hadronization difficult?  
It has been conjectured    \cite{stech} 
that the $\Theta^{+}$  was in a diquark ($u d$) - diquark ($u d$) - 
antidiquark ($\overline{s}$)  configuration, such 
that, due to the  not very well understood repulsive and/or attractive 
diquark interactions, the two diquarks were far 
apart from each other with the antiquark standing in the middle.  In this 
configuration it was difficult for one diquark 
to capture the missing $d$ quark (to form a neutron), which was very far.  
These words were implemented in a quantitative 
model and this configuration was baptized ``peanut'', since  the three bodies 
were nearly aligned. 
Another pentaquark spatial configuration with small decay width was the  
equilateral tetrahedron with four quarks located in 
the corners and the antiquark in the center \cite{bgs}. An  other interesting 
conjecture based on lattice arguments
was that, due to the dynamics of string rearrangement, the pentaquark  
constructed with three string junctions had, during 
the decay process, to pass through a very energetic configuration. Since its 
initial energy would be much less than that, this 
passage would  have to proceed through tunneling and would therefore be 
strongly  suppressed  \cite{suganuma}.

These ideas are relevant for the new charmonium physics because if they are 
multiquark states, as it seems to be the case, 
their spatial configuration will play a more important role both in 
production and decay.

The production of some of these states (such as, for example,  the $X$) has 
been observed 
both in $e^+ \, e^-$ and $p \, \bar{p}$ colliders, the latter being much more 
energetic. 
The $X$ produced in electron - positron collisions comes  from $B$ decays, 
whereas in  
$p \, \bar{p}$ reactions it must come  from a hard gluon splitting into a 
$c \, \bar{c}$ pair, 
which then undergoes some complicated fragmentation process. In theoretical 
simulations   \cite{grinstein}, 
it was shown that it is very difficult to produce $X$ if it is composed 
by two bound mesons.

Some reviews about these states can be found in the literature  
\cite{swanson,bauer,rosre,kz,zhure,ss,volre,egmr,olsen}. 
There are at least two reasons for writting a report on the subject. The 
first one is because this is a rapidly evolving field with enormous amount 
of new experimental information coming from the analysis of BELLE and BABAR 
accumulated data and also from CLEO and BES which are still running and 
producing new data. New data are also expected to come from the LHCb,
 which will start to operate 
soon and will be generating data with very high statistics for the next 
several years. In the present report we include and discuss 
data which were not yet available to the previous reviewers. 

The other reason 
which motivates us to write this report is that, on the theory side each 
review is biased and naturally emphasizes the approach followed by the authors. 
Thus, in Refs. \cite{swanson} and \cite{volre} the authors present the available 
data and then discuss their  theoretical interpretations making a sketch of the 
existing theories. In  \cite{swanson} attention is given to the conventional
quark antiquark potential model and to models of the interaction between mesons 
which form a molecular state.  In \cite{volre} a very nice overview of different 
theoretical approaches such as potential models, QCD sum rules and lattice QCD 
was presented. The author works out some pedagogical examples, using general 
considerations and simplified  assumptions. The present review is the first one 
written from the perspective of QCD sum rules, where we present the main steps 
of concrete calculations and compare the results with other approaches and with 
experimental data.

In the next sections we discuss the experimental data and the possible 
interpretations for the recently observed $X$, $Y$ and $Z$ mesons.

\markboth{\sl Interpretation of the New Charmonium States }{\sl Introduction}
\section{\label{SR} QCD sum rules }

\subsection{Correlation functions}

QCD sum rules are discussed in many reviews \cite{svz,rry,SNB,col}
emphasizing
various aspects of the method. The basic idea of the QCD sum rule formalism
is to approach the bound state problem in QCD from the asymptotic freedom
side, {\it i.e.}, to start at short distances, where the quark-gluon
dynamics is essentially perturbative, and move to larger distances
where the hadronic states are formed, including non-perturbative effects
``step by step'', and using some approximate procedure to extract
information on hadronic properties.

The QCD sum rule calculations are based on
the correlator of two hadronic currents. A generic two-point
correlation function is given by
\beq
\Pi(q)\equiv i\int d^4 x\, e^{iq\cdot x}
\lag{0}| T [j(x)j^\dagger(0)]|0\rag\ ,
\label{cor}
\enq
where $j(x)$ is a current with the quantum numbers of the hadron we want to
study.

The fundamental hypothesis of the sum rules approach is the assumption 
that there is an interval in $q$
over which the correlation function may be equivalently
described at both the quark and the hadron levels.
The former is called QCD or OPE side and the latter is called phenomenological
side. By matching the two sides of the sum rule we obtain information on
hadron properties. 

\subsection{The OPE side}

At the quark level the complex structure
of the QCD vacuum leads us to employ the Wilson's operator product
expansion (OPE) \cite{ope}.

In QCD we only know how to work analytically in the perturbative regime.
Therefore, the perturbative part of $\Pi(q)$ in Eq.(\ref{cor}) can
be reliably calculated. However, this does not yet imply that all
important contributions to the QCD side of the sum rule have been taken into
account. The complete calculation has to include the effects due to
the fields of soft gluons and quarks populating the QCD vacuum. A
practical way to calculate the vacuum-field contributions to the
correlation function is through a generalized Wilson OPE. To apply
this method to the correlation function (\ref{cor}), one
has to expand the product of two currents in a series of local operators:
\beq
\Pi(q)= i\int d^4 x\, e^{iq\cdot x}
\lag{0}| T[j(x)j^\dagger(0)|0\rag\ =\sum_nC_n(Q^2)\hat{O}_n\;,
\label{cope}
\enq
where the set $\{\hat{O}_n\}$ includes all local gauge invariant
operators expressible in terms of the gluon fields and the fields of light
quarks. Eq.~(\ref{cope}) is a concise form of the Wilson OPE. The
coefficients $C_n(Q^2)~(Q^2=-q^2)$, by construction, include only the
short-distance domain and can, therefore, be evaluated
perturbatively. Non-perturbative long-distance effects are contained
only in the local operators. In this expasion, the operators are
ordered according to their dimension $n$. The lowest-dimension
operator with $n=0$ is the unit operator associated with the perturbative
contribution: $C_0(Q^2)=\Pi^{per}(Q^2)$, $\hat{O}_0=1$. The QCD
vacuum fields are represented in (\ref{cope}) in the form of 
vacuum condensates. The lowest dimension condensates are the
quark condensate of dimension three: $\hat{O}_3=\langle\bar{q}q\rangle$,
and the gluon condensate of dimension four: $\hat{O}_4=
\langle g^2 G^2\rangle$. For non exotic mesons, {\it i.e.} normal 
quark-antiquark states, the contributions of condensates with
dimension higher than four are suppressed by large powers of 
$\Lambda_{QCD}^2/Q^2$, where $1/\Lambda_{QCD}$ is the typical long-distance 
scale. Therefore, even at intermediate values of $Q^2~(\sim1\GeV^2)$, the 
expansion in Eq.~(\ref{cope}) can be safely truncated after dimension four 
condensates. However, for molecular or tetraquark states, the 
mixed-condensate
of dimension five: $\hat{O}_5=\lag\bar{q}g\si.Gq\rag$, the four-quark 
condensate
of dimension six:  $\hat{O}_6=\lag\bar{q}q\bar{q}q\rag$ and even the
quark condensate times the mixed-condensate of dimension eight:
$\hat{O}_8=\lag\bar{q}q\bar{q}g\si.Gq\rag$, can play an important role.
The three-gluon condensate of dimension-six: $\hat{O}_6=\lag g^3 G^3\rag$ 
can be safely neglected, since it is suppressed by the loop factor 
$1/16\pi^2$. 

In the case of the (dimension six) four-quark condensate and the (dimension 
eight) quark condensate times the mixed-condensate, in general
factorization  assumption is assumed and  their vacuum saturation values 
are given by:
\beq
\lag\bar{q}q\bar{q}q\rag=\lag\bar{q}q\rag^2,\;\;\;\lag\bar{q}q\bar{q}
g\si.Gq\rag=\lag\bar{q}q\rag\lag\bar{q}g\si.Gq\rag.
\enq
Their precise evaluation requires more 
involved analysis including a non-trivial choice of factorization scheme 
\cite{BAGAN}. In order to account for deviations of the factorization
hypothesis, we will use the parametrization:
\beq
\lag\bar{q}q\bar{q}q\rag=\rho\lag\bar{q}q\rag^2,\;\;\;\lag\bar{q}q\bar{q}
g\si.Gq\rag=\rho\lag\bar{q}q\rag\lag\bar{q}g\si.Gq\rag,
\label{visa}
\enq
where $\rho=1$ gives the vacuum saturation values and $\rho=2.1$ indicates
the violation of the factorization assumption \cite{SNB,LNT,SNTAU}.

\subsection{The phenomenological side}

The generic correlation function in Eq.~(\ref{cor}) has a dispersion
representation
\beq
\Pi(q^2)=-\int ds\, {\rho(s)\over q^2-s+i\epsilon}\,+\,\cdots\,,
\label{disp}
\enq
through its discontinuity, $\rho(s)$, on the physical cut.
The dots in Eq.~(\ref{disp}) represent subtraction terms.

The discontinuity can be written as the imaginary part of the correlation
function:
\beq
\rho(s)={1\over\pi}Im[\Pi(s)]\;.
\enq
The evaluation of the spectral density ($\rho(s)$) is simpler than the
evaluation of the correlation function itself, and the knownledge of
$\rho(s)$ allows one to recover the whole function $\Pi(q^2)$ through the
integral in Eq.~(\ref{disp}).

The calculation of the phenomenological side proceeds by inserting
intermediate states for the hadron, $H$, of interest.
The current $j~(j^\dagger)$ is an operator that
annihilates (creates) all hadronic states that have the same quantum
numbers as $j$. Consequently, $\Pi(q)$ contains information about
all these hadronic states, including the low mass hadron of interest.
In order for the QCD sum rule technique to be useful, one must parametrize
$\rho(s)$ with a small number of parameters. The lowest resonance
is often fairly narrow, whereas higher-mass states are broader.
Therefore, one can parameterize the spectral density as a single sharp
pole representing the lowest resonance of mass $m$, plus a smooth continuum 
representing higher mass states:
\beq
\rho(s)=\lambda^2\delta(s-m^2) +\rho_{cont}(s)\,,
\label{den}
\enq
where $\lambda$ gives the coupling of the current with
the low mass hadron, $H$:
\beq
\lag 0 |
j|H\rag =\lambda.
\label{cou}
\enq

For simplicity, one often assumes that the continuum contribution to the
spectral density, $\rho_{cont}(s)$ in Eq.~(\ref{den}), vanishes bellow a
certain continuum threshold $s_0$. Above this threshold, it is assumed to be
given by the result obtained with the OPE.  Therefore, one uses the ansatz
\beq
\rho_{cont}(s)=\rho^{OPE}(s)\Theta(s-s_0)\;.
\enq
\subsection{Choice of currents}

Mesonic currents for open charm mesons are given in Table~\ref{tab1}.

\begin{table}[h]

  \begin{center}
    \caption{Currents for the $D$ mesons}
  \label{tab1}
    \begin{tabular}{|c|c|c|c|} \hline
      state    & symbol    & current  & $J^P$ \\ \hline
     scalar meson & $D_0$  &  $\bar{q}{c}$ & $0^+$ \\
     pseudoscalar meson & $D$ &  $i\bar{q}\gamma_5{c}$ & $0^-$ \\
     vector meson  & $D^*$ & $\bar{q}\gamma_\mu{c}$ & $1^-$ \\
     axial-vector meson & $D_1$ & $\bar{q}\gamma_\mu\gamma_5{c}$ & $1^+$ \\
\hline
     \end{tabular}
  \end{center}
\end{table}

From these currents we can construct molecular currents which can be
eigenstates of charge conjugation $C$ and $G$-parity. Let us consider, as
an example, a current with $J^{PC}=1^{++}$ for the molecular $D^0\bar{D}^{*0}$
system. It can be written as a combination of two currents
\cite{liuliu,stancu0}:
\beq
j_\mu^1(x)=[\bar{u}(x)\gamma_5 c(x)][\bar{c}(x)\gamma_\mu u(x)],
\enq
and
\beq
j_\mu^2(x)=[\bar{u}(x)\gamma_\mu c(x)][\bar{c}(x)\gamma_5 u(x)].
\enq
Since the charge conjugation  transformation is defined as: $(\bar{q})_C=
-q^TC^{-1}=q^TC$ and $(q)_C=C\bar{q}^T$, we get
\beq
(j_\mu^1)_C=-(\bar{c}\gamma_5u)(\bar{u}\gamma_\mu c)=-j_\mu^2,
\enq
\beq
(j_\mu^2)_C=-(\bar{c}\gamma_\mu u)(\bar{u}\gamma_5 c)=-j_\mu^1.
\enq
Therefore, the current
\beq
j_{\mu}(x) =  {1 \over \sqrt{2}}\left(j_\mu^1(x)-j_\mu^2(x)\right),
\label{molX}
\enq
has positive $C$. However, this current is not a $G$-parity eigenstate.
The $G$-parity transformation is an isospin rotation of the
charge conjugated current:
\beq
(j_\mu^1)_G=-(\bar{c}\gamma_5d)(\bar{d}\gamma_\mu c),
\enq
\beq
(j_\mu^2)_G=-(\bar{c}\gamma_\mu d)(\bar{d}\gamma_5 c).
\enq

In the case of a charged molecular $D_1\bar{D}^*$ current with $J^P=0^-$, it
can also be written as a combination  of two currents:
\beq
j_1=(\bar{c}\gamma_\mu\gamma_5u)(\bar{d}\gamma^\mu c),
\enq
\beq
j_2=(\bar{c}\gamma^\mu u)(\bar{d}\gamma_\mu\gamma_5c).
\enq
The charge conjugation transformation in these currents leads to
\beq
(j_1)_C=-(\bar{u}\gamma_\mu\gamma_5c)(\bar{c}\gamma^\mu d),
\enq
\beq
(j_2)_C=-(\bar{u}\gamma^\mu c)(\bar{c}\gamma_\mu\gamma_5d),
\enq
and the isospin rotation gives
\beq
(j_1)_G=(\bar{d}\gamma_\mu\gamma_5c)(\bar{c}\gamma^\mu u)=j_2,
\enq
\beq
(j_2)_G=(\bar{d}\gamma^\mu c)(\bar{c}\gamma_\mu\gamma_5u)=j_1.
\enq
Therefore, the current
\beq
j=  {1 \over \sqrt{2}}\left(j_1+j_2\right),
\enq
has positive $G$-parity.

In the case of tetraquark $[cq][\bar{c}\bar{q}]$ currents, they can be
constructed in terms of color anti-symmetric diquark states:
$\varepsilon_{abc}[q_{a}^TC\Gamma c_{b}]$,
where $a,~b,~c$ are color indices of the $SU(3)$ color group, $C$ is the
charge conjugation matrix, and
$\Gamma$ stands for Dirac matrices. The quantum numbers of the diquark
states are given in Table~\ref{tab2}.

\begin{table}[h]

  \begin{center}
    \caption{Currents for charmed diquark states.}
  \label{tab2}
    \begin{tabular}{|c|c|c|c|} \hline
      state        & current  & $J^P$ \\ \hline
     scalar diquark   &  $q_{a}^TC\gamma_5c_{b}$ & $0^+$ \\
     pseudoscalar diquark  &  $q_{a}^TCc_{b}$ & $0^-$ \\
     vector diquark   & $q_{a}^TC\gamma_5\gamma_\mu c_{b}$ & $1^-$ \\
     axial-vector diquark  & $q_{a}^TC\gamma_\mu c_{b}$ & $1^+$ \\
\hline
     \end{tabular}
  \end{center}
\end{table}

From these diquarks we can  construct  tetraquark
currents which can be eigenstates of charge conjugation $C$ and $G$-parity.
In the case of a $J^{PC}=1^{++}$ current it can be written as a combination
of scalar and axial-vector diquarks:
\beq
j_\mu^1=\epsilon_{abc}\epsilon_{dec}(q_a^TC\gamma_5c_b)(\bar{q}_d\gamma_\mu
C\bar{c}_e^T),
\label{j1}
\enq
and
\beq
j_\mu^2=\epsilon_{abc}\epsilon_{dec}(q_a^TC\gamma_\mu c_b)
(\bar{q}_d\gamma_5C\bar{c}_e^T).
\label{j2}
\enq
It is interesting to notice that the structure of the currents in
Eqs.~(\ref{j1}) and (\ref{j2}) relates the spin of the charm quark with the
spin of the light quark. This is very different from the spin structure of
the heavy quark effective theory \cite{isgur}. Heavy quark effective
theory, in leading order in 1/M, possesses a heavy-quark  spin symmetry.
Therefore, hadrons can be classified according to the angular momentum and
parity of the light fields only. This gives $0^-$ and $1^-$ mesons with
identical properties.

Using the charge conjugation transformations one gets:
\beq
(j_\mu^1)_C=\epsilon_{abc}\epsilon_{dec}(\bar{q}_a\gamma_5C\bar{c}_b^T)
(q_d^TC\gamma_\mu c_e)=j_\mu^2,
\enq
\beq
(j_\mu^2)_C=\epsilon_{abc}\epsilon_{dec}(\bar{q}_a\gamma_\mu C\bar{c}_b^T)
(q_d^TC\gamma_5c_e)=j_\mu^1.
\enq
Therefore, the current
\beq
j_{\mu}(x) =  {i \over \sqrt{2}}\left(j_\mu^1(x)+j_\mu^2(x)\right),
\label{tetraX}
\enq
has positive $C$. The $i$ was used in Eq.~(\ref{tetraX}) to insure that
$j_\mu^\dagger=j_\mu$. As in the case of the $D^0\bar{D}^{*0}$ molecular current,
the current in Eq.~(\ref{tetraX}) is not a $G$-parity eigenstate. However,
other combinations of tetraquark currents can be constructed to be
$G$-parity eigenstates.

In general,  there is no one to one correspondence between
the current and the state, since the current
in Eq.~(\ref{tetraX}) can be rewritten in terms of a sum over molecular type
currents through the Fierz transformation.  In the appendix, we provide 
the general expressions for the Fierz transformation of the tetraquark 
currents into molecular type of currents as given in Eq.~(\ref{molX}).
However, as shown in the appendix, in the Fierz transformation of a
 tetraquark current, each molecular component contributes with suppression 
factors that originate from picking up the correct Dirac and color indices.
This means that if the physical state is a molecular state, it would be 
best to choose a molecular type of current 
so that it has a large overlap with the physical state. Similarly for a 
tetraquark state it would be best to choose a tetraquark current.   If the 
current is found to have a large overlap with the physical state, the range
 of Borel parameters where the pole dominates over the continuum would be 
larger, and the OPE for the mass would have a better convergence.  These 
conditions will lead to a better sum rule; this means that the Borel curve 
has an extremum or is flat, and the calculated mass is close  to the 
physical value.
Therefore, if the sum rule gives a mass and width consistent with the 
physical values, we can infer that the physical state has a structure well 
represented by the chosen current. In this way, we can indirectly discriminate
 between  the tetraquark and the molecular structures of the recently 
observed states.  However, it is very important to notice that since
the molecular currents, as the one in Eq.~(\ref{molX}), are local, they do
not represent extended objects, with two mesons separated in space, but
rather a very compact object with two singlet quark-antiquark pairs.

When the current is fixed, we proceed by inserting it into Eq.~(\ref{cor}).
Contracting all the quark anti-quark pairs, we can rewrite the correlation 
function in terms of the quark propagators, and then we can perform the  
OPE expansion of these propagators. For the light quarks,  keeping terms which 
are linear in the light quark mass $m_q$,
this expansion reads: 
\beqa
S_{ab}(x)&=&\lag{0} T[q_a(x)\overline{q}_b(0)]\rag{0}={i\delta_{ab}\over2
\pi^2x^4}\xsla-{m_q\delta_{ab}\over4\pi^2x^2}
\nonumber\\
&-&{t^A_{ab}g G^A_{\mu\nu}\over32\pi^2}\left({i\over x^2}
(\xsla\sigma^{\mu\nu}+\sigma^{\mu\nu}\xsla)-{m_q}\sigma^{\mu\nu}\ln(-x^2)\right)
\nonumber\\
&-&{\delta_{ab}\over12}\qq
+{i\delta_{ab}\over48}m_q\qq\xsla-{x^2\delta_{ab}
\over2^6\times3}\mix
\nonumber\\
&+&{ix^2\delta_{ab}\over2^7\times3^2}m_q\mix\xsla,
\label{proleve}
\enqa
where we have used the fixed-point gauge. For heavy quarks, it is 
more convenient 
to work in the momentum space. In this case the expansion is given by: 
\beqa
S_{ab}(p)&=&
i \frac{\psla + m}{p^2 - m^2} \delta_{ab}
-{i\over4} \frac{t^A_{ab}g G^A_{\mu\nu} 
[\sigma^{\mu\nu} (\psla + m) + (\psla + m) \sigma^{\mu\nu}]}{(p^2 - m^2)^2}
\nonumber\\
&+&\frac{i \delta_{ab}}{12} m\Gd\frac{p^2 + m\psla}{(p^2 - m^2)^4}
\label{propesado}
\enqa

\subsection{The mass sum rule}

Now one might attempt to match the two descriptions of the correlator:
\beq
\Pi^{phen}(Q^2)\leftrightarrow\Pi^{OPE}(Q^2)\;.
\enq
However, such a matching is not yet practical. The OPE side is
only valid at sufficiently large spacelike $Q^2$. On the other hand, the
phenomenological description is significantly dominated by the
lowest pole only for sufficiently small $Q^2$, or better yet, timelike
$q^2$ near the pole. To improve the overlap between the two sides
of the sum rule, one applies the Borel transformation
%
\begin{equation}
{\cal B}_{M^2}[\Pi(q^2)]=
\lim_{\stackrel{\scriptstyle -q^2,n\rightarrow\infty}
{\scriptstyle -q^2/n=M^2}}
{(-q^2)^{n+1}\over n!}\left(d\over d q^2\right)^n
\Pi(q^2)\;.
\label{borel}
\end{equation}
%
Two important examples are:
\beq
{\cal B}_{M^2}\left[{q^2}^n\right]=0\,,
\label{po}
\enq
and
\beq
{\cal B}_{M^2}\left[\frac{1}{(m^2-q^2)^n}\right] = \frac{1}{(n-1)!}
\frac{e^{-m^2/M^2}}{(M^2)^{n-1}}\;,
\label{fra}
\enq
for $n>0$. From these two results, (\ref{po}) and  (\ref{fra}), one can
see that the Borel transformation removes the subtraction terms
in the dispersion relation, and
exponentially suppresses the contribution from excited resonances and
continuum states in the phenomenological side. In the OPE side the Borel
transformation suppresses the contribution from higher dimension condensates
by a factorial term.

After making a Borel transform on both sides of the sum rule, and
transferring the continuum contribution to the OPE side, the sum rule
can be written as
\beq \lambda^2e^{-m^2/M^2}=\int_{s_{min}}^{s_0}ds~
e^{-s/M^2}~\rho^{OPE}(s)\;. \lb{sr} \enq

If both sides of the sum rule were calculated to arbitrary high accuracy,
the matching would be independent of $M^2$. In practice, however,
both sides are represented imperfectly. The hope is that there exists
a range of $M^2$, called Borel window, in which the two sides have a good
overlap and information on the lowest resonance can be extracted.
In general, to determine the allowed Borel window, one analyses the OPE
convergence and the pole contribution: the minimum value of the Borel mass
is fixed by considering the convergence of the OPE, and the maximum value of
the Borel mass is determined by imposing the condition that the pole 
contribution should be bigger than the continuum contribution.

In order to extract the mass $m$ without worrying about the value of
the coupling $\lambda$, it is possible to take the derivative of 
Eq.~(\ref{sr}) with respect to $1/M^2$, and divide the result by 
Eq.~(\ref{sr}). This gives:
\beq
m^2={\int_{s_{min}}^{s_0}ds ~e^{-s/M^2}~s~\rho^{OPE}(s)\over\int_{s_{min}}^{
s_0}ds ~e^{-s/M^2}~\rho^{OPE}(s)}\;.
\lb{m2}
\enq
This quantity has the advantage to be less sensitive to the perturbative 
radiative corrections than the individual sum rules.
Therefore, we expect that our results obtained  to leading order 
in $\alpha_s$ will be quite accurate.

\subsection{Numerical inputs}

In the following sections we will present numerical results.
In the quantitative aspect, QCDSR is not  like a model which contains 
free parameters to be adjusted by fitting data. The inputs for numerical 
evaluations are the following: {\it i}) the vacuum matrix elements of composite 
operators involving quarks and gluons which appear in the operator product 
expansion. These numbers, known as  condensates, contain all the 
non-perturbative component of the approach. They could in principle, be 
calculated in lattice QCD. In  practice they are estimated 
phenomenologically.  They are universal and, once 
adjusted to fit, for example,  the mass of a particle, they must have always 
that same value. They are the analogue for spectroscopy of the parton 
distribution functions in deep inelastic scattering; {\it ii}) quark masses, 
which are extracted from many  different phenomenological analyses and used 
in our calculation; {\it iii})  the threshold parametr $s_0$ is the energy  
(squared) 
which characterizes the beginning of the continuum. Typically the quantity 
$\sqrt{s_0} - m$  (where $m$ is the mass of the ground state particle) is 
the  
energy needed to excite the  particle to its first excited state with the 
same 
quantum  numbers. This number is not well known, but should lie between 
$0.3$ and
$0.8$ GeV. If larger deviations from this interval are needed, the 
calculation becomes less reliable. 

All in all, in QCDSR  we do not have much freedom for choosing  numbers. In 
the calculations discussed in the next sections we will use 
\cite{SNB,SNCB,narpdg,SNG}: 
\beqa\label{qcdparam}
&m_s=(0.13\pm0.03)~\GeV,\nnb\\
&m_c(m_c)=(1.23\pm 0.05)\,\GeV,\nnb\\
&m_b(m_b)=(4.24\pm 0.06)\,\GeV,\nnb\\
&\lag\bar{q}q\rag=\,-(0.23\pm0.03)^3
\,\GeV^3,\nnb\\
&\lag\bar{s}s\rag=(0.8\pm0.2)\lag\bar{q}q\rag,\nnb\\
&\lag g^2G^2\rag=0.88~\GeV^4,\nnb\\
&\lag\bar{q}g\si.Gq\rag=m_0^2\lag\bar{q}q\rag,\;\;
m_0^2=0.8\,\GeV^2.
\enqa

\markboth{\sl Interpretation of the New Charmonium States }{\sl The $X(3872)$ meson}
\section{\label{X} The $X(3872)$ meson}

\begin{figure}[h]
\scalebox{0.45}{\includegraphics[angle=0]{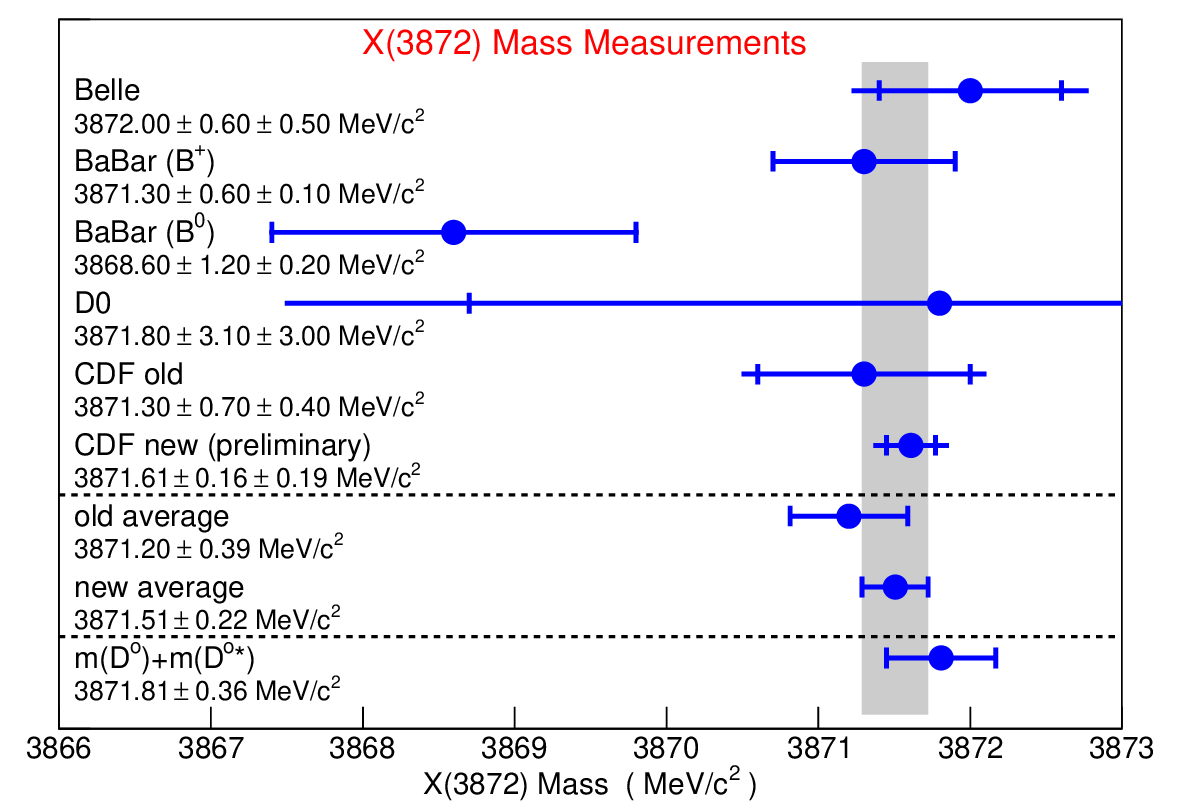}}
\caption{An overview of the measured $X(3872)$ masses from ref.~\cite{cdfx2}.
\label{massx}}
\end{figure}

Since its first observation in August 2003 by  Belle Collaboration
\cite{belle1},
the $X(3872)$ represents a puzzle and, up to now, there is
no consensus about its structure. The $X(3872)$ has been confirmed by CDF,
D0 and BaBar \cite{cdf,d0,babarx}. Besides the discovery production mode
$B^+\!\rightarrow\!X(3872)K^+\rightarrow\!J/\psi\pi^+\pi^- K^+$,
the $X(3872)$ has been observed in pronpt $p\bar{p}$ production 
\cite{cdf,d0}.
However, searches in prompt $e^+e^-$ production \cite{babaree} and in
$e^+e^-$ or $\gamma\gamma$ formation \cite{cleox} have given negative results
so far. The current world average mass is
\beq
M_X=(3871.20\pm0.39)\MeV\;,
\enq
and the most precise measurement to date is $M_X=(3871.61\pm0.16\pm0.19)
\MeV$, as can be seen by Fig.\ref{massx} \cite{cdfx2}.

The mass of the $X(3872)$ is at the threshold for the production of the
charmed
meson pair $m(D^0{\bar D}^{0\ast})=3871.81\pm0.36\MeV$ \cite{pdg}, and
this state is extremely narrow: its
width is less than 2.3 MeV at 90\% confidence level.

\subsection{Experiment versus theory}

Both Belle and Babar collaborations reported the radiative decay
mode $X(3872)\to \gamma J/\psi$ \cite{belleE,babar2}, which
determines $C=+$.  Belle Collaboration reported the branching ratio:
\beq            
\frac{\Gamma(X\to\,J/\psi\gamma)}{\Gamma(X\to
  J/\psi\,\pi^+\pi^-)}=0.14\pm0.05. 
\lb{gamarate} 
\enq

Recent studies from Belle and CDF that combine
angular information and kinematic properties of the $\pi^+\pi^-$ pair,
strongly favor the quantum numbers $J^{PC}=1^{++}$ \cite{belleE,cdf2,cdf3}.
In particular, in ref.~\cite{cdf3} it was shown that only the hypotheses
$J^{PC}=1^{++}$ and $2^{-+}$ are compatible with data. All other possible
quantum numbers are ruled out by more than three standard deviations. 
The possibility $2^{-+}$ is disfavored
by the  observation of the decay into $\psi(2S)\gamma$ \cite{babar09}
and also by the observation of the decays into
$D^0\bar{D}^0\pi^0$ by Belle and BaBar Collaborations \cite{belleD,babar3}.
On the other hand, the possibility $1^{++}$ is disfavored by the
observation of the decay into $J/\psi\omega$ by BaBar Collaborations 
\cite{babarate}.  
In the following we will asume the quantum numbers of the $X(3872)$
to be $1^{++}$.

From constituent quark models \cite{bg} the masses of the possible charmonium
states with $J^{PC}=1^{++}$ quantum numbers are: $2~^3P_1(3925)$ and
$3~^3P_1(4270)$, and lattice QCD calculations give $2~^3P_1(4010)$ \cite{lat1}
and $2~^3P_1(4067)$ \cite{lat2}. In all cases the predictons for the mass
of the $2~^3P_1$ charmonium state are much bigger than the observed mass.
However, a more recent lattice QCD calculation gives $2~^3P_1(3853)$
\cite{lat3} and, therefore, this interpretation can not be totally discarded
\cite{meng1}. In any case, the strongest fact against the $c\bar{c}$
assignment for $X(3872)$ \cite{bg,elq}, is the fact that from
the study of the dipion mass distribution in the $X(3872)\to J/\psi\pi^+\pi^-$
decay, Belle \cite{belle1} and CDF \cite{cdf2} concluded that it proceeds
through the $X(3872)\to J/\psi\rho^0$ decay.
Since a charmonium state has isospin zero, it can not decay easily into a
$J/\psi\rho$ final state.

The proximity between the $X$ mass and the $D^{*0}D^0$ threshold inspired
the proposal that the $X(3872)$ could be a molecular
bound state with small binding energy
\cite{torn,close,swan1,vol,wong,bira,faes}.
As a matter of fact, Tornqvist, using a meson potential
model \cite{torn2}, essentially predicted the $X(3872)$ in 1994, since he
found that there should be molecules near the $D^*\bar{D}$ threshold in the
$J^{PC}=0^{-+}$ and $1^{++}$ channels. The only other molecular state that is
 predicted in the potential model updated by Swanson \cite{swanson} is a
$0^{++}~~D^*\bar{D}^*$ molecule at 4013 MeV.

In ref.~\cite{swan1} Swanson proposed that the $X(3872)$ could be mainly a
$D^0\bar{D}^{*0}$ molecule with a small but
important admixture of $\rho J/\psi$ and $\omega J/\psi$ components.
With this picture the decay mode $X(3872) \to J/\psi\,\pi^+\pi^-\pi^0$
was predicted at a rate comparable to the $X(3872)\rightarrow\!J/\psi
\pi^+\pi^-$ mode. Soon after this prediction, Belle Coll.  \cite{belleE}
reported the observation of these two decay modes at a rate:
\beq
{{\cal B}(X \to J/\psi\,\pi^+\pi^-\pi^0)\over {\cal B}(X\to\!J/\psi\pi^+
\pi^-)}=1.0\pm0.4\pm0.3.
\label{rate}
\enq
This observation establishes strong isospin and G parity violation and
strongly favors a molecular assignment for $X$. However, it still does not
completely exclude a $c\bar{c}$ interpretation for $X$ since 
the isospin and G parity non-conservation in Eq.~(\ref{rate}) could be
of dynamical origin due to $\rho^0-\omega$ mixing \cite{tera} 
or even due to 
final state interactions (FSI) containing $D$ loops, such as 
$X \rightarrow J/\psi \omega \rightarrow D \bar{D} \rightarrow  J/\psi \rho$. 
Although all the ingredients 
(specially the charm form factors \cite{mnnr02,bcnn05})
for the relevant effective field theory are available, 
there are no quantitative results in the FSI approach yet. 

The decay $X \to J/\psi\omega$ was also observed by BaBar 
Collaboration \cite{babarate} at a rate:
\beq
{{\cal B}(X \to J/\psi\pi^+\pi^-\pi^0)\over {\cal B}(X\to\!J/\psi\pi^+
\pi^-)}=0.8\pm0.3,
\label{barate}
\enq
which is consistent with the result in Eq.~(\ref{rate}).

It is also important to notice that, although a $D^0\bar{D}^{*0}$ molecule is 
not
an isospin eingenstate, the ratio in Eq.~(\ref{rate}) can not be reproduced
by  a pure $D^0\bar{D}^{*0}$ molecule. In ref.~\cite{x24} it was shown that
for a pure $D^0\bar{D}^{*0}$ molecule
\beq
{\Gamma(X(D^0\bar{D}^{*0})\to J/\psi\,\pi^+\pi^-\pi^0)\over \Gamma(X(
D^0\bar{D}^{*0})\to J/\psi\,\pi^+\pi^-)}
\simeq0.15.
\label{rationum}
\enq

In refs.~\cite{belleD,babar3} Belle and BaBar Collaborations reported the
observation of a near threshold enhancement in the $D^0\bar{D}^0\pi^0$
system. The peak mass values for the two observations are in good agreement
with each other: $(3875.2\pm1.9)$ MeV for Belle and $(3875.1\pm1.2)$ MeV
for BaBar, and are higher than in the  mass of the $X(3872)$ observed in
the $J/\psi\pi^+\pi^-$ channel
by $(3.8\pm1.1)$ MeV. Since this peak lies about 3 MeV above the $D^{*0}
\bar{D}^0$ threshold, it is very awkward to treat it as a $D^{*0}\bar{D}^0$
bound state. According to Braaten \cite{braaten}, the larger mass of the $X$
measured in the $D^0\bar{D}^0\pi^0$ decay channel could be explained by the
difference between the line shapes of the $X$ into the two decays:
$D^0\bar{D}^0\pi^0$ and  $J/\psi\pi^+\pi^-$. In the decay of a narrow $X$ 
molecular state into its constituents $D^{*0}\bar{D}^0$, 
the width of $D^{*0}$ distorts the decay line shape of the $X(3872)$ 
\cite{bralu}. Therefore, the peak observed in
the $B\to K~D^0\bar{D}^0\pi^0$ decay channel could be  a combination of a
resonance below the $D^{*0}\bar{D}^0$ threshold from the
$B\to K~X$ decay and a threshold enhancement above the
$D^{*0}\bar{D}^0$ threshold. In this case, fitting the $D^0\bar{D}^0\pi^0$
invariant mass to a Breit-Wigner does not give reliable values for the mass
and width. However, in a new measurement \cite{belle0810}
Belle has obtained a mass $(3872.6\pm0.6)$ MeV in the
$\bar{D}^0{D}^{*0}$ invariant mass spectrum, which is consistent with the
current world average mass for $X(3872)$. Using this new data and taking into
account the universal features of the $S$-wave threshold resonance
Braaten and Stapleton  concluded that the $X(3872)$ is a extremely
weakly-bound charm meson molecule \cite{brasta}.

Other interesting possible interpration of the $X(3872)$, first proposed by
Maiani {\it et al.} \cite{maiani}, is that it could be a tetraquark state
resulting from the binding of a diquark and an antidiquark \cite{tera}. This 
construction
is based in the idea that diquarks can form bound-states, which can be
treated as confined particles, and used as degrees of freedom in parallel 
with quarks thenselves \cite{clo2,wilc,fried}. Therefore,
the tetraquark interpretation differs from the molecular interpretation in
the way that the quarks are organized in the state, as shown in Fig. 1.
\begin{figure}[h]
\scalebox{0.6}{\includegraphics[angle=0]{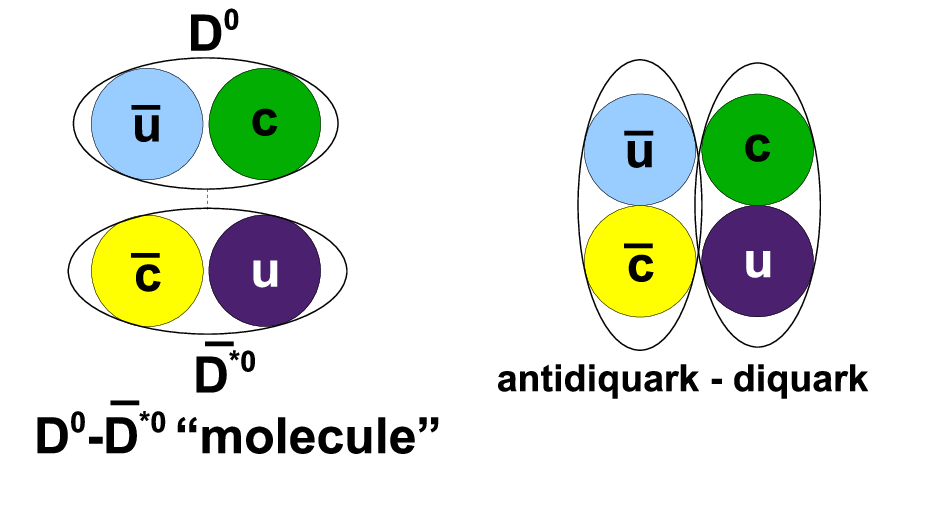}}
\caption{Cartoon representation for the molecular and tetraquark
interpretations of $X(3872)$.
\label{figure1}}
\end{figure}
The drawback of the tetraquark picture is the proliferation of the predicted
states \cite{maiani} and the lack of selection rules that could explain why
many of these states are not seen \cite{dfp}.

The authors of ref.~\cite{maiani} have considered diquark-antidiquark
states with $J^{PC}=1^{++}$  and symmetric spin distribution:
\beq
X_q=[cq]_{S=1}[\bar{c}\bar{q}]_{S=0}+[cq]_{S=0}[\bar{c}\bar{q}]_{S=1}.
\enq
The most general states that can decay into $2\pi$ and $3\pi$ are:
\beq
X_l=\cos{\theta}X_u+\sin{\theta}X_d,\;\;\;\;X_h=\cos{\theta}X_d-\sin{\theta}
X_u.
\enq
Imposing the rate in Eq.(\ref{rate}), they get $\theta\sim20^0$. It is
important to notice that a similar mixture between  $D^0\bar{D}^{*0}$
and $D^+D^{*-}$ molecular states with the same mixing angle $\theta\sim20^0$
\cite{x24}, would also reproduce the decay rate in Eq.(\ref{rate}).

The authors of ref.~\cite{maiani} also
argue that if $X_l$ dominates $B^+$ decays, then $X_h$ dominates the $B^0$
decays and vice-versa. They have also predicted that the mass difference
between the $X$ particle in $B^+$ and $B^0$ decays should be
\cite{maiani,polosa}
\beq
M(X_h)-M(X_l)=(8\pm3)\MeV.
\label{2X}
\enq

\begin{figure}[h]
\scalebox{0.45}{\includegraphics[angle=0]{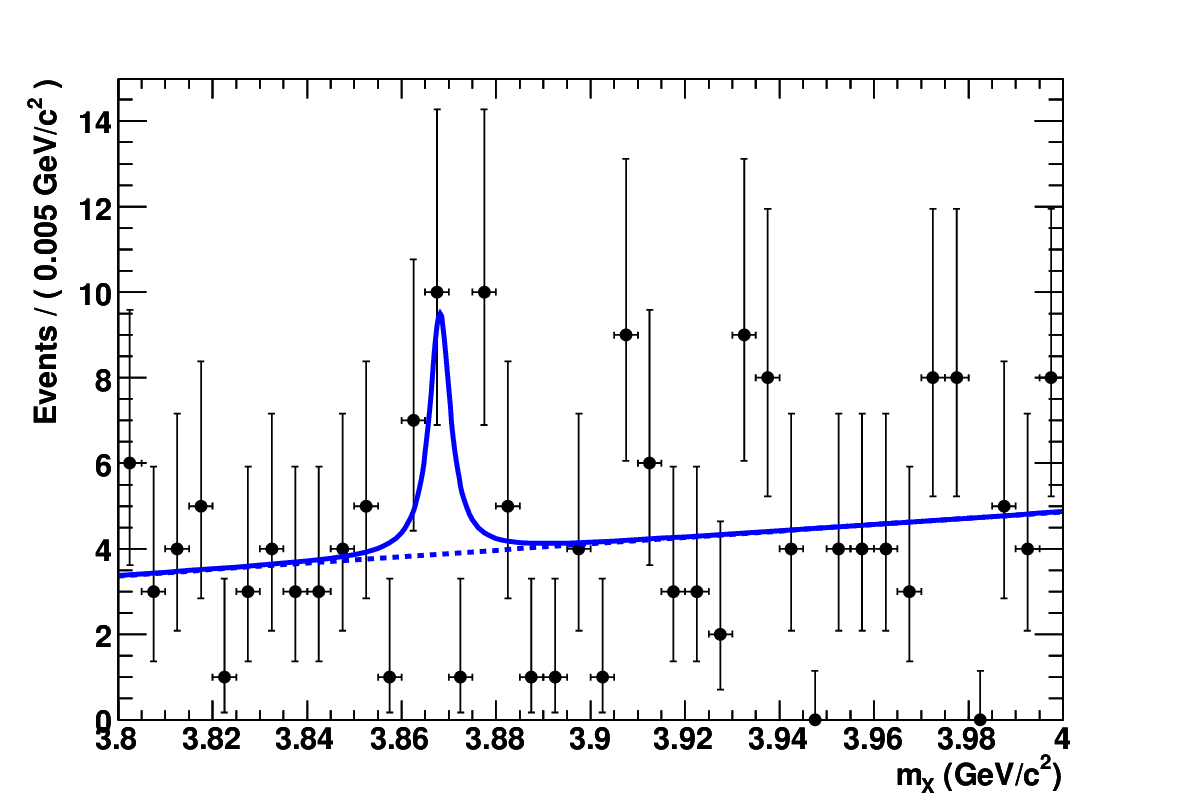}}
\scalebox{0.45}{\includegraphics[angle=0]{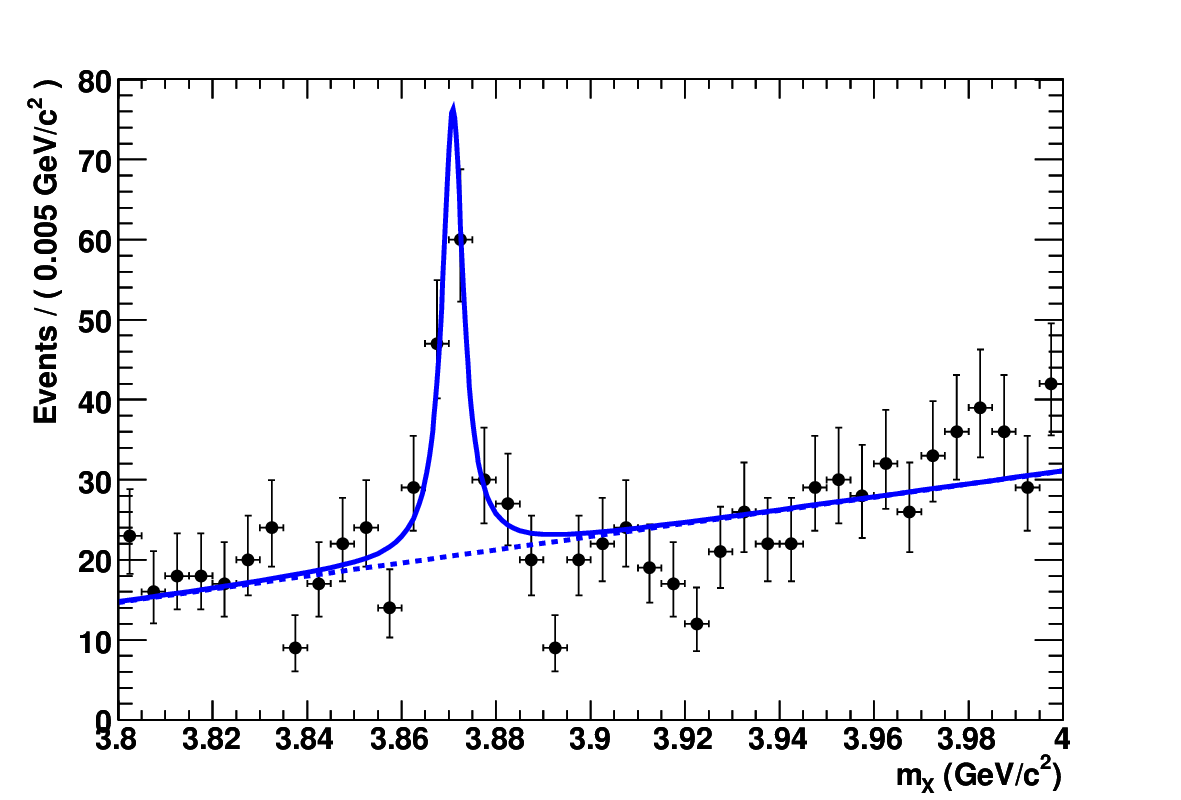}}
\caption{Fits to the $m(J/\psi\pi^+\pi^-)$ distribution for the decays
$B^0\to K^0~X$ (top) and $B^+\to K^+~X$ (botton) from ref.\cite{babarB0}.
\label{babarb0+}}
\end{figure}

\begin{figure}[h]
\scalebox{0.4}{\includegraphics[angle=0]{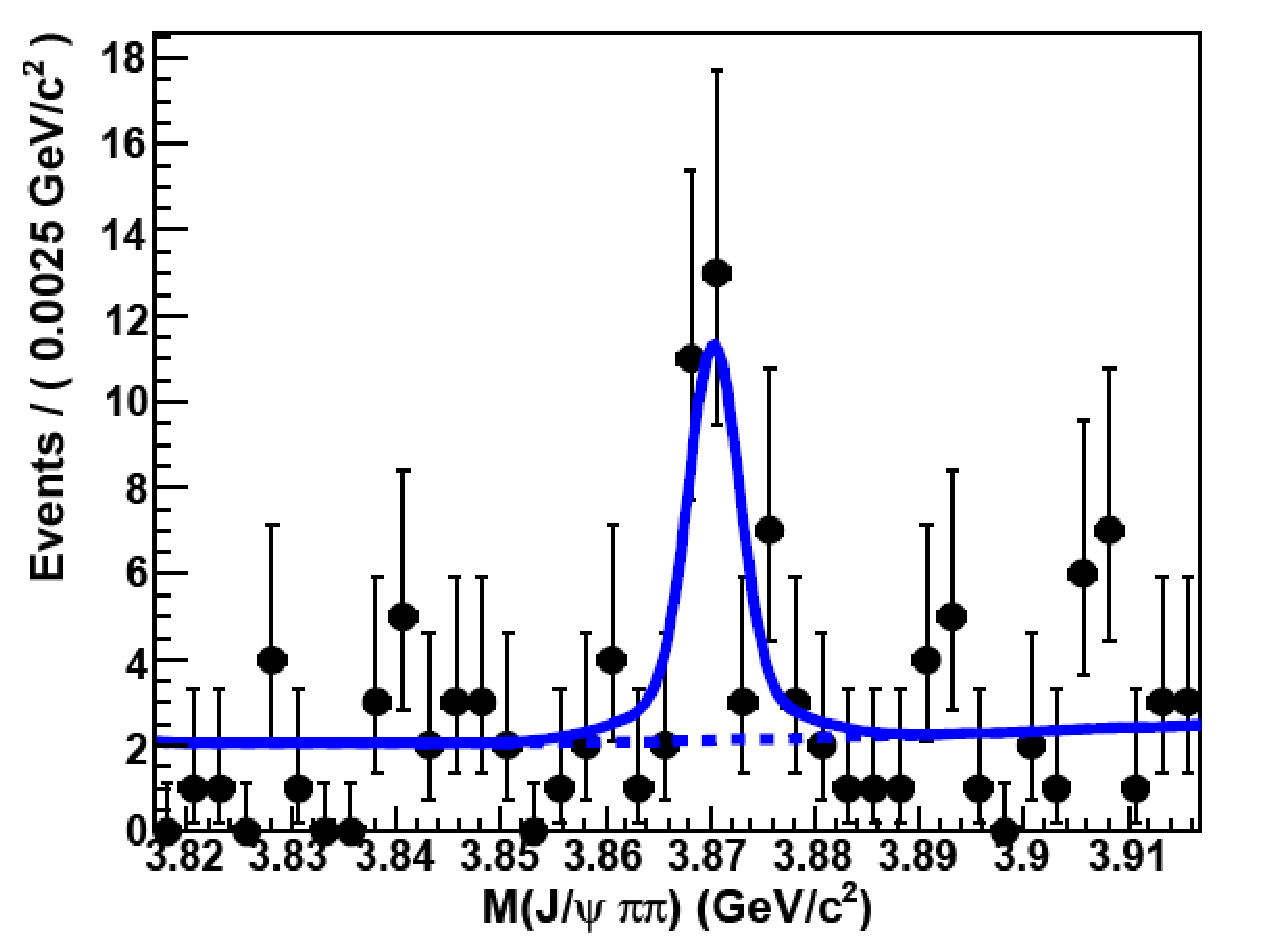}}
\scalebox{0.4}{\includegraphics[angle=0]{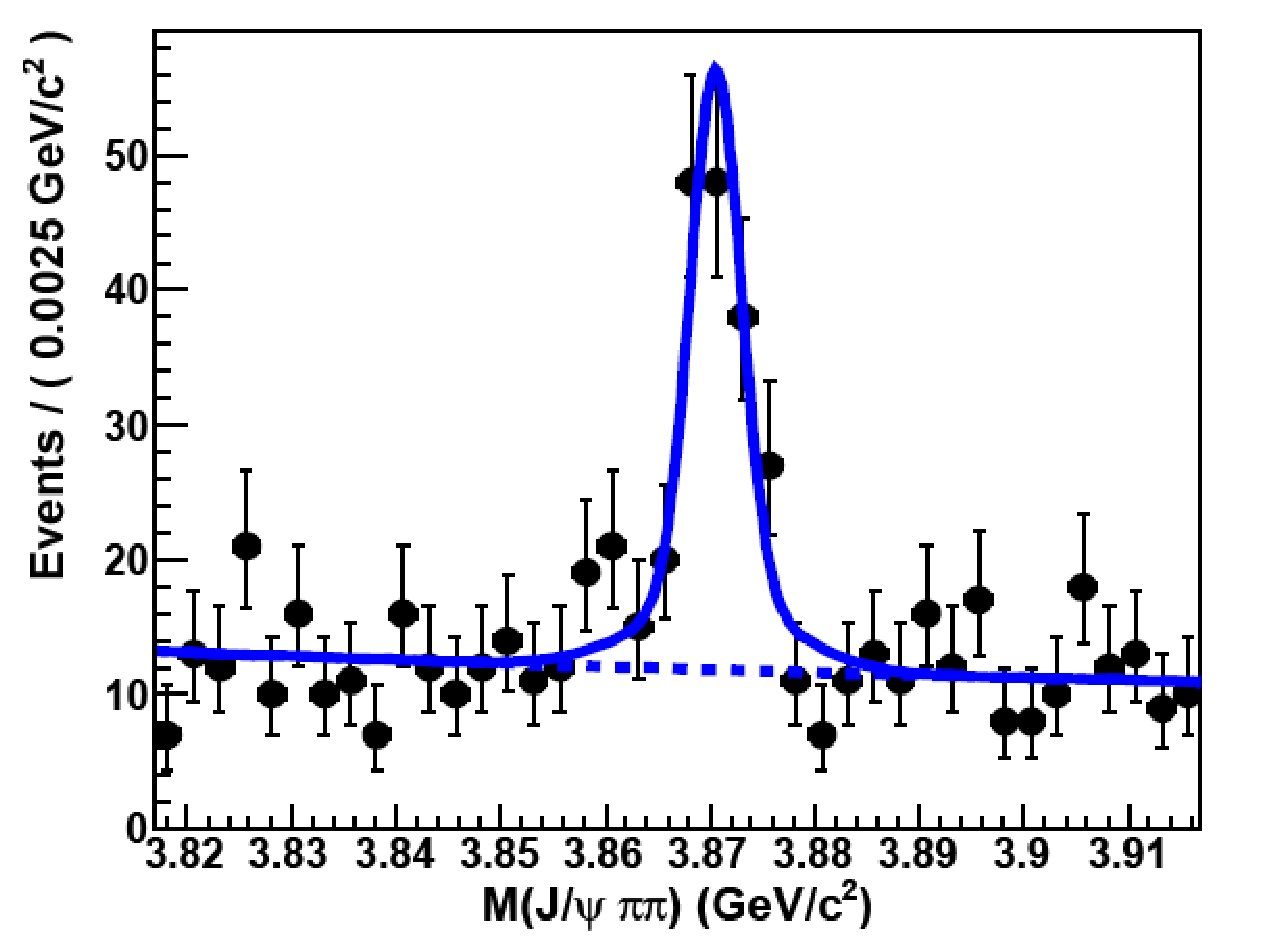}}
\caption{Fits to the $m(J/\psi\pi^+\pi^-)$ distribution for the decays
$B^0\to K^0~X$ (top) and $B^+\to K^+~X$ (botton) from ref.\cite{belleB0}.
\label{belleb0+}}
\end{figure}

There are two reports from Belle \cite{belleB0} and Babar  \cite{babarB0}
Collaborations for the observation of the $B^0\to K^0~X$
and $B^+\to K^+~X$ decays that we show in Figs. \ref{babarb0+} and
\ref{belleb0+}.

Although the identification of the $X(3872)$ from the decay $B^+\to K^+~X$
in these two figures is very clear, this is not the case for the
$B^0\to K^0~X$ decay, where the evidence for the existence of such a state is
not so clear.

In any case, if one accepts the existence of the $X$ in the decay
$B^0\to K^0~X$, these reports are not consistent with each other. While
Belle measures \cite{belleB0}:
\beq
{{\cal B}(B^0\to X K^0)\over {\cal B}(B^+\to X K^+)}=0.82\pm0.22\pm0.05,
\enq
and
\beq
M(X)_{B^+}-M(X)_{B^0}=(0.18\pm0.89\pm0.26)\MeV,
\label{difbel}
\enq
BaBar measures \cite{babarB0}:
\beq
{{\cal B}(B^0\to X K^0)\over {\cal B}(B^+\to X K^+)}=0.41\pm0.24\pm0.05,
\enq
and
\beq
M(X)_{B^+}-M(X)_{B^0}=(2.7\pm1.6\pm0.4)\MeV.
\label{difba}
\enq

In both measurements, the mass difference between the two states is much
smaller than the prediciton in Eq.(\ref{2X}).

If $X(3872)$ is a loosely bound molecular state, the branching ratio for the
decay $B^0\to K^0~X$ is suppresed by more than one order of magnitude
compared to the decay $B^+\to K^+~X$. The prediction in refs.
\cite{swanson,braaten2} gives:
\beq
0.06\leq{\Gamma(B^0\to X K^0)\over \Gamma(B^+\to X K^+)}\leq0.29,
\enq
which is, considering the errors, still consistent with the data from BaBar.

Recentely BaBar has reported the observation of the decay $X(3872)\to \psi(2S)
\gamma$ \cite{babar09} at a rate:
\beq
{{\cal B}(X \to \psi(2S)\,\gamma)\over {\cal B}(X\to\psi\gamma)}=3.4\pm1.4,
\label{rategaexp}
\enq
while the prediction from ref.~\cite{swan1} gives
\beq
{\Gamma(X \to \psi(2S)\,\gamma)\over\Gamma(X\to\psi\gamma)}\sim4\times10^{-3}.
\label{ratega}
\enq
While this difference could be interpreted as a strong point against the
molecular model and as a point in favor of a conventional charmonium
interpretation \cite{lichao}, it can also be  interpreted as an indication
that there is a significant  mixing of the $c\bar{c}$ component with
the $D^0\bar{D}^{*0}$ molecule. As a matter of fact, the necessity of
mixing a $c\bar{c}$ component with the $D^0\bar{D}^{*0}$ molecule was
already pointed out in some works \cite{elq,mechao,braaten3,suzuki,dong,li1}. 
In particular, in ref.~\cite{suzuki} it was phenomenologically shown that, 
because of the very loose binding of the molecule, the production rates of a 
pure molecule 
$X(3872)$ should be at least one order of magnitude smaller than what is seen
experimentally. 

In ref.~\cite{grinstein}, a Monte Carlo simulation of the production of a 
bound $D^0\bar{D}^{*0}$ state with binding energy as small as 0.25 MeV, 
obtained a cross section of about two orders of magnitude smaller than the 
prompt production cross section for the $X(3872)$ observed by the CDF 
Collaboration. The authors of ref.~\cite{grinstein} concluded that 
$S$-wave resonant scattering is unlikely to allow the formation of a 
loosely bound $D^0\bar{D}^{*0}$ molecule, thus calling for alternative 
(tetraquark) explanation of CDF data. However, it was pointed out in 
ref.~\cite{Artoisenet:2009wk} that a consistent analysis of $D\bar{D}^*$ 
molecule production requires taking into account the effect of final state 
interactions of the $D$ and $D^*$ mesons.  This observation changes the 
results of the Monte Carlo calculations bringing the theoretical value of 
the cross section very close to the observed one. 
Thus, the question of interpretation of $X(3872)$ production at hadronic 
machines is not yet settled. 

Other interpretations for the $X(3872)$ like cusp \cite{bugg0}, hybrids
\cite{li,closefrey}, or glueball \cite{seth} have already been
covered in refs.~\cite{swanson,bauer,rosre,kz,zhure,ss,volre,egmr}. Here we
would like to focus on the QCD sum rules studies of this meson.

\subsection{QCDSR studies of $X(3872)$}

Considering the $X(3872)$ as a $J^{PC}=1^{++}$ state we can construct a
current based on diquarks  in the
color triplet configuration, with symmetric spin distribution: $[cq]_{S=1}
[\bar{c}\bar{q}]_{S=0}+[cq]_{S=0}[\bar{c}\bar{q}]_{S=1}$, as proposed in
ref.~\cite{maiani}. Therefore, the corresponding
lowest-dimension interpolating operator for describing $X_q$ as a tetraquark
state is given by:
\beqa
j^{(q,di)}_\mu&=&{i\epsilon_{abc}\epsilon_{dec}\over\sqrt{2}}[(q_a^TC
\gamma_5c_b)(\bar{q}_d\gamma_\mu C\bar{c}_e^T)
\nn\\
&+&(q_a^TC\gamma_\mu c_b)
(\bar{q}_d\gamma_5C\bar{c}_e^T)]\;,
\label{cur-di}
\enqa
where $q$ denotes a $u$ or $d$ quark.

On the other hand, we can construct a current describing $X_q$ as a molecular
$D\bar{D}^*$ state:
\beqa
j^{(q,mol)}_{\mu}(x) & = & {1 \over \sqrt{2}}
\bigg[
\left(\bar{q}_a(x) \gamma_{5} c_a(x)
\bar{c}_b(x) \gamma_{\mu}  q_b(x)\right) \nonumber \\
& - &
\left(\bar{q}_a(x) \gamma_{\mu} c_a(x)
\bar{c}_b(x) \gamma_{5}  q_b(x)\right)
\bigg]
\lb{cur-mol}
\enqa

In general, other four-quark operators with $1^{++}$ are possible.  For 
example, starting from the simple charmed diquark states given in Table II, 
another tetraquark current with $J^{PC}=1^{++}$ can be constructed by 
combining the pseudo scalar $0^-$ and vector $1^-$ diquark.  Equivalently, 
additional current can be constructed for the meson type currents.  The 
number of currents increases further, if one allows for additional color 
states; color sextet for the diquark and color octet for the molecular 
states.  An extensive study has been carried out for the  $0^{++}$ light 
mesons\cite{chz}, with their mixing under
renormalizations \cite{TARRACH} from which one can form  renormalization
group invariant physical currents.  
 The choice of the current does not matter too much 
provided that we can work with quantities less affected by radiative
corrections and where the OPE converges quite well. This is borne out in the 
well-known case of
baryon sum rules, where a simple choice of operator \cite{io1} and a more 
general choice \cite{DOSCH} have been studied. Even 
though apparently different, mainly in the region of convergence of the 
OPE, the two choices of interpolating currents have provided the same 
predictions for the proton mass.
In some cases however, particular choices might be preferable over the 
others. 

\subsubsection{Two-point function}

For the present case, the two currents in Eqs.~(\ref{cur-di}) and 
(\ref{cur-mol}) were used,
in refs.~\cite{x3872} and \cite{lnw} respectively, to study the $X(3872)$.
The correlator for these currents can be written as:
\beqa
\Pi_{\mu\nu}(q)&=&i\int d^4x ~e^{iq.x}\lag 0
|T[j_\mu^{(q)}(x)j^{(q)\dagger}_\nu(0)]
|0\rag
\nn\\
&=&-\Pi_1(q^2)(g_{\mu\nu}-{q_\mu q_\nu\over q^2})+\Pi_0(q^2){q_\mu
q_\nu\over q^2},
\lb{2po}
\enqa
where, since the axial vector current is not conserved, the two functions,
$\Pi_1$ and $\Pi_0$, appearing in Eq.~(\ref{2po}) are independent and
have respectively the quantum numbers of the spin 1 and 0 mesons.

Using the current in Eq.~(\ref{cur-di}), as an example, the correlation 
function
in Eq.~(\ref{2po}) can be written in terms of the quark propagators as:
\beq
\Pi_{\mu\nu}(q)=-{i\varepsilon_{abc}\varepsilon_{a'b'c'}\varepsilon_{dec}
\varepsilon_{d'e'c'}\over2(2\pi)^8}\int~d^4xd^4p_1d^4p_2~e^{ix.(q-p_1-p_2)}
\bigg\{
\nn
\enq
\beq
\Tr\left[S^c_{bb'}(p_1)\gamma_5CS^{qT}_{aa'}(x)C
\gamma_5\right]\Tr\left[S^q_{d'd}(-x)\gamma_\mu CS^{cT}_{e'e}(-p_2)C
\gamma_\nu\right]+\nn
\enq
\beq
+\Tr\left[S^c_{bb'}(p_1)\gamma_\nu CS^{qT}_{aa'}(x)C
\gamma_\mu\right]\times\nn
\enq
\beq
\times\Tr\left[S^q_{d'd}(-x)\gamma_5CS^{cT}_{e'e}(-p_2)C
\gamma_5\right]\bigg\}.
\label{cor-pro}
\enq

In the OPE side, we work at leading order in $\alpha_s$ and consider the
contributions of condensates up to dimension eight. To evaluate the correlation
function in Eq.~(\ref{cor-pro}),
we use the momentum-space expression for the charm-quark
propagator given in Eq.~(\ref{propesado}).  The light-quark part of the 
correlation function is calculated in the coordinate-space, using the 
propagator given in Eq.~(\ref{proleve}). The resulting light-quark part
is combined with the charm-quark part before it is dimensionally
regularized at $D=4$.

The correlation function, $\Pi_1$, in the OPE side can be written as a
dispersion relation:
\beq
\Pi_1^{OPE}(q^2)=\int_{4m_c^2}^\infty ds {\rho(s)\over s-q^2}\;,
\lb{ope}
\enq
where the spectral density is given by the imaginary part of the
correlation function: $\pi \rho(s)=\mbox{Im}[\Pi_1^{OPE}(s)]$.
For the current in Eq.~(\ref{cur-di}) we obtain \cite{x3872}:
\beqa
\rho^{OPE}(s)&=&\rho^{pert}(s)+\rh^{m_q}(s)+\rh^{\qq}(s)+\rh^{\lag G^2\rag}
(s)\nn\\
&+&\rh^{mix}(s)+\rh^{\qq^2}(s)\;,
\lb{rhoeq}
\enqa
with
\beqa\label{eq:pert}
&&\rho^{pert}(s)={1\over 2^{10} \pi^6}\int\limits_{\almi}^{\alma}
{d\al\over\alpha^3}
\int\limits_{\bemi}^{1-\al}{d\be\over\be^3}(1-\al-\be)(1+\al\nn\\
&&
+\be)\left[(\al+\be)
m_c^2-\al\be s\right]^4,
\nn\\
&&\rho^{m_q}(s)=-{m_q \over 2^3 \pi^4} \int\limits_{\almi}^{\alma}
{d\al\over\al} 
\bigg\{ -{\qq\over 2^2}{[m_{c}^2-\al(1-\al)s]^2 \over (1-\al)}\nn\\
&&+ \int\limits_{\bemi}^{1-\al}{d\be\over\be}\kab \bigg[ 
- m_{c}^{2} \qq+{\qq\over 2^2} \left[(\al+\be)m_c^2\right.\nn\\
&-&\left.\al\be s\right]
+{m_c\over 2^5 \pi^2 \al \be^2}(3+\al+\be)(1-\al-\be)\kab^2 
\bigg] \bigg\},\nn\\
&&\rho^{\qq}(s)=-{m_c\qq\over 2^{5}\pi^4}\int\limits_{\almi}^{\alma}
{d\al\over\al^2}
\int\limits_{\bemi}^{1-\al}{d\be\over\be}(1+\al+\be)
\nn\\
&&\times\left[(\al+\be)m_c^2-
\al\be s\right]^2,\nn
\enqa
\beqa
&&\rho^{\lag
G^2\rag}(s)={\Gd\over2^{9}3\pi^6}\int\limits_{\almi}^{\alma} d\al\!\!
\int\limits_{\bemi}^{1-\al}{d\be\over\be^2}\left[(\al+\be)m_c^2-\al\be
s\right]
\nn\\
&&\times
\left[{m_c^2(1-(\al+\be)^2)\over\be}-
{(1-2\al-2\be)\over2\al}\left[(\al+\be)m_c^2-\al\be s\right]
\right], \nn
\enqa
\beqa
&&\rho^{mix}(s)={m_c\mix\over 2^{6}\pi^4}\int\limits_{\almi}^{\alma}
d\al
\bigg[-{2\over\al}(m_c^2-\al(1-\al)s)\nn\\
&&
+\int\limits_{\bemi}^{1-\al}d\be\left[(\al+\be)m_c^2-\al\be
s\right]\left({1\over
\al}+{\al+\be\over\be^2}\right)\bigg],
\enqa
where the integration limits are given by $\almi=({1-\sqrt{1-
4m_c^2/s})/2}$, $\alma=({1+\sqrt{1-4m_c^2/s})/2}$ and $(\bemi={\al
m_c^2)/( s\al-m_c^2)}$.
We have included the contribution of the dimension-six four-quark condensate:
\beqa
\rho^{\qq^2}(s)={m_c^2\rho\qq^2\over 12\pi^2}\sqrt{s-4m_c^2\over
s},
\label{dim6}
\enqa
and (for completeness) a part of the dimension-8 condensate contributions: 

\beqa
\Pi_1^{mix\qq}(M^2)&=&-{m_c^2\rho\mix\qq\over 24\pi^2}\int_0^1
d\al\,\bigg[1+{m_c^2
\over\al(1-\al) M^2}
\nn\\
&-&{1\over2(1-\al)}\bigg]\,\exp\!\left[{-{m_c^2
\over\al(1-\al)M^2}}\right].
\label{dim8}
\enqa

In Eqs.~(\ref{dim6}) and (\ref{dim8}) the parametrization in Eq.~(\ref{visa})
was assumed.

Parametrizing the coupling of the axial vector meson
$1^{++}$, $X$, to the current, $j_\mu^{(q)}$, in Eqs.~(\ref{cur-di}) and 
(\ref{cur-mol}) as:
\beq\label{decayla}
\lag 0 |
j_\mu^{(q)}|X\rag =\lambda^q\epsilon_\mu~, 
\enq
the phenomenological side
of Eq.~(\ref{2po}) can be written as 
\beq
\Pi_{\mu\nu}^{phen}(q^2)={|\lambda^{(q)}|^2\over
M_X^2-q^2}\left(-g_{\mu\nu}+ {q_\mu q_\nu\over M_X^2}\right)
+\cdots\;, \lb{phe} \enq
where the Lorentz structure projects out the $1^{++}$ state.  The dots
denote higher axial-vector resonance contributions that will be
parametrized, as explained in Sec.~2.3, through the introduction of a 
continuum threshold parameter $s_0$.
After making a Borel transform of both sides, and
transferring the continuum contribution to the OPE side, the sum rule
for the axial vector meson $X$ 
up to dimension-eight condensates can
be written as: 
\beq |\lambda^{(q)}|^2 e^{-M_X^2/M^2}=\int_{4m_c^2}^{s_0}ds~
e^{-s/M^2}~\rho^{OPE}(s)\; +\Pi_1^{mix\qq}(M^2)\;, \lb{sr} \enq
%

\begin{figure}[h]
\scalebox{0.8}{\includegraphics[angle=0]{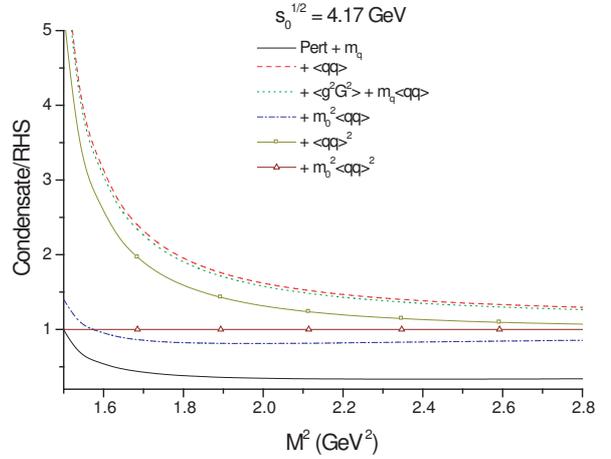}}
\caption{The $j^{(q-di)}_\mu$ OPE convergence in the region $1.6 \leq M^2 \leq
2.8~\GeV^2$ for $\sqrt{s_0} = 4.17 \GeV$ (taken from ref.\cite{x3872}).
\label{convx}}
\end{figure}
For the current in Eq.~(\ref{cur-di}) we show, in Fig.~\ref{convx},
the relative contribution of each term on the OPE expansion of
the sum rule. One can see that for $M^2 > 1.9 \GeV^2$, the  addition
of a subsequent term of the expansion brings  the curve (representing the
sum) closer to an asymptotic value (which was normalized to 1).
Furthermore the changes in  this  curve become smaller with
increasing dimension. These are the requirements for a good OPE convergence
and this fixes the lower limit of the Borel window to be $M^2 \geq 2\GeV^2$.

\begin{figure}[h]
\scalebox{0.45}{\includegraphics[angle=0]{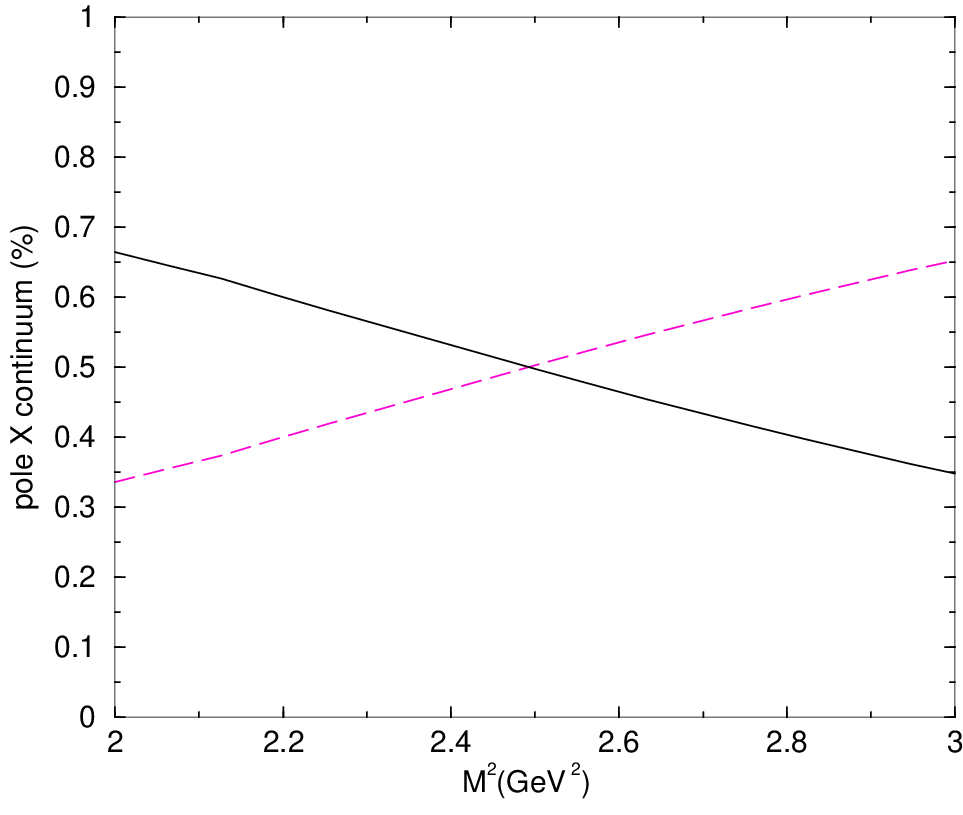}}
\caption{The solid line shows the relative pole contribution (the
pole contribution divided by the total, pole plus continuum,
contribution) and the dashed line shows the relative continuum
contribution for $\sqrt{s_0}=4.14$ GeV and $\rho=2.1$.
\label{pvcx}}
\end{figure}

We obtain an upper limit for $M^2$ by imposing the
constraint that the QCD continuum contribution should be smaller than the
pole contribution. The maximum value of $M^2$ for which this constraint is
satisfied depends on the value of $s_0$.  The comparison between pole and
continuum contributions for $\sqrt{s_0} = 4.15$ GeV  and $\rho=2.1$ is shown in
Fig.~\ref{pvcx}.

\begin{figure}[h]
\scalebox{0.45}{\includegraphics[angle=0]{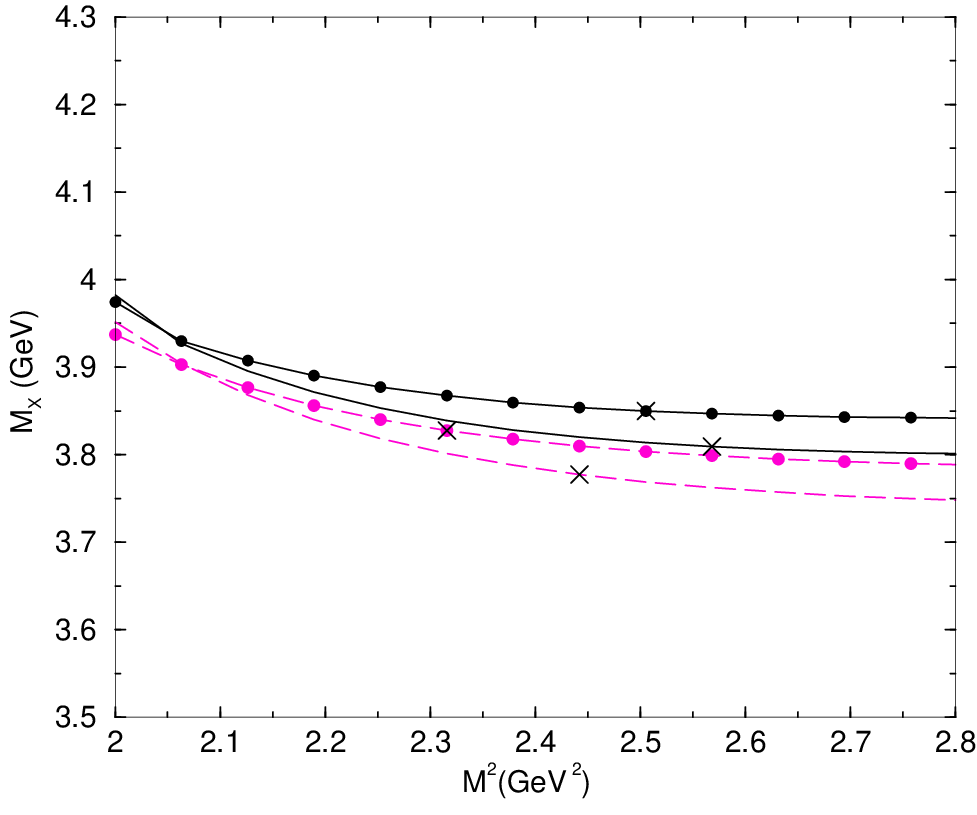}}
\caption{The $X$ meson mass as a function of the Borel parameter
for different values of $\sqrt{s_0}$ and $\rho$. The solid line and the solid 
line with dots show the results for $\sqrt{s_0} = 4.2$ GeV using $\rho=1$
and $\rho=2.1$ respectively. The dashed lines are the same for $\sqrt{s_0} = 
4.1$ GeV. The crosses indicate uper limit of the allowed Borel window.
\label{mxrho}}
\end{figure}

Having the Borel window fixed, the mass is obtained by using Eq.~(\ref{m2}).
 In Fig.~\ref{mxrho} we show the obtained $X$ mass for different values of
$\sqrt{s_0}$ and $\rho$. We can see by this figure that the effect of the
violation of the factorization assumption, given by $\rho$, is similar to
the effect of changing the continuum threshold.

\begin{figure}[h] 
\centerline{\epsfig{figure=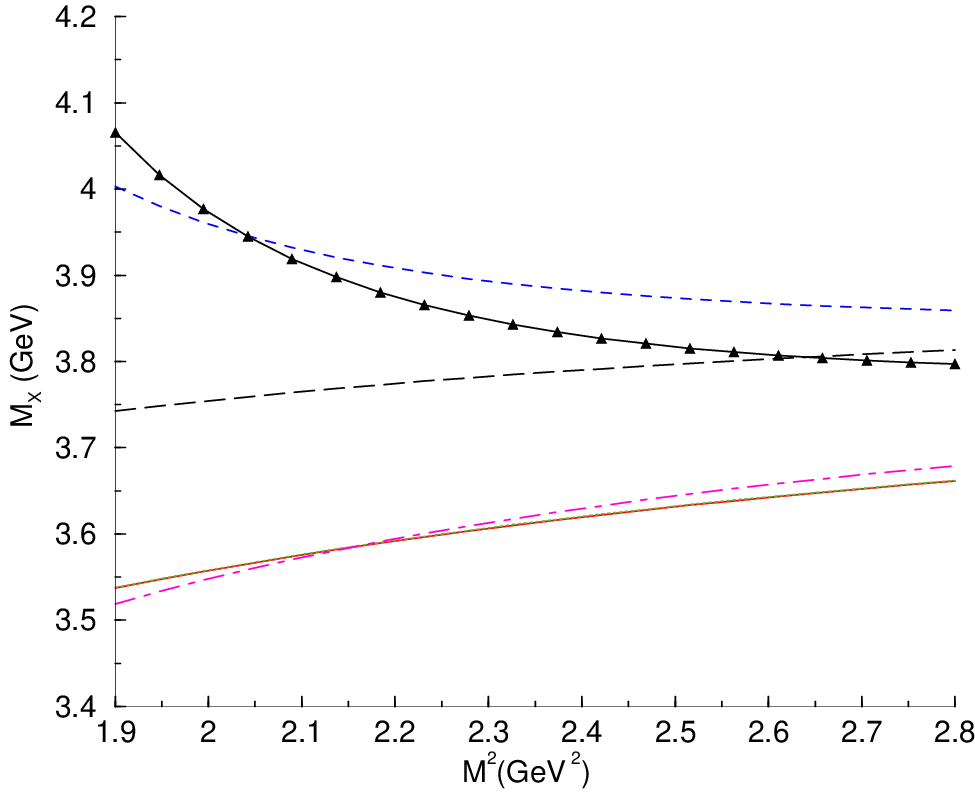,height=70mm}}
\caption{The OPE convergence for $M_X$ in the region $1.9 \leq M^2 \leq
2.8~\GeV^2$ for $s_0^{1/2} = 4.1$ GeV. We start with the perturbative
contribution plus a very small $m_q$ contribution (long-dashed line) and each
subsequent line represents the addition of one extra condensate of higher
dimension in the expansion: $+\langle \bar{q}q\rangle$ (solid line),
$+\langle g^2G^2\rangle+ m_q\langle \bar{q}q\rangle$ (dotted-line on top 
the solid line), $+ m_0^2\langle \bar{q}q
\rangle$ (dashed line),
$+ \langle \bar{q}q\rangle^2$ (dot-dashed line), $+ m_0^2\langle \bar{q}q
\rangle^2$ (solid line with triangles) (taken from ref.\cite{x3872}).}
\label{figconvmx} 
\end{figure} 

To study the effect of higher
dimension condensates in the mass we show, in Fig.~\ref{figconvmx}, the 
contributions of the individual condensates  to $M_X$ obtained from 
Eq.~(\ref{m2}). From this figure we see that the results oscillate around
the perturbative result, and that the results obtained up to dimension-5 are 
very close to the ones obtained up to dimension-8. For definiteness,  the 
value of 
$M_X$ obtained by including the dimension-5 mixed condensate will be considered
as the final  prediction from the SR, and
the effects of the higher condensates as the error due to the truncation of 
the OPE. 

The final result for $M_X$, obtained in \cite{x3872} considering the allowed 
Borel window and the uncertaities in the parameters, is
\beq
M_X=(3.92\pm0.13)~\GeV\;,
\enq
which is compatible with the experimental value of the mass of the $X(3872)$.
In ref.~\cite{x3872}
the mass difference in Eq.(\ref{2X}) was also evaluated giving:
\beq
M(X_h)-M(X_l)=(3.3\pm0.7)\MeV,
\enq
in agreement with the BaBar measurement in Eq.~(\ref{difba}).

\begin{figure}[h]
\scalebox{0.45}{\includegraphics[angle=0]{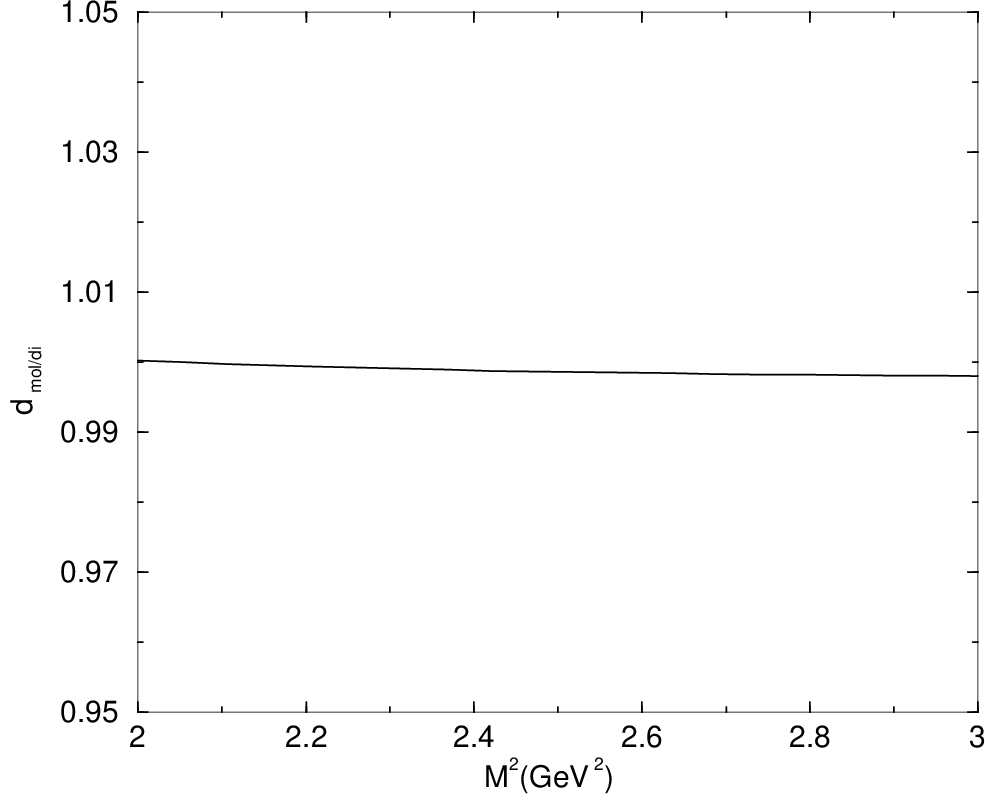}}
\caption{The double ratio $d_{mol/di}$ obtained from the QCDSR results for 
the $X$ mass using the currents in Eqs.~(\ref{cur-di}) and (\ref{cur-mol}).
\label{ratiox}}
\end{figure}

In the case of the current in Eq.~(\ref{cur-mol}), the OPE convergence
and the pole contribution determine a similar Borel window \cite{lnw}.
The result for the mass obtained in ref.~\cite{lnw} is $M_X=(3.87\pm0.07)$
GeV, in an even better agreement with the experimental mass. However, due to 
the uncertainties inherent to the QCDSR method, we can not really say that
the $X(3872)$ is better described with a molecular type of current than
with a tetraquark  type of current. To see that we show, in Fig.~\ref{ratiox},
the double ratio of the sum rules
\beq
d_{mol/di}={M_{X_{mol}}\over M_{X_{di}}},
\enq
where $M_{X_{mol}}$ and $M_{X_{di}}$ are the QCDSR results obtained by using the
currents in Eqs.~(\ref{cur-di}) and (\ref{cur-mol}). The ratio plotted in this
figure was obtained by using $\sqrt{s_0} = 4.15$ GeV and by considering the 
OPE up to dimension-6. 

We see from Fig.~\ref{ratiox} that the deviations between the two QCDSR
results in the allowed Borel windon are smaller than 0.01\%.

In ref.~\cite{width}
the same currents given in Eqs.~(\ref{cur-di}) and (\ref{cur-mol}) were used
to study the importance of including the width of the state, in a sum rule
calculation. This can be done by replacing the delta function in Eq.~(\ref{den})
by the relativistic Breit-Winger function:
\begin{equation}
 \delta(s-m^2) \to \frac{1}{\pi}\frac{\Gamma
  \sqrt{s}}{(s-m^2)^2+s\Gamma^2}.\label{eq:BW}
\end{equation}
The mass and width are determined by looking at the stability of the obtained
 mass against varying the  Borel parameter $M^2$, as usual.

\begin{figure}[h]
\scalebox{0.5}{\includegraphics[angle=0]{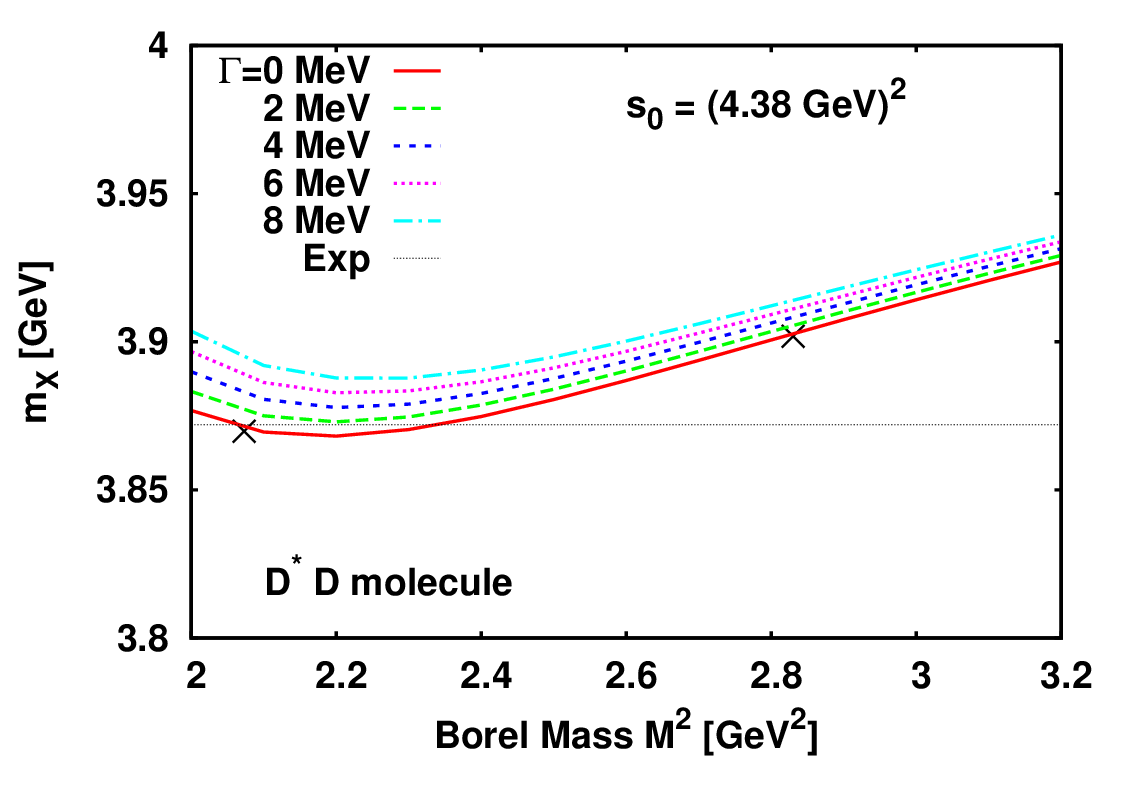}}
\caption{Results for the $X(3872)$ meson mass with a molecular current.
The crosses indicate lower and upper limit of the Borel window, respectively
(taken from ref.\cite{width}).
\label{widthmol}}
\end{figure}

Although the effect of the width is not large, in the case of the
molecular current, Eq.~(\ref{cur-mol}), it was possible
to fit the experimental mass, 3872 MeV, and the width
$\Gamma < 2.3$ MeV simultaneously with a continum threshold,
$\sqrt{s_0}=4.38$ GeV, as can be seen by Fig.~\ref{widthmol}.

\subsubsection{Three-point function}

One important question, when proposing a multiquark structure for the $X(3872)$, 
is whether with a tetraquark or molecular structure for the
$X(3872)$, it is possible to explain a total width smaller than 2.3 MeV.
As a matter of fact, a large partial decay width for the decay $X\to J/\psi
\pi^+\pi^-$ should be expected in this case.
The initial state already contains all the four quarks needed for the decay,
and there is no selection rules prohibiting the decay. Therefore, the decay
is allowed as in the case of the light scalars $\sigma$ and $\kappa$
studied in \cite{sca}, where widths of the order of 400 MeV were found.
The decay width $\Gamma(X\to J/\psi\pi\pi)$, with $X(3872)$ considered as a
tetraquark state, was also studied in ref.~\cite{maiani}. The authors
arrived at
$\Gamma(X\to J/\psi\pi\pi)\sim5~\MeV$. In order to have such small
decay width the authors had to make a bold guess about the order of magnitude
of the coupling constant in the vertex $XJ/\psi V$ (where $V$ stands for the
$\rho$ or $\omega$ vector meson): 
\beq
g_{X\psi V}=0.475.
\label{gma}
\enq

The decay width for the decay $X\to J/\psi V\to J/\psi F$ where $F=\pi^+\pi^-
(\pi^+\pi^-\pi^0)$ for $V=\rho(\omega)$ is given by \cite{maiani,decayx}
\beqa
&&{d\Gamma\over ds}(X\to J/\psi F)={1\over8\pi m_X^2}|{\cal{M}}|^2B_{V\to F}
\nn\\
&\times&
{\Ga_V m_V\over\pi}{p(s)\over(s-m_V^2)^2+(m_V\Ga_V)^2},
\label{de1}
\enqa
where
\beq
p(s)={\sqrt{\la(m_X^2,m_\psi^2,s)}\over2m_X},
\enq
with $\la(a,b,c)=a^2+b^2+c^2-2ab-2ac-2bc$. The
invariant amplitude squared is given by:
\beq
|{\cal M}|^2=g_{X\psi V}^2f(m_X,m_\psi,s),
\enq
where $g_{X\psi V}$ is the coupling constant in the vertex $XJ/\psi V$ and
\beqa
&&f(m_X,m_\psi,s)={1\over3}\left(4m_X^2-{m_\psi^2+s\over2}
+{(m_X^2-m_\psi^2)^2\over2s}\right.
\nn\\
&+&\left.{(m_X^2-s)^2\over2m_\psi^2}\right){m_X^2-m_\psi^2+s\over2m_X^2}.
\label{fm2}
\enqa
Therefore, the ratio in Eq.~(\ref{rate}) is given by:
\beq
{\Gamma(X\to J/\psi\,\pi^+\pi^-\pi^0)\over \Gamma(X\to J/\psi\,\pi^+\pi^-)}
={g_{X\psi\omega}^2m_\omega\Gamma_\omega B_{\omega\to\pi\pi\pi} I_\omega\over
g_{X\psi\rho}^2m_\rho\Gamma_\rho B_{\rho\to\pi\pi} I_\rho},
\label{ratioga}
\enq
where 
\beqa
I_V&=&\int_{(nm_\pi)^2}^{(m_X-m_\psi)^2}ds\bigg(f(m_X,m_\psi,s)
\nn\\
&\times&{p(s)\over(s-m_V^2)^2+(m_V\Ga_V)^2}\bigg).
\enqa

Using $B_{\omega\to\pi\pi\pi}=0.89$, $\Gamma_\omega=8.49\GeV$, $m_\omega=782.6
\MeV$, $B_{\rho\to\pi\pi}=1$, $\Gamma_\rho=149.4\GeV$ and $m_\rho=775.5
\MeV$ we get
\beq
{\Gamma(X\to J/\psi\,\pi^+\pi^-\pi^0)\over \Gamma(X\to J/\psi\,\pi^+\pi^-)}
=0.118\left({g_{X\psi\omega}\over g_{X\psi\rho}}\right)^2.
\label{rationum}
\enq

The coupling constant at the $XJ/\psi V$ vertex can be evaluated
directly  from a QCD sum rule calculation, which is based on the three-point 
correlation function:
\beq
\Pi_{\mu\nu\al}(p,\pli,q)=\int d^4x d^4y ~e^{i\pli.x}~e^{iq.y}
\Pi_{\mu\nu\al}(x,y),
\enq
with
\beq
\Pi_{\mu\nu\al}(x,y)=\lag 0 |T[j_\mu^{\psi}(x)j_{\nu}^{V}(y){j_\al^{X}}^
\dagger(0)]|0\rag,
\lb{3po}
\enq
where $p=\pli+q$ and the interpolating fields are given by:
\beqa
j_{\mu}^{\psi}&=&\bar{c}_a\gamma_\mu c_a,\nn\\
j_{\nu}^{\rho}&=&{1\over2}(\bar{u}_a\gamma_\nu u_a-\bar{d}_a\gamma_\nu
d_a),,\nn\\
j_{\nu}^{\omega}&=&{1\over6}(\bar{u}_a\gamma_\nu u_a+\bar{d}_a\gamma_\nu
d_a).
\enqa
For $j_{\al}^{X}$, it was shown in ref.~\cite{x24} that if we use
the currents in Eqs.(\ref{cur-di}) or (\ref{cur-mol}), the QCDSR result for the
ratio between the coupligs is given by:
\beq
{g_{X\psi \omega}\over g_{X\psi \rho} }=1.14.
\enq
Using this result in Eq.~(\ref{rationum}) we finally get
\beq
{\Gamma(X\to J/\psi\,\pi^+\pi^-\pi^0)\over \Gamma(X\to J/\psi\,\pi^+\pi^-)}
\simeq0.15.
\label{ratiofi}
\enq
Therefore, to be able to reproduce the experimental result in Eq.~(\ref{rate})
it is necessary to use a mixed current:
\beq
j_{\al}^{X}=\cos{\theta}j_\al^{(u)}+\sin{\theta}j_\al^{(d)},
\label{mixrate}
\enq
with $j_\al^{(q)}$  given by Eq.(\ref{cur-di}) for a tetraquark current or
by Eq.(\ref{cur-mol}) for a molecular current. Using the above definitions 
in Eq.(\ref{3po}), and considering the quarks $u$ and $d$ as degenerated,
 we arrive at
\beq
\Pi_{\mu\nu\al}(x,y)={-iN_V\over2\sqrt{2}}\left(\cos{\theta}~
+(-1)^{I_V}\sin{\theta}\right)~\Pi^q_{\mu\nu\al}(x,y),
\label{piAI}
\enq
where $N_\rho=I_\rho=1$ and  $N_\omega=1/3$, $I_\omega=0$.

To evaluate the phenomenological side of the sum rule we  
insert, in Eq.(\ref{3po}), intermediate states for $X$, $J/\psi$ and $V$. 
Using the definitions: 
\beqa
\lag 0 | j_\mu^\psi|J/\psi(\pli)\rag &=&m_\psi f_{\psi}\epsilon_\mu(\pli),
\nn\\
\lag 0 | j_\nu^V|V(q)\rag &=&m_{V}f_{V}\epsilon_\nu(q),
\nn\\
\lag X(p) | j_\al^X|0\rag &=&\la_X \epsilon_\al^*(p),
\lb{fp}
\enqa
where $\la_X=(\cos{\theta}+\sin{\theta})\la^q$, with $\lambda^q$ defined in 
Eq.~(\ref{decayla}),
we obtain the following relation:
\beqa
\Pi_{\mu\nu\al}^{(phen)} (p,\pli,q)={i(\cos\theta+\sin\theta)\lambda^q m_{\psi}
f_{\psi}m_Vf_{V}~g_{X\psi V}
\over(p^2-m_{X}^2)({\pli}^2-m_{\psi}^2)(q^2-m_V^2)}
\nn\\
\times\bigg(-\epsilon^{\al
\mu\nu\si}(\pli_\si+q_\si)-\epsilon^{\al\mu\si\ga}{\pli_\si q_\ga q_\nu
\over m_V^2}
-\epsilon^{\al\nu\si\ga}{\pli_\si q_\ga\pli_\mu\over m_\psi^2}
\bigg)+\cdots\;,
\lb{phen3}
\enqa
where the dots stand for the contribution of all possible excited states, 
and the coupling constant, $g_{X\psi V}$, is defined by the  matrix element, 
$\lag J/\psi V|X\rag$: 
\beq
\lag J/\psi(\pli) V(q)|X(p)\rag=g_{X\psi V}\epsilon^{\si\al\mu\nu}p_\si
\epsilon_\al(p)\epsilon_\mu^*(\pli)\epsilon_\nu^*(q),
\label{coup}
\enq
which can be extracted from the effective Lagrangian that describes the
coupling between two vector mesons and one axial vector meson \cite{maiani}:
\beq
{\cal{L}}=ig_{X\psi V}\epsilon^{\mu\nu\al\si}(\partial_\mu X_\nu)\Psi_\al
V_\si.
\enq

With the current in Eq.~(\ref{mixrate}), the ratio between the coupling 
constants is now given by \cite{x24}:
\beq
{g_{X\psi \omega}\over g_{X\psi \rho}}=1.14{\big(\cos\theta+\sin
\theta\big)\over \big(\cos\theta-\sin\theta\big)}.
\enq
 Using the previous result in Eq.~(\ref{rationum}) we get
\beq
{\Gamma(X\to J/\psi\,\pi^+\pi^-\pi^0)\over \Gamma(X\to J/\psi\,\pi^+\pi^-)}
\simeq0.15\left({\cos\theta+\sin\theta\over\cos\theta-\sin\theta}\right)^2.
\label{ratioal}
\enq
This is exactly the same relation obtained in refs.~\cite{maiani,decayx}, that
imposes $\theta\sim20^0$ to reproduce the experimental result in
Eq.(\ref{rate}). A similar relation was obtained in ref.~\cite{oset} where
the decay of the $X$ into two and three pions goes through a $D$ $D^*$ loop.

 Using $\theta=20^0$,
the $XJ/\psi\omega$ coupling constant was estimated from the sum
rule with a tetraquark current to be \cite{decayx}: 
\beq
g_{X\psi \om}=13.8\pm2.0,
\label{gnos}
\enq
which is much bigger than the number in Eq.(\ref{gma}), and leads to a
much bigger partial decay width:
\beq
\Ga(X\to J/\psi \pi^+ \pi^- \pi^0)=(50\pm15)~\MeV.
\label{width}
\enq
A similar width was obtained in ref.~\cite{x24} by using a molecular current
like the one in Eq.~(\ref{cur-mol}). Therefore, from a QCDSR calculation it
is not possible to explain the small width of the $X(3872)$ if it is a pure
four-quark state.

Considering the fact that also the relation in Eq.~(\ref{rategaexp}) can not
be reproduced with a pure molecular state, in ref.~\cite{x24} the $X(3872)$
was treated as a mixture between a $c\bar{c}$ current and a molecular
current, similar to the mixing considered in ref.~\cite{oka24} to study the
light scalar mesons:
\beq
J_{\mu}^q(x)= \sin(\alpha) j^{(q,mol)}_{\mu}(x) + \cos(\alpha)
j^{(q,2)}_{\mu}(x),
\lb{cur24}
\enq
with $j^{(q,mol)}_{\mu}(x)$ given in Eq.~(\ref{cur-mol}) and
\beq
j^{(q,2)}_{\mu}(x) = {1 \over 6 \sqrt{2}} \qq [\bar{c}_a(x) \gamma_{\mu}
\gamma_5 c_a(x)].
\enq

There is no problem in reproducing the experimental mass of the $X(3872)$
with this current for a large range of the mixture angle $\alpha$.  Considering
$\al$ in the region $5^\circ \leq \al \leq 13^\circ$ they get \cite{x24}:
\beq\label{sr.mx}
M_X = (3.77 \pm 0.18) \GeV,
\enq
which is in a good agreement with the experimental value. The value obtained
for the mass grows with the value of the mixing angle $\al$, but for 
$\al\geq30^\circ$ it reaches a stable value being completely determined
by the molecular part of the current. 

Considering
 also a mixture of $D^+D^{*-}$ and $D^-D^{*+}$ components,  the most general
current is given by
\beq
j_{\mu}^X(x)= \cos\theta J_{\mu}^u(x)+\sin\theta J_{\mu}^d(x),
\label{4mix}
\enq
with $J_{\mu}^u(x)$ and $J_{\mu}^d(x)$ given by Eq.~(\ref{cur24}).

Using the current in Eq.~(\ref{4mix}), the $XJ/\psi\omega$ coupling constant
obtained in ref.~\cite{x24} 
for $\theta=20^0$ and a mixing angle in Eq.~(\ref{cur24}) $\alpha=9^0\pm4^0$ 
is:
\beq
g_{X \psi \omega} = 5.4 \pm 2.4
\label{sr.coupl}
\enq 
which gives:
\beq
\Gamma\left(X\rightarrow J/\psi \pi^+ \pi^- \pi^0 \right) = (9.3 \pm 6.9)~\MeV.
\label{xpipipi}
\enq

The result in Eq.~(\ref{xpipipi}) is in  agreement with the experimental
upper limit. It is important to notice that the width and the mass grow with 
the mixing angle $\al$. Therefore, there is only a small range for the values
of this angle that can provide simultaneously good agreement with the
experimental values of the mass and the decay width, and this range is 
$5^\circ \leq \al \leq 13^\circ$. This means that the $X(3872)$ is basically 
a $c\bar{c}$ state with a small, but fundamental, admixture of molecular 
$DD^*$ states. By molecular states we mean an admixture between  
$D^0\bar{D}^{*0},~\bar{D}^0{D}^{*0}$ and $D^+D^{*-},~D^-D^{*+}$ states, as 
given by Eq.~(\ref{4mix}).

In ref.~\cite{gutsche} a similar mixture between a $c\bar{c}$ state and
molecular states (including $J/\psi\rho$ and $J/\psi\omega$) was considered
to study the $X(3872)$ decays into $J/\psi\gamma$ and $\psi(2S)\gamma$,
using  effective Lagrangians. In this approach the authors only needed a small
admixture of the $c\bar{c}$ component ( equivalent to $\alpha=78^0\pm2^0$
in Eq.~(\ref{cur24})) to reproduce the ratio in Eq.~(\ref{rategaexp}).
It is not clear, however, if with this small $c\bar{c}$ admixture it is 
possible to obtain the prompt production cross section for the $X(3872)$ 
observed by the CDF Collaboration \cite{grinstein}.

With the mixing angles $\alpha$ and $\theta$ fixed, it is possible
to  evaluate the 
width of the radiative decay $X(3872)\to J/\psi\gamma$, to check if it
is possible to reproduce the experimental result in Eq.~(\ref{gamarate}).
To do that one has just to consider the current in Eq.~(\ref{4mix}) for
the $X(3872)$, and exchange the $j_\nu^V$ current, in 
Eq.~(\ref{3po}), by the photon current $j_\nu^\gamma$:
\beq
j_{\nu}^{\gamma}=\sum_{q=u,d,c} e_q\,\bar{q}\gamma_\nu q\,,
\lb{jgamma}
\enq
with $e_q=\frac{2}{3}e$ for quarks $u$ and $c$, and $e_q=-\frac{1}{3}e$ 
for quark $d$  ($e$ is the modulus  of the electron  charge). 

The phenomenological side of the sum rule is given by \cite{niza}
\beqa
\Pi_{\mu\nu\al}^{\mathrm{phen}} (p,\pli,q)&=&-\frac{(\cos\theta+\sin\theta)
\lambda^q m_{\psi}f_{\psi}\epsilon_\mu(\pli)\epsilon_\al^*(p)}
{(p^2-m_{X}^2)({\pli}^2-m_{\psi})}
\nn\\
&\times&\langle\psi(\pli)\vert j_\nu^\gamma\vert X(p)\rangle
\,,
\lb{phenrad}
\enqa
where
\beqa\langle\psi(\pli)\vert
j_\nu^\gamma(q)\vert X(p)\rangle=i\,\epsilon_\nu^\gamma(q)\,
\mathcal{M}(X(p)\to\gamma(q)J/\psi(\pli))\,,
\enqa
with \cite{dong}
\beqa
&&\mathcal{M}(X(p)\to\gamma(q)J/\psi(\pli))= e\,\varepsilon^{\kappa
\lambda\rho\sigma}\epsilon_X^\alpha(p)\epsilon^\mu_\psi(p^\prime)
\epsilon^\rho_\gamma(q)\times\nn\\& &\times \frac{q_\sigma}{m_X^2}(A\,
g_{\mu\lambda}g_{\al\kappa}p\cdot q+B g_{\mu\lambda}p_\kappa q_\al+C 
g_{\al\kappa}p_\lambda q_\mu).
\lb{matrix}
\enqa
In Eq.~(\ref{matrix}), $A,~B$ and $C$ are  dimensionless couplings that are
determined by the sum rule. Using this  relation in
Eq.(\ref{phenrad}), the phenomenological  side of  the sum
rule becomes:
 \beqa  &&\Pi_{\mu\nu\al}^{\mathrm{phen}}   (p,\pli,q)=\frac{i  e
(\cos\theta+\sin\theta)  \lambda^q m_{\psi}f_{\psi}}{m_X^2(p^2-m_{X}^2)
({\pli}^2-m_{\psi})}
\nn\\
&\times&\bigg(\epsilon^{\al\mu\nu\si}q_\si
A+\epsilon^{\mu\nu\la\si}\pli_\la        q_\si       q_\alpha       B-
\epsilon^{\al\nu\la\si}q_\mu q_\si\pli_\la C
\nn\\
&+&\epsilon^{\al\nu\la\si}\pli_\la\pli_\mu      q_\si(C-A)\frac{p\cdot
  q}{m_\psi^2}
\nn\\
&                                     -&\epsilon^{\mu\nu\la\si}\pli_\la
q_\si(q_\al+\pli_\al)(A+B)\frac{p\cdot  q}{m_X^2}\bigg)\,.  \lb{phenmix}
\enqa

The values obtained in ref.~\cite{niza} for the couplings, using
$\theta=20^o$ and varying  $\alpha$ in the range $5^o\leq\al\leq13^o$
are:

\beqa
A&=&18.65\pm0.94\,,\nn\\
A+B&=&-0.24\pm0.11\,,\nn\\
C&=&-0.843\pm0.008\,.\lb{resabc}
\enqa

The decay width is given in terms of these couplings through:
\beq
\Gamma(X\to J/\psi~\gamma)=\frac{\alpha}{3}\frac{p^{*5}}{m_X^4}
\bigg((A+B)^2+\frac{m_X^2}{m_\psi^2}(A+C)^2\bigg)\,,\lb{width}\nn\\
\enq  
where $p^*=(m_X^2-m_\psi^2)/(2m_X)$. Using the result  for the decay width 
of the channel $J/\psi\pi^+\pi^-$ in Eq.~(\ref{xpipipi}): 
$\Gamma(X\to J/\psi~\pi\pi)=9.3\pm6.9\MeV$,  we get
\beq
\frac{\Gamma(X\to J/\psi~\gamma)}{\Gamma(X\to J/\psi~\pi^+\pi^-)}=0.19
\pm0.13\,,\lb{brfinal}
\enq
which is in complete agreement with the experimental result in 
Eq.~(\ref{gamarate}). Therefore, from a QCDSR point of view,
 the $X(3872)$ is a mixture of a $c\bar{c}$ state ($\sim$97\%) and
molecular $D^0\bar{D}^{*0},~\bar{D}^0{D}^{*0}$ ($\sim$2.6\%) 
and $D^+D^{*-},~D^-D^{*+}$ ($\sim$0.4\%) states. 

\subsection{Summary  for $X(3872)$}

To summarize, there is an emerging consensus that the $X(3872)$ is not a
pure $c\bar{c}$ state neither a pure multi-quark state. From the ratio in
Eq.~(\ref{rate}),  we know that $X(3872)$ is not an isospin
eigenstate, therefore, it can not be a pure $c\bar{c}$ state. On the other
hand, the binding energy, the production rates and the observed
ratio in Eq.~(\ref{rategaexp}) are not compatible with a pure molecular state.
Considering all the available experimental information, it is very probable
that the $X(3872)$ is a admixture of a charmonium state with other
multi-quark states: molecular or tetraquark states.

\subsection{Predictions for $X_b$, $X^s$, $X^s_b$}

It is straightforward to extend the analysis done for the $X(3872)$ to the
case of the bottom quark. Using the same interpolating field of
Eq.~(\ref{cur-di}) with the charm quark replaced by the bottom one, the
analysis done for $X(3872)$ was repeated for $X_b$ in ref.\cite{x3872}.
In this case there is also a good Borel window and the prediction
for the mass of the state that couples with a tetraquark $(bq)(\bar{b}\bar{q}
)$ with $J^{PC}=1^{++}$ current is:
\beq
\lb{massXb}
M_{X_b}=(10.27\pm0.23)~\GeV~.
\enq

The central value in Eq.~(\ref{massXb}) is close to the mass of
$\Upsilon(3S)$, and appreciably below the $B^*\bar{B}$ threshold at
about $10.6\;$GeV. For comparison, the molecular model predicts for
$X_b$ a mass which is about $50-60\;$MeV below this threshold
\cite{swanson}, while a relativistic quark model without explicit
$(b\bar{b})$ clustering predicts a value of about 133 MeV below this
threshold \cite{EBERT}.  A future
discovery of this state, e.g. at LHCb, will certainly test the different
theoretical models of this state and clarify, at  the same time, the
nature of the $X(3872)$.

In the case of $X^s$ ($[cs][\bar{c}\bar{s}]$) and $X^s_b$ ($[bs][\bar{b}
\bar{s}]$), one has to replace the light quarks in the currents for $X$
and $X_b$ by strange quarks.
To extract the relatively small mass-splitting, it is appropriate
to use the double ratio of moments \cite{SNB,SNBB}:
\beq\label{doubleratio}
{d}_{c}^s\equiv  {M^2_{X^s}\over M^2_X},
\enq
which suppresses different systematic errors and the dependence on the sum
rule parameters like $s_0$ and  $M^2$. The result obtained for this ratio in
ref.~\cite{x3872} is:
\beq
\sqrt{d^s_c}= 0.984\pm 0.007~.
\enq
This leads to the mass splitting:
\beq
M_{X^s}-M_X\simeq-(61\pm 30)~\MeV~.
\label{mxs}
\enq
Similar methods used in \cite{SNB,SNBB} have predicted successfully the
values of $M_{D_s}/M_D$ and $M_{B_s}/M_B$, which is not quite surprising,
as in the double ratios, all irrelevant sum rules systematics cancel out.

It is interesting to notice that the $X^s$ mass prediction from
ref.~\cite{x3872} is slightly smaller
than the $X(3872)$ mass, which is quite unusual. Such a small and negative
mass-splitting is rather striking and needs to be checked using alternative
methods. The (almost) degenerate value of the $X$ and of the $X^s$
masses may suggest that the physically observed $X(3872)$ state can also have
a $c\bar c s\bar s$ component.

A similar analysis can be done for the $X_b^s~([bs][\bar{b}\bar{s}])$ giving
\cite{x3872}:
\beq
\sqrt{d_b^S}\equiv  {M_{X^s_b}\over M_{X_b}}=0.988\pm 0.018~,
\enq
and
\beq
{M_{X^s_b}- M_{X_b}} = -(123\pm 182) ~{\rm MeV}~.
\label{eq:split3b}
\enq

We expect that the $X_b$-family
will show up at LHCb in the near future, which will serve as a test of this
predictions.

\markboth{\sl Interpretation of the New Charmonium States }{\sl The $Y(J^{PC}=
1^{--})$ family}
\section{\label{Y} The $Y(J^{PC}=1^{--})$ family}

\subsection{Experiment versus theory}
 
The $e^+e^-$ annihilation through initial state radiation (ISR) is a 
powerful tool to search for $1^{--}$ states at the $B$-factories.
The first state in the $1^{--}$ family discoverd using this process was the 
$Y(4260)$. It was first observed in 2005, by the BaBar 
Collaboration \cite{babar1}, in the reaction:
\beq
e^+e^-\to\gamma_{ISR}J/\psi\pi^+\pi^-,
\enq
with mass $M=(4259\pm10)\MeV$ and width $\Gamma=(88\pm24)\MeV$. The 
$Y(4260)$ was confirmed by CLEO and Belle Collaborations \cite{yexp}. 

\begin{figure}[h]
\scalebox{0.4}{\includegraphics[angle=0]{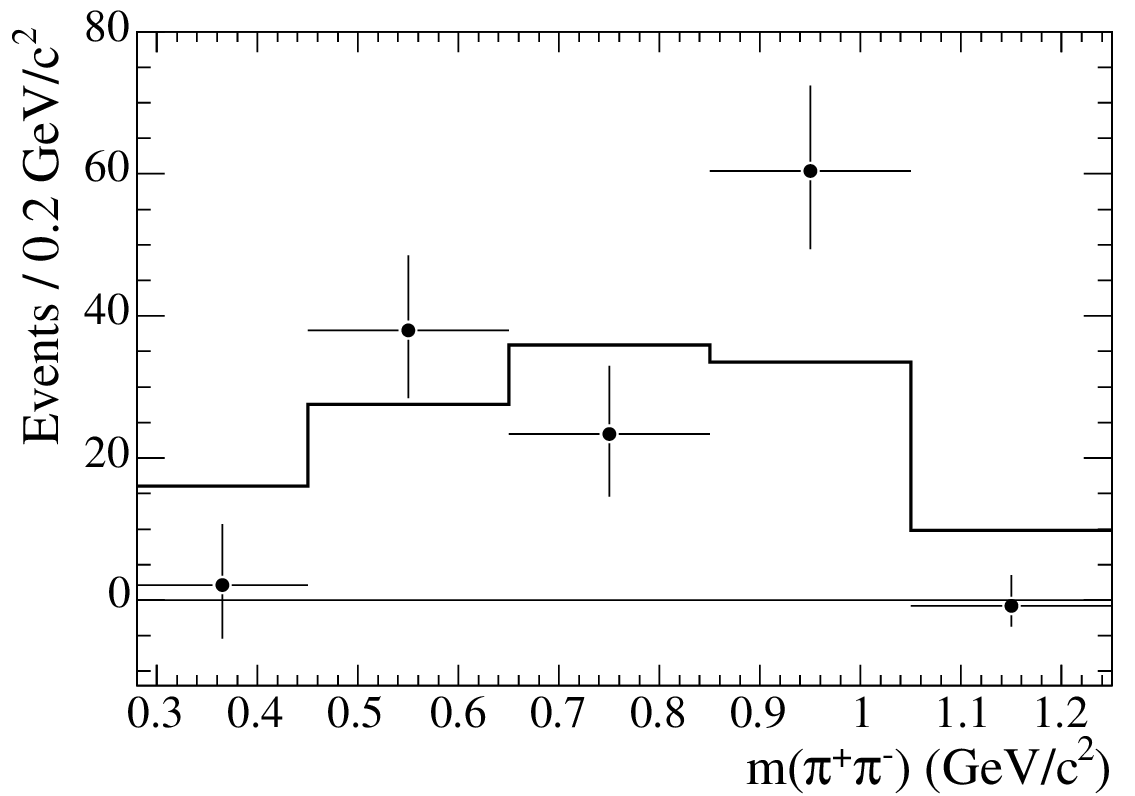}}
\caption{Dipion mass distribution for $Y(4260)\to J/\Psi\pi^+\pi^-$ from 
ref.~\cite{babar1}.
\label{figurey1}}
\end{figure}

The BaBar  Collaboration also reported a $\pi\pi$ mass distribution that 
peaks 
near 1 GeV, as can be seen in Fig.~\ref{figurey1} \cite{babar1}, and this 
information was interpreted as being consistent with the $f_0(980)$ decay. 

The $Y(4260)$ was also observed in the 
$B^-\to Y(4260)K^-\to J/\Psi\pi^+\pi^-K^-$ decay \cite{babary2}, and CLEO
reported two additional decay channels: $J/\Psi\pi^0\pi^0$ and
$J/\Psi K^+K^-$ \cite{yexp}. 

\begin{figure}[h]
\scalebox{0.4}{\includegraphics[angle=0]{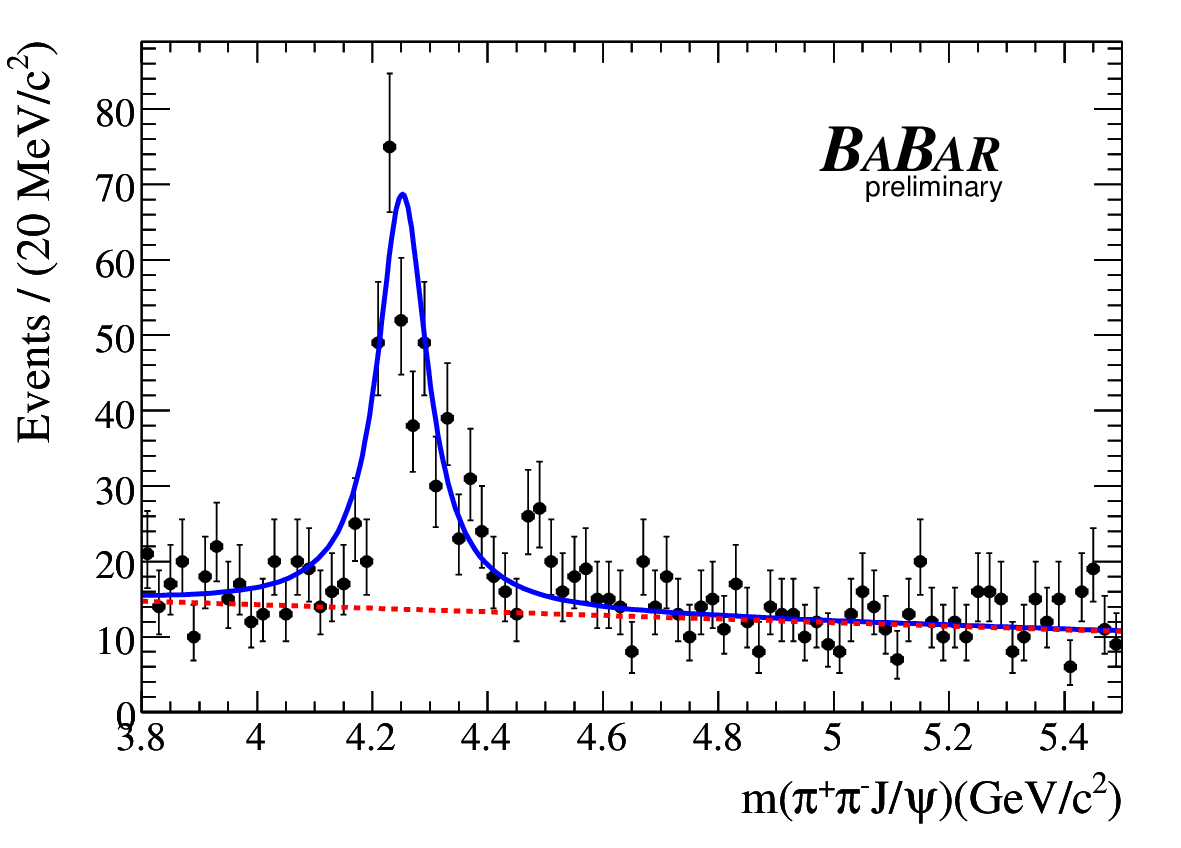}}
\caption{The $\pi^+\pi^-J/\Psi$ invariant mass spectrum from 
ref.~\cite{babary}.
\label{figurey2}}
\end{figure}

\begin{figure}[h]
\scalebox{0.4}{\includegraphics[angle=0]{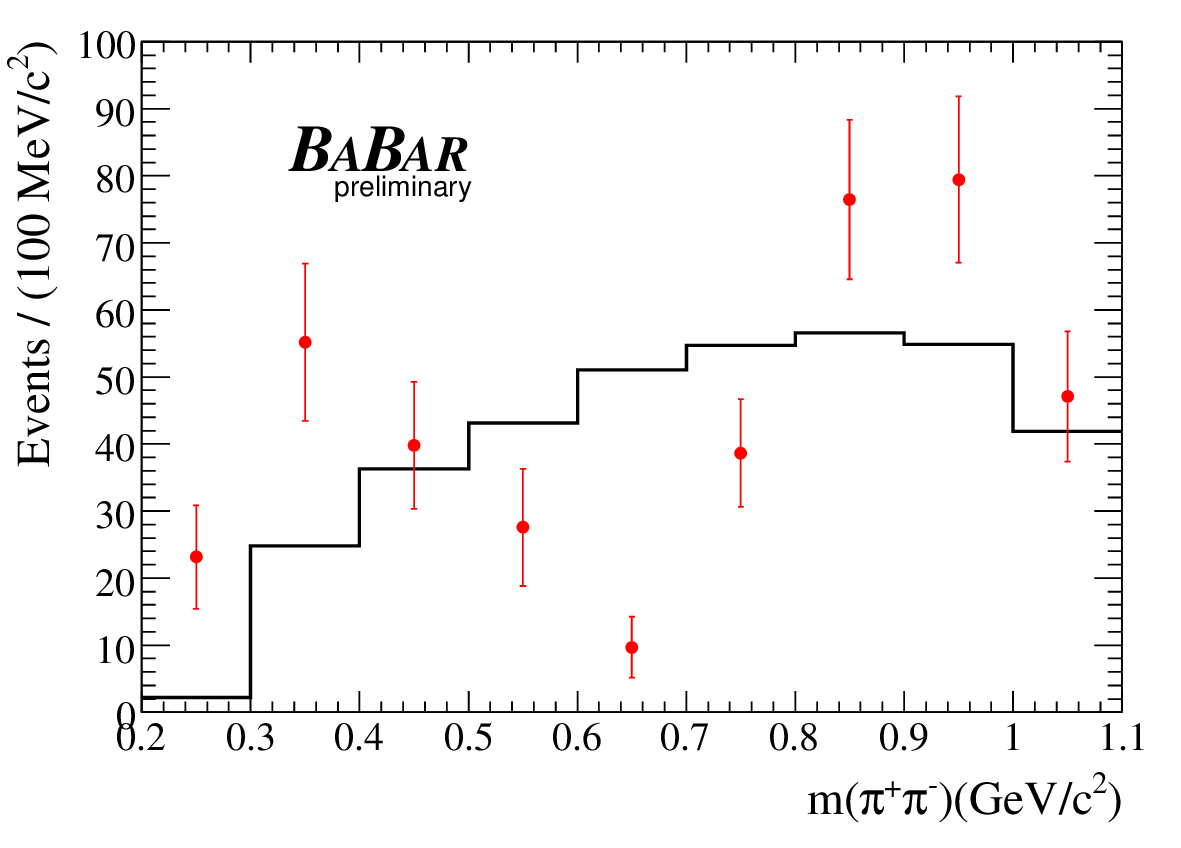}}
\caption{Dipion mass distribution for $Y(4260)\to J/\Psi\pi^+\pi^-$ from 
ref.~\cite{babary}.
\label{figurey3}}
\end{figure}

In an updated
report \cite{babary}, BaBar has confirmed the observation of the $Y(4260)$,
shown in Fig.~\ref{figurey2}. However, the new $\pi\pi$ mass distribution 
shows a more complex structure, as can be seem in Fig.~\ref{figurey3}.

\begin{figure}[h]
\scalebox{0.6}{\includegraphics[angle=0]{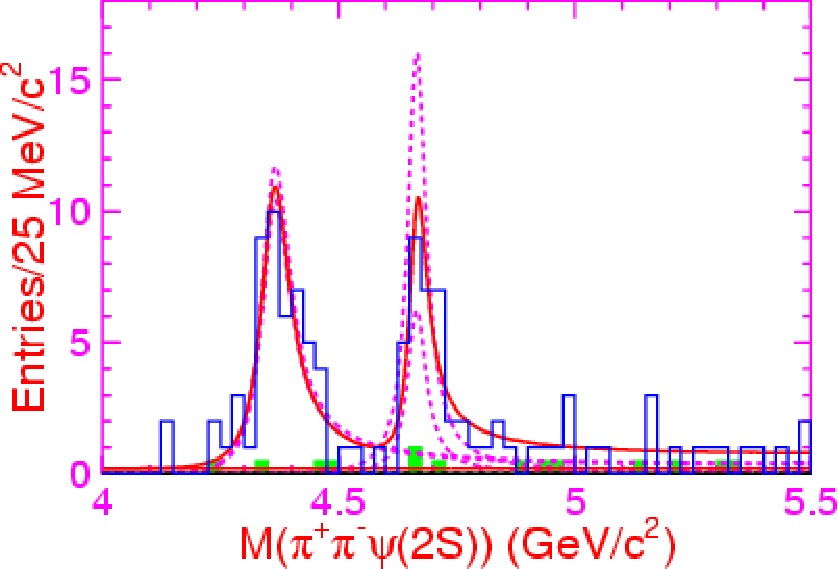}}
\caption{Evidence for the vector states $Y(4360)$ and $Y(4660)$ from 
ref.~\cite{belle4}.
\label{figurey4}}
\end{figure}

Repeating the same kind of analysis leading to the observation of the
$Y(4260)$ state, in the channel $e^+e^-\to\gamma_{ISR}\Psi(2S)\pi^+\pi^-$,
BaBar \cite{babar4} has identified another broad peak at a mass around
4.32 GeV, which was confirmed by Belle \cite{belle4}. Belle found that the 
$\psi^\prime\pi^+\pi^-$ enhancement observed by BaBar was, in fact, produced 
by two distinct peaks, as can be seen in Fig.~\ref{figurey4}. The masses 
and widths obtained by Belle and BaBar from fits to Breit-Wigner resonant 
shapes are summarized in Table~\ref{tabY}

\begin{table}[h]

  \begin{center}
    \caption{Masses and widths of $Y( 1^{--})$ states, measured by Belle 
      and BaBar}
  \label{tabY}
    \begin{tabular}{|cccc|} \hline
                                  &             
      & \multicolumn{2}{c}{Belle} \\ \hline 
      state                     & Decay mode 
      & $M$ ($\MeV)$ & $\Gamma$($\MeV$) \\ \hline
      $Y(4260)$                              & $J/\psi \pi^+ \pi^-$ 
      & $4247 \pm 12{}^{+17}_{-32}$ & $108 \pm 19 \pm 10$ \\ 
      $Y(4360)$                         & $\psi(2S) \pi^+ \pi^-$ 
      &  $4361 \pm 9 \pm 9$ & $74 \pm 15 \pm 10 $    \\
      $Y(4660)$                             & $\psi(2S) \pi^+ \pi^-$ 
      & $4664 \pm 11 \pm 5$          & $48 \pm 15 \pm 3$ \\ \hline
                                  &             
      & \multicolumn{2}{c}{BaBar}                          \\ \hline 
      state                     & Decay mode 
      & $M$ ($\MeV)$ & $\Gamma$($\MeV$) \\ \hline
      $Y(4260)$                              & $J/\psi \pi^+ \pi^-$ 
      & $4259 \pm 6 {}^{+2}_{-3}$   & $105 \pm 18 {}^{+4}_{-6}$ \\ 
      $Y(4360)$                         & $\psi(2S) \pi^+ \pi^-$ 
      &  $4324 \pm 24$                & $172 \pm 33 $         \\ \hline
     \end{tabular} 
  \end{center}
\end{table}

\begin{figure}[h]
\scalebox{0.6}{\includegraphics[angle=0]{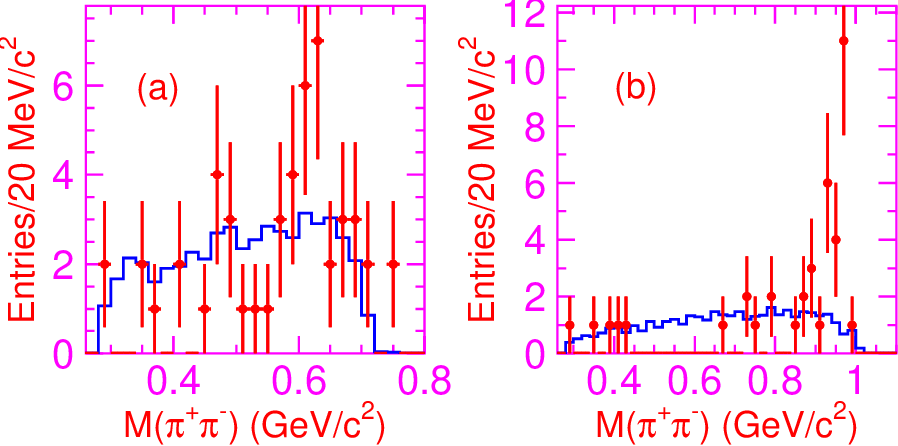}}
\caption{Dipion mass distribution  for the vector states $Y(4360)$ (a) and 
$Y(4660)$ (b) from ref.~\cite{belle4}.
\label{figurey5}}
\end{figure}

As it is evident from Fig.~\ref{figurey4}, there is no sign of the $Y(4260)$
in the $\psi^\prime\pi^+\pi^-$ mass spectrum.
The $\pi\pi$ mass distribution reported in \cite{belle4} for these two 
resonances can be seen in Fig.~\ref{figurey5}.

All these three states share the following properties: they are vector 
states with
the photon quantum numbers, they have large total widths, and they have only 
been observed in charmonium decay modes.

\begin{figure}[h]
\scalebox{0.5}{\includegraphics[angle=0]{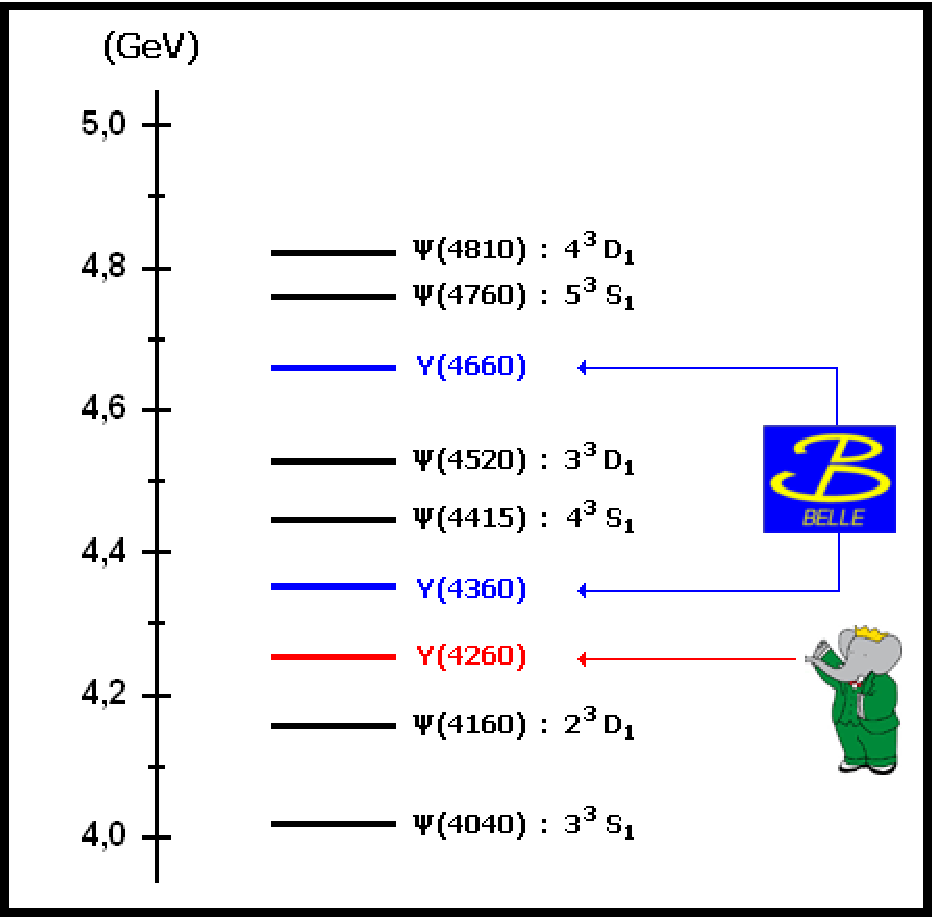}}
\caption{Masses of the $1^{--}$ $c\bar{c}$ states from quark models and the
masses of the $Y$ states. 
\label{figurey6}}
\end{figure}

Since the masses of these states are higher than the $D^{(*)}\bar{D}^{(*)}$
threshold, if they were $1^{--}$ charmonium states
they should decay mainly to $D^{(*)}\bar{D}^{(*)}$. However, the
observed $Y$ states do not match the peaks in $e^+e^-\to D^{(*)\pm}D^{(*)
\mp}$ cross sections measured  by Belle \cite{belle5} and BaBar 
\cite{babar5,babar6}.
Besides, the $\Psi(3S),~\Psi(2D)$ and $\Psi(4S)$ $c\bar{c}$ states have been 
assigned to the well stablished $\Psi(4040),~\Psi(4160),~$ and $\Psi(4415)$ 
mesons respectively. The predictions from quark models for the $\Psi(3D)$ 
and $\Psi(5S)$ charmonium states are 4.52 GeV and 4.76 GeV respectively. 
Therefore, the masses and widths of these three new $Y$ states are not 
consistent with any of the $1^{--}$ $c\bar{c}$ states \cite{kz,zhure,zhu}, 
and their discovery represents an overpopulation of the charmonium states, 
as it can be seen in Fig.~\ref{figurey6}.

An interesting interpretation is that the $Y(4260)$ is a charmonium hybrid. 
Hybrids are hadrons in which the gluonic degree of freedom has been excited.
The nature of this gluonic excitation is not well understood, and has been
described by various models. The
spectrum of charmonium hybrids has been calculated using lattice gauge theory
\cite{lattice}. The result for the mass is approximately 4200 MeV, which
is consistent with the flux tube model predictions \cite{bsc}. However, more 
recent lattice simulations \cite{latticenew}  and QCD string models 
calculations \cite{kala},
predict that the lightest charmonium hybrid has a mass of 4400 MeV, which is 
closer to the mass of the $Y(4360)$. In any case, a prediction of the hybrid
 hypothesis
is that the dominant open charm decay mode would be a meson pair with one
$S$-wave $D$ meson $(D,~D^*,~D_s,~D_s^*)$ and one $P$-wave $D$ meson $(D_1,~
D_{s1}$) \cite{closey}. In the case of the $Y(4260)$ this suggests dominance
of the decay mode $D\bar{D}_1$. Therefore, a large $D\bar{D}_1$ signal 
could be understood as a strong evidence in favor of the hybrid 
interpretation for the $Y(4260)$. In the case of the $Y(4360)$ and $Y(4660)$,
since their masses are well above the $D\bar{D}_1$ threshold, their decay 
into $D\bar{D}_1$ should be very strong if they were charmonium hybrids.

A critical information for understanding the structure of these states is 
wether the pion pair comes from a resonance state. From the di-pion 
invariant 
mass spectra shown in Figs.~\ref{figurey3} and \ref{figurey5}, there is some 
indication that only the $Y(4660)$ has a well defined intermediate state 
consistent with $f_0(980)$ \cite{babarconf}. Due to this fact and the 
proximity of the mass of the $\psi'-f_0(980)$ system with the mass of the 
$Y(4660)$ state, in ref.~\cite{ghm},
the $Y(4660)$ was considered as a $f_0(980)~\psi'$ bound state. 
If this interpretation of the $Y(4660)$ is correct, heavy quark spin symmetry
implies that there should be a $\eta^\prime_c-f_0(980)$ bound state 
\cite{ghm2}. This state would decay mainly into $\eta^\prime_c\pi\pi$, and  
the authors of ref.~\cite{ghm2} predicted the mass of such state to be
$4616^{+5}_{-6}$ MeV. The observation of this new state would clearly
determine the structure of the $Y(4660)$.
The $Y(4660)$
was also suggested to be a baryonium state \cite{qiao}, a canonical  
5 $^3$S$_1$ $c\bar{c}$ state \cite{dzy}, and a tetraquark with a $sc$-scalar-
diquark and a $\bar{s}\bar{c}$-scalar-antidiquark in a $2P$-wave state 
\cite{efg}.

In the case of $Y(4260)$, in ref.~\cite{maiani2} it was considered as a 
$sc$-scalar-diquark $\bar{s}\bar{c}$-scalar-antidiquark in a $P$-wave state. 
Maiani {\it et al.} \cite{maiani2} tried different ways to determine the 
orbital term and they arrived at $M=(4330\pm70)\MeV$, which is more 
consistent with $Y(4360)$. However, from the $\pi\pi$ mass 
distribution in refs.~\cite{babary,belle4}, none of these two states,  
$Y(4260)$ and $Y(4360)$ has a decay
with an intermediate state consistent with $f_0(980)$ and, therefore, it is 
not clear that they should have an $s\bar{s}$ pair in their structure.
Besides, in ref.~\cite{efg} the authors show that the mass of a
$[sc]_{S=0}[\bar{s}\bar{c}]_{S=0}$ tetraquark in  a $P$-wave state
would be 200 MeV higher than the $Y(4260)$ mass. The authors of 
ref.~\cite{efg}, found that a more natural interpretation for the $Y(4260)$
would be a $[qc]_{S=0}[\bar{q}\bar{c}]_{S=0}$ tetraquark in  a $P$-wave state.
The $Y(4260)$ was 
also interpreted as a baryonium $\Lambda_c-\bar{\Lambda}_c$ 
state \cite{qiao2}, a $S$-wave threshold effect \cite{rosnery}, as a resonance
due to the interaction between the three, $J/\psi\pi\pi$ and $J/\psi K\bar{
K}$, mesons \cite{osety} and as
a 4$S$ charmonium state \cite{estra}. Although there are some arguments 
against the molecular interpretation \cite{swanson,zhuy1}, the $Y(4260)$ 
was also considered as a molecular state bound by meson exchange 
\cite{ding,lzl,ywm}. The three $Y$ states were also interpreted as 
non-resonant manifestations of the Regee zeros \cite{eefy}.

\subsection{QCDSR studies for the $Y(J^{PC}=1^{--})$ states}

The $Y(J^{PC}=1^{--})$ states can be described by molecular or tetraquark
currents, with or without a $s\bar{s}$ pair.
In refs.~\cite{rapha,z12} a QCD sum rule calculation was performed  
using these kind of currents.

The lowest-dimension interpolating operator to describe  a $J^{PC}=1^{--}
$ state with the symmetric spin distribution: $[cs]_{S=0}[\bar{c}\bar{s}]_{S
=1}+[cs]_{S=1}[\bar{c}\bar{s}]_{S=0}$ is given by: 
\beqa
j_\mu&=&{\epsilon_{abc}\epsilon_{dec}\over\sqrt{2}}[(s_a^TC\gamma_5c_b)
(\bar{s}_d\gamma_\mu\gamma_5 C\bar{c}_e^T)
\nn\\
&+&(s_a^TC\gamma_5\gamma_\mu c_b)
(\bar{s}_d\gamma_5C\bar{c}_e^T)]\;.
\label{fieldyt}
\enqa

\begin{figure}[h]
\scalebox{0.8}{\includegraphics[angle=0]{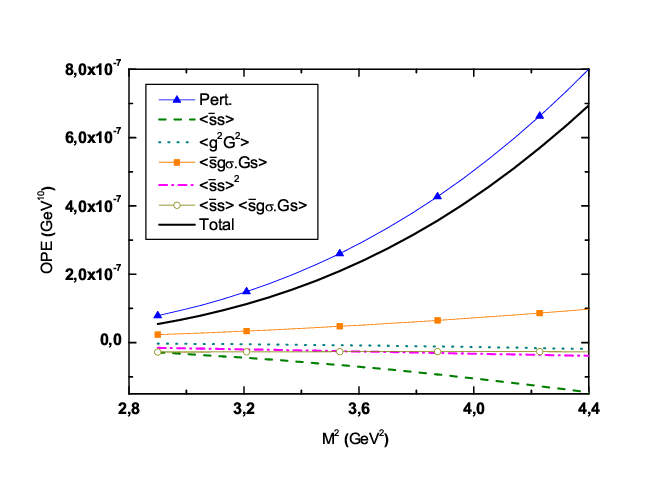}}
\caption{The $j_\mu$ OPE convergence in the region $2.8 \leq M^2 \leq
4.5~\GeV^2$ for $\sqrt{s_0} = 5.1$ GeV (taken from ref.\cite{rapha}).
\label{convy}}
\end{figure}

From Fig.~\ref{convy} we see that the OPE convergence for $M^2\geq 3.2$ 
GeV$^2$ is better than the one shown in Fig.~\ref{convx}, since
the perturbative contribution is the dominant one in the whole Borel region.
Using the dominance of the pole contribution to fix the upper value of
the Borel window, the result for the mass of the state described by
the current in Eq.~(\ref{fieldyt}) obtained in ref.~\cite{rapha} is:
\beq
m_Y = (4.65\pm0.10)~\GeV,
\label{massy46}
\enq
in excellent agreement with the mass of the $Y(4660)$ meson. Therefore, the 
conclusion in ref.~\cite{rapha} is that the meson $Y(4660)$ can be described
with a diquark-antidiquark tetraquark current with a spin configuration 
given by scalar and vector diquarks. The quark content in the current
in Eq.~(\ref{fieldyt}) is also consistent with the di-pion invariant mass 
spectra shown in Fig.~\ref{figurey5}, which shows that there is some 
indication that the $Y(4660)$ has a well defined di-pion intermediate state 
consistent with $f_0(980)$. It is also very interesting to notice that
the $f_0(980)$ meson may be itself considered as a tetraquark state \cite{hoo}.
This aspect should play an important role in the $Y(4660)$ decay since, as 
shown in ref.~\cite{decayx}, it is very hard to explain a small decay width
when the intital four-quark state decays into two final two-quark states.

Replacing the strange quarks in Eq.~(\ref{fieldyt}) by a generic light 
quark $q$ the mass obtained for a $1^{--}$ state described with 
the symmetric spin distribution: $[cq]_{S=0}[\bar{c}\bar{q}]_{S
=1}+[cq]_{S=1}[\bar{c}\bar{q}]_{S=0}$ is \cite{rapha}:
\beq
\lb{massy43}
m_Y=  (4.49\pm0.11)~\GeV~,
\enq
which is bigger than the $Y(4350)$ mass, but is consistent with it 
considering the uncertainty.

The $Y$ mesons can also be described by molecular-type currents.
In particular a $D_{s0}(2317)\bar{D}_s^*(2110)$ molecule with $J^{PC}=
1^{--}$, could also decay into $\psi^\prime\pi^+\pi^-$ with a dipion mass 
spectra consistent with $f_0(980)$. A current with $J^{PC}=1^{--}$ and a 
symmetrical combination of  scalar and vector mesons is given by:
\beq
j_\mu={1\over\sqrt{2}}[(\bar{s}_a\gamma_\mu c_a)(\bar{c}_b{s}_b)+(\bar{c}_a
\gamma_\mu s_a)(\bar{s}_b{c}_b)]\;.
\label{mol}
\enq

The mass obtained in ref.~\cite{rapha} for the current in Eq.~(\ref{mol})
is
\beq
\lb{massmols}
m_{D_{s0}\bar{D}_s^*}=  (4.42\pm0.10)~\GeV~,
\enq
which is more in agreement with the $Y(4350)$ mass than with the $Y(4660)$
mass.

To consider a molecular $D_{0}\bar{D}^*$ current with $J^{PC}=1^{--}$, one 
has only to change the strange quarks in Eq.(\ref{mol}) by a generic light 
quark $q$. The mass obtained with such current is \cite{rapha}
\beq
\lb{massmol}
m_{D_{0}\bar{D}^*}=  (4.27\pm0.10)~\GeV~,
\enq
in excellent agreement with the mass of the meson $Y(4260)$.
Again, in order to conclude if we
can associate this molecular state with the meson $Y(4260)$ we need a better
understanding of the di-pion invariant mass spectra for the pions in the 
decay $Y(4260)\to J/\psi\pi^+\pi^-$. From the spectra given in 
Fig.~\ref{figurey3}, it seems that the $Y(4260)$ is  consistent
with a non-strange molecular state ${D_{0}\bar{D}^*}$.
Using a $D_0$ mass \cite{pdg} $m_{D_0}=2352\pm50~\MeV$, the ${D_{0}
\bar{D}^*}$ threshold is around 4360 MeV  and it is 100 MeV above the 
mass in Eq.~(\ref{massmol}), indicating the possibility of a bound state.

A $J^{PC}=1^{--}$ molecular current can also be constructed with
pseudoscalar and  axial-vector mesons. A molecular $D\bar{D}_1$ 
current was used in ref.~\cite{z12}. The mass obtained with this current is:
\beq
\lb{massDD1}
m_{D\bar{D}_1}=(4.19\pm0.22)~\GeV~.
\enq
Therefore, considering the errors, the molecular $D\bar{D}_1$ assignement 
for the meson $Y(4260)$ is also possible, in agreement with the
findings of ref.~\cite{ding}, where a meson exchange model was used to
study the $Y(4260)$ meson. the $D\bar{D}_1$ threshold is around 4285 MeV  
and very close to the  $Y(4260)$ mass, indicating the possibility of a 
loosely bound molecular state.

\subsection{Summary  for $Y(J^{PC}=1^{--})$ states}

To summarize, the discovery of the $Y(4260),~Y(4360)$ and $Y(4660)$ appears
to represent an overpopulation of the expected charmonium $1^{--}$ states.
The absence of open charm production is also inconsistent with a conventional
$c\bar{c}$ explanation. Possible explanations for these states include 
charmonium hybrid, tetraquark state and $D_0\bar{D}^*$ or $D\bar{D}_1$ 
molecular state for $Y(4260)$. The $Y(4360)$ could be a charmonium hybrid,  
or a tetraquark state (with two axial $[cs]$ diquarks in $P$-wave or with two 
scalar $[cs]$ diquarks in $P$-wave). The $Y(4660)$ could be a canonical  
5 $^3$S$_1$ $c\bar{c}$ state, or a tetraquark state (with 
symmetrical $[cs]_{S=1}[\bar{c}\bar{s}]_{S=0}$ or with $[cs]_{S=0}[\bar{c}
\bar{s}]_{S=0}$ in a $2P$-wave).

From the  QCDSR results described is this Section, 
the $Y(4260)$ could be a $D_0\bar{D}^*$ or a $D\bar{D}_1$ molecular state
and the  $Y(4660)$ could be a tetraquark state with symmetrical spin  
distribution: $[cs]_{S=1}[\bar{c}\bar{s}]_{S=0}+[cs]_{S=0}[\bar{c}
\bar{s}]_{S=1}$. It is not 
possible to describe the $Y(4360)$ neither as a $D_{s0}\bar{D}^*_{s}$ molecular
state, nor as a tetraquark state with symmetrical spin  
distribution: $[cq]_{S=1}[\bar{c}\bar{q}]_{S=0}+[cq]_{S=0}[\bar{c}
\bar{q}]_{S=1}$.

\markboth{\sl Interpretation of the New Charmonium States }{\sl The $Z^+(4430)$
meson}
\section{\label{Z} The $Z^+(4430)$ meson}

All states discussed so far are electrically neutral. The real turning point
in the discussion about the structure of the new observed charmonium states
was the observation by Belle Collaboration of a charged state decaying into
$\psi'\pi^+$, produced in $B^+\to K\psi'\pi^+$ \cite{bellez}.

\subsection{Experiment versus theory}

The measured mass and width of this state are $M=(4433\pm4\pm2)\MeV$ and
$\Gamma=(45^{+18+30}_{-13-13})\MeV$ \cite{bellez}. The $B$ meson decay rate to
this
state is similar to that for the decays to the $X(3872)$ and $Y(3930)$ mesons.
There are no reports of a $Z^+$ signal in the $J/\psi\pi^+$ decay channel.
Since the minimal quark content of this state is $c\bar{c}u\bar{d}$, this
state is a prime candidate for a multiquark meson. The $Z^+(4430)$ was
observed in the $\psi'\pi^+$ channel, therefore, it is an isovector state
with positive $G$-parity: $I^G=1^+$.

\begin{figure}[h]
\scalebox{0.45}{\includegraphics[angle=0]{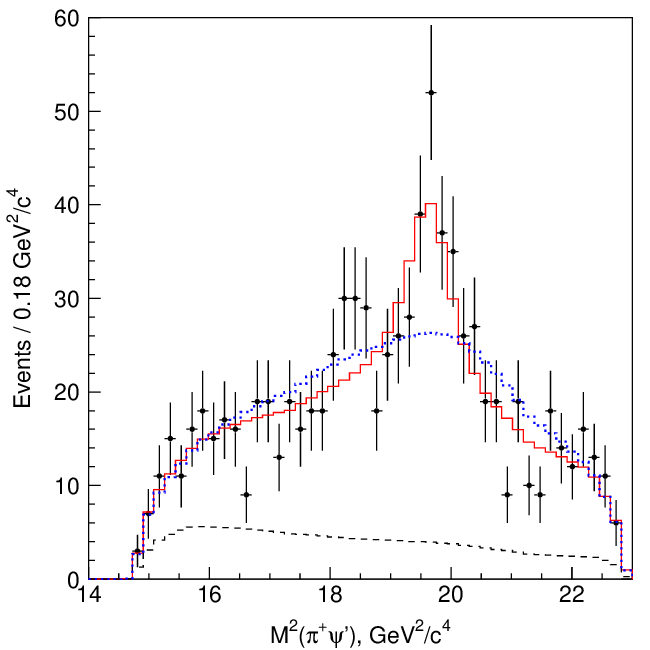}}
\caption{Dalitz plot projection for the $\psi'\pi^+$ invariant mass from
ref.~\cite{bellez2}. The solid (dotted) histograms shows the fit result with
(without) a single $\psi'\pi^+$ state. The dashed histogram represents the
background.
\label{massz}}
\end{figure}

Using the same data sample as in ref.~\cite{bellez}, Belle also performed a
full Dalitz plot analysis \cite{bellez2} and  has confirmed the observation of
the $Z^+(4430)$ signal with a 6.4$\sigma$ peak significance, as can be seen
in Fig.~\ref{massz}. The updated $Z^+(4430)$ parameters are:
$M=(4433^{+15+19}_{-12-13})\MeV$ and $\Gamma=(109^{+86+74}_{-43-56})\MeV$.

\begin{figure}[h]
\scalebox{0.45}{\includegraphics[angle=0]{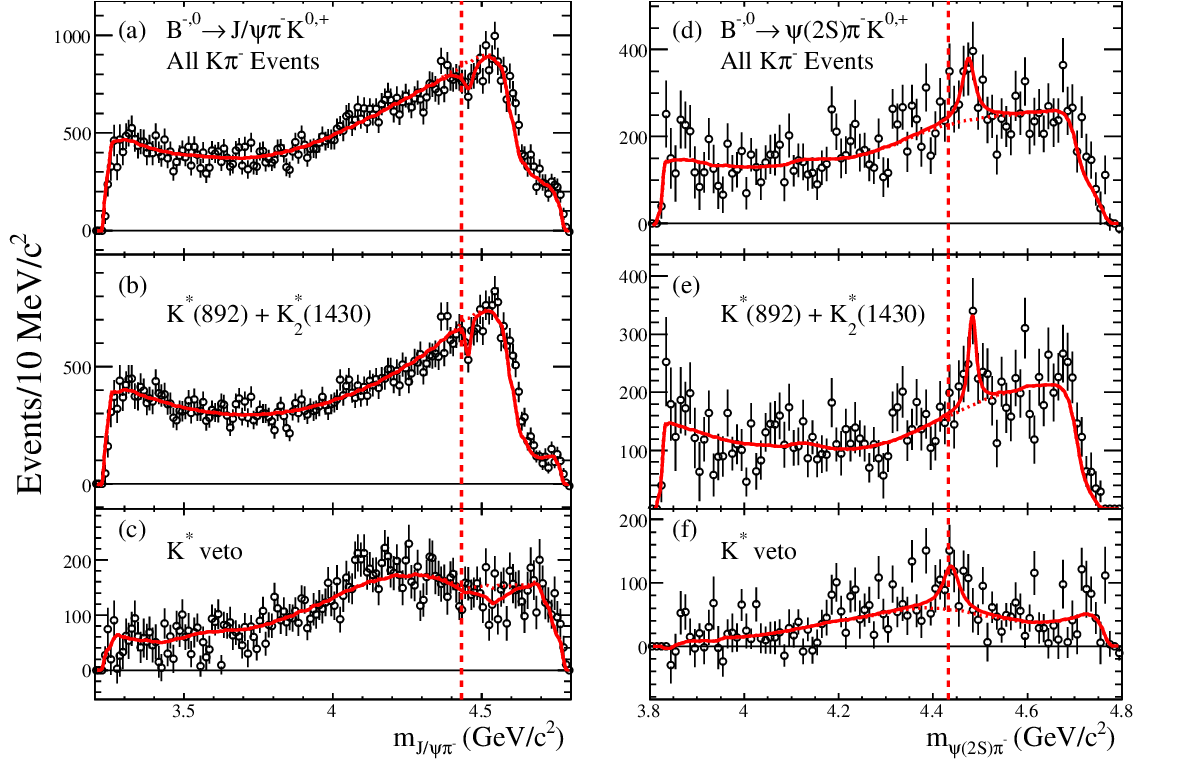}}
\caption{The $J/\psi\pi^-$ (left) and $\psi'\pi^-$ (right) invariant mass from
ref.~\cite{babarz}.
\label{figbabarz}}
\end{figure}

Babar Collaboration  \cite{babarz} also searched the $Z^-(4430)$ signature in
four decay modes: $B\to\psi\pi^-K$, where $\psi=J/\psi$ or $\psi^\prime$ and
$K=K_S^0$ or $K^+$. No significant evidence for a signal peak was found in any
of the processes investigated, as it can be seen in Fig.~\ref{figbabarz}.

The Babar result gives no conclusive evidence of the $Z^+(4430)$ seen by 
Belle.

There are many theoretical interpretations of the $Z^+(4430)$ structure
\cite{rosner,meng,nos,zhuz,zhuz2,dingz,z10,maiani3,glp,bugg,cong,mms,zbraa,cz1,cz2,cz3,biz,zclose}.
Because its mass is close to the $D^*D_1$ threshold, Rosner \cite{rosner}
suggested that it is an $S$-wave threshold effect, while others considered it
to be a strong candidate for a $D^*D_1$ molecular state
\cite{meng,nos,zhuz,zhuz2,dingz,z10}.
Other possible interpretations are tetraquark state \cite{maiani3,glp}, a cusp
in the $D^*D_1$ channel \cite{bugg},  a baryonium state \cite{cong}, a 
radially
excited $c\bar{s}$ state \cite{mms}, or a hadro-charmonium state \cite{volre}.
The tetraquark hypothesis implies that
the $Z^+(4430)$ will have neutral partners decaying into $\psi'\pi^0/\eta$.
Besides the spectroscopy, there are discussions about its production
\cite{rosner,cz1,cz2,cz3} and decay \cite{biz}.

Considering the  $Z^+(4430)$ as a loosely bound $S$-wave $D^*D_1$ molecular
state, the allowed angular momentum and parity are $J^P=0^-,~1^-,~2^-$,
although the $2^-$  assignment is probably suppressed in the $B^+\to Z^+K$
decay by the small phase space. Among the remaining possible $0^-$ and $1^-$
states, the former will be more stable as the later can also decay to
$DD_1$ in $S$-wave. Moreover, one expects a bigger mass for the $J^P=1^-$
state as compared to a $J^P=0^-$ state.
There is also a quenched lattice QCD calculation that finds attractive
interatiction in the $D^*D_1$ system in the $J^P=0^-$ channel \cite{laz}.
The authors of ref.~\cite{laz} also find positive scattering lenght. Based
on these findings, they conclude that although the interaction between the two 
charmed mesons is attractive in this channel, it is unlikely that they can 
form a genuine bound state right below the threshold. 

\subsection{QCDSR calculations for $Z^+(4430)$}

Considering $Z^+(4430)$ as a $D^*D_1$ molecule with $J^P=0^-$, a possible
current describing such state, considered in  ref.~\cite{nos}, is given by:
\beq
j={1\over\sqrt{2}}\left[(\bar{d}_a\gamma_\mu c_a)(\bar{c}_b\gamma^\mu\gamma_5
u_b)+(\bar{d}_a\gamma_\mu\gamma_5 c_a)(\bar{c}_b\gamma^\mu u_b)\right]\;.
\label{zmol}
\enq
This current corresponds to a symmetrical state $D^{*+}\bar{D}_1^0+\bar{D}^{*0}
D_1^+$, and has positive $G$-parity,
which is consistent with the observed decay $Z^+(4430)\to\psi^\prime\pi^+$.

The mass obtained in a QCDSR calculation using such a current was \cite{nos}:
\beq
m_{D^*D_1}=(4.40\pm0.10)\GeV,
\label{masszmol}
\enq
in an excellent agreement with the experimental mass.

To check if the $Z^+(4430)$ could also be described as
a diquark-antidiquark state with $J^P=0^-$ , the following current was
considered in ref.~\cite{z10}:
\beq
j_{0^-}={i\epsilon_{abc}\epsilon_{dec}\over\sqrt{2}}[(u_a^TC\gamma_5c_b)
(\bar{d}_dC\bar{c}_e^T)-(u_a^TC c_b)(\bar{d}_d\gamma_5C\bar{c}_e^T)]\;.
\label{z0-}
\enq

The mass obtained with this current was \cite{z10}
\beq
m_{Z_{(0^-)}} = (4.52\pm0.09)~\GeV,
\label{massz0}
\enq
which is a little bigger than the experimental value \cite{bellez}, but still
consistent with it, considering the uncertainties. Comparing this result with
the result in Eq.~(\ref{masszmol}), we see that the result obtained using a
molecular-type current is in a better agreement with the experimental value.
As mentioned in  Sec. II, this strongly suggests that the state is better 
explained as a molecular state than as a diquark-antidiquark
state.  This is explicitly borne out in the calculation for the coupling, 
$\lambda$, between the state and the current defined in
Eq.~(\ref{cou}) of  Sec. II. We get:
\beq
\lambda_{Z_{(0^-)}} = \left(3.75\pm0.48\right)\times 10^{-2}~\GeV^5,
\label{ladi}
\enq
\beq
\lambda_{D^*D_1} = \left(5.66\pm1.26\right)\times 10^{-2}~\GeV^5.
\label{lamo}
\enq
Therefore, one can conclude that the physical particle with
$J^P=0^-$ and quark content $c\bar{c}u\bar{d}$ couples with a larger
strength with the molecular $D^*D_1$ type current than with the
current in Eq.~(\ref{z0-}).

In ref.~\cite{z10} it was also considered as a diquark-antidiquark interpolating
operator  with $J^P=1^-$ and positive $G$ parity:
\beqa
j_\mu^{1^-}&=&{\epsilon_{abc}\epsilon_{dec}\over\sqrt{2}}[(u_a^TC\gamma_5c_b)
(\bar{d}_d\gamma_\mu\gamma_5 C\bar{c}_e^T)
\nn\\
&+&(u_a^TC\gamma_5\gamma_\mu c_b)
(\bar{d}_d\gamma_5C\bar{c}_e^T)]\;.
\label{z1-}
\enqa
In this case the Borel stability obtained is worse than
for the $Z^+$ with $J^P=0^-$ \cite{z10}, and the obtained mass was
\beq
\lb{massz1}
m_{Z_{(1^-)}}=  (4.84\pm0.14)~\GeV~,
\enq
which is much bigger than the experimental value and bigger than the
result obtained using the current with $J^P=0^-$ in Eq.~(\ref{massz0}).
From these results it is possible to conclude that, while it is also 
possible to
describe the $Z^+(4430)$ as a  diquark-antidiquark state or a molecular state
with $J^P=0^-$, the $J^P=1^-$ configuration is disfavored.

\subsection{Summary  for $Z^+(4430)$}

A confirmation of the existence of the $Z^\pm(4430)$ is critical
before a complete picture can be drawn. If confirmed,
the only open options for the $Z^+(4430)$ structure are
tetraquark or molecule. The favored quantum numbers are $J^P=0^-$.

\subsection{Sum rule predictions for $B^*B_1$ and $D_s^*D_1$ molecules}

It is straightforward to extend the analysis done for the $D^*D_1$ molecule
to the case of the bottom quark. Using the same interpolating field of
Eq.~(\ref{zmol}) with the charm quark replaced by the bottom one, the
analysis done for $Z^+(4430)$ was repeated for $Z_b$ in ref.\cite{nos}.
The OPE convergence in this case is even better than the one for $Z^+(4430)$
and the predicted mass is
\beq
\lb{masszb}
m_{Z_{B^*B_1}}=  (10.74\pm0.12)~\GeV~,
\enq
in a very good agreement with the prediction in ref.~\cite{cheung}.

In the case of the strange analogous meson $Z_{s}^+$ considered as a
pseudoscalar $D_s^*D_1$ molecule, the current
is obtained by exchanging the $d$ quark in Eq.~(\ref{zmol}) by the $s$ quark.
The predicted mass is \cite{nos}:
\beq
\lb{masszs}
m_{D_s^*D_1}=  (4.70\pm0.06)~\GeV~,
\enq
which is bigger than the $D_s^{*}D_1$ threshold $\sim4.5~\GeV$, indicating
that this state is probably a very broad one and, therefore, it might be very
dificult to be seen experimentally.

\markboth{\sl Interpretation of the New Charmonium States }{\sl The $Z_1^+(4050)$
and $Z_2^+(4250)$ states}
\section{\label{Z12} The $Z_1^+(4050)$ and $Z_2^+(4250)$ states}

The $Z^+(4430)$ observation motivated studies of other $\bar{B}^0\to K^-\pi^+
(c\bar{c})$ decays. In particular, the Belle Collaboration has 
reported the observation of two resonance-like structures in the $\pi^+
\chi_{c1}$ mass distribution \cite{belle3}.

\subsection{Experiment versus theory}

The two resonance-like structures, called $Z_1^+(4050)$ and $Z_2^+(4250)$, 
were observed in the exclusive process $\bar{B}^0\to K^-\pi^+\chi_{c1}$.
The significance of each of the
$\pi^+\chi_{c1}$ structures exceeds 5$\sigma$ and, if they are interpreted
as meson states, their minimal quark content must be $c\bar{c}u\bar{d}$.
Since they were observed in the $\pi^+\chi_{c1}$ channel,
the only quantum numbers that are known about them are $I^G=1^-$.

\begin{figure}[h]
\scalebox{0.75}{\includegraphics[angle=0]{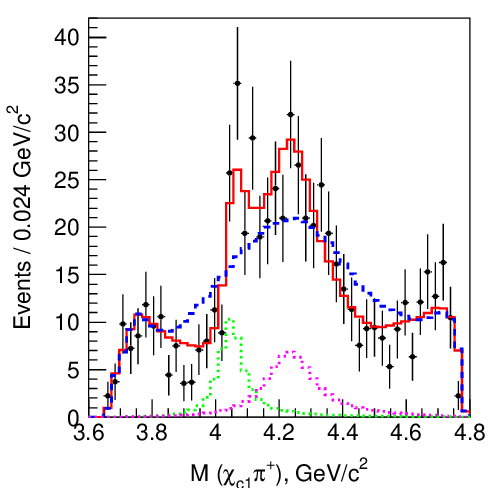}}
\caption{The $\pi^+\chi_{c1}$ invariant mass distribution from 
ref.~\cite{belle3}. The solid (dashed) histograms shows the fit result with two
(without any) $\pi^+\chi_{c1}$ resonance. The dots with error represent data.
\label{massz12}}
\end{figure}

When fitted with two Breit-Wigner reconance amplitudes, the
resonance parameters are:
\beqa
M_1&=&(4051\pm14^{+20}_{-41})~\MeV,\nn\\
\Gamma_1&=&(82^{+21+47}_{-17-22})~\MeV,\nn\\
M_2&=&(4248^{+44+180}_{-29-~35})~\MeV,\nn\\
\Gamma_2&=&(177^{+54+316}_{-39-~61})~\MeV.
\enqa 
The invariant mass distribution, where the contribution of the structure
in the $\pi^+\chi_{c1}$ channel is most clearly seen, is shown in 
Fig.~\ref{massz12}

Before this observation, it was suggested, in ref.~\cite{volz12}, that 
resonances decaying into $\chi_{c1}$ and one or two pions are expected in the 
framework of the hadro-charmonium model.

Due to the
closeness of the $Z_1^+(4050)$ and $Z_2^+(4250)$ masses to the
$D^{*}\bar{D}^*(4020)$ and $D_1\bar{D}(4285)$ thresholds, these states could 
also be interpreted as molecular states or threshold effects. Liu et al. 
\cite{llz}, using a meson exchange model find strong attraction for the
$D^{*}\bar{D}^*$ system with $J^{P}=0^+$, while using a boson exchange model, 
the author of ref.~\cite{ding3} concluded that the interpretation of 
$Z_1^+(4050)$ as a $D^{*}\bar{D}^*$ molecule is not favored. In any case,
it is very difficult to understand a bound molecular state which mass is above 
the $D^{*}\bar{D}^*$ threshold. In the case of $Z_2^+(4250)$, using a meson 
exchange model, it was shown in ref.~\cite{ding} that its interpretation as a 
$D_1\bar{D}$ or $D_0\bar{D}^*$ molecule is disfavored.

\subsection{QCDSR calculations}

In ref. \cite{z12}, the QCD sum rules formalism was used to study
the $D^{*}\bar{D}^*$ and $D_1\bar{D}$ molecular states with $I^GJ^P=1^-0^+$
and $1^-1^-$ respectively. The currents used in both cases are: 
\beq
j_{D^{*}{D}^*}=(\bar{d}_a\gamma_\mu c_a)(\bar{c}_b\gamma^\mu u_b)
\;,
\label{curz1}
\enq
and
\beq
j_\mu={i\over\sqrt{2}}\left[(\bar{d}_a\gamma_\mu\gamma_5 c_a)(\bar{c}_b
\gamma_5u_b)+(\bar{d}_a\gamma_5 c_a)(\bar{c}_b\gamma_\mu\gamma_5 u_b)
\right]\;.
\label{curz2}
\enq

The mass obtained with the current in Eq.~(\ref{curz1}) is \cite{z12}:
\beq
m_{D^*D^*} = (4.15\pm0.12)~\GeV
\enq
where the central value is around 130 MeV above the $D^*D^*(4020)$ threshold,
indicating the existence of  repulsive interactions between the two $D^*$ 
mesons.
Strong interaction effects might lead to a repulsion that
could result in a virtual state above the threshold. Therefore,
this structure may or may not indicate a resonance. 

For the current in Eq.~(\ref{curz2}), the mass obtained is \cite{z12}:
\beq
m_{D_1D} = (4.19\pm0.22)~\GeV,
\enq
where the central value is around 100 MeV below the $D_1D(4285)$ threshold, 
and, considering the errors, consistent with the mass of the $Z_2^+(4250)$ 
resonance. Therefore, in this
case, there is an attractive interaction between the mesons $D_1$ and $D$
which can lead to a molecular state.

\begin{figure}[h]
\scalebox{0.3}{\includegraphics[angle=0]{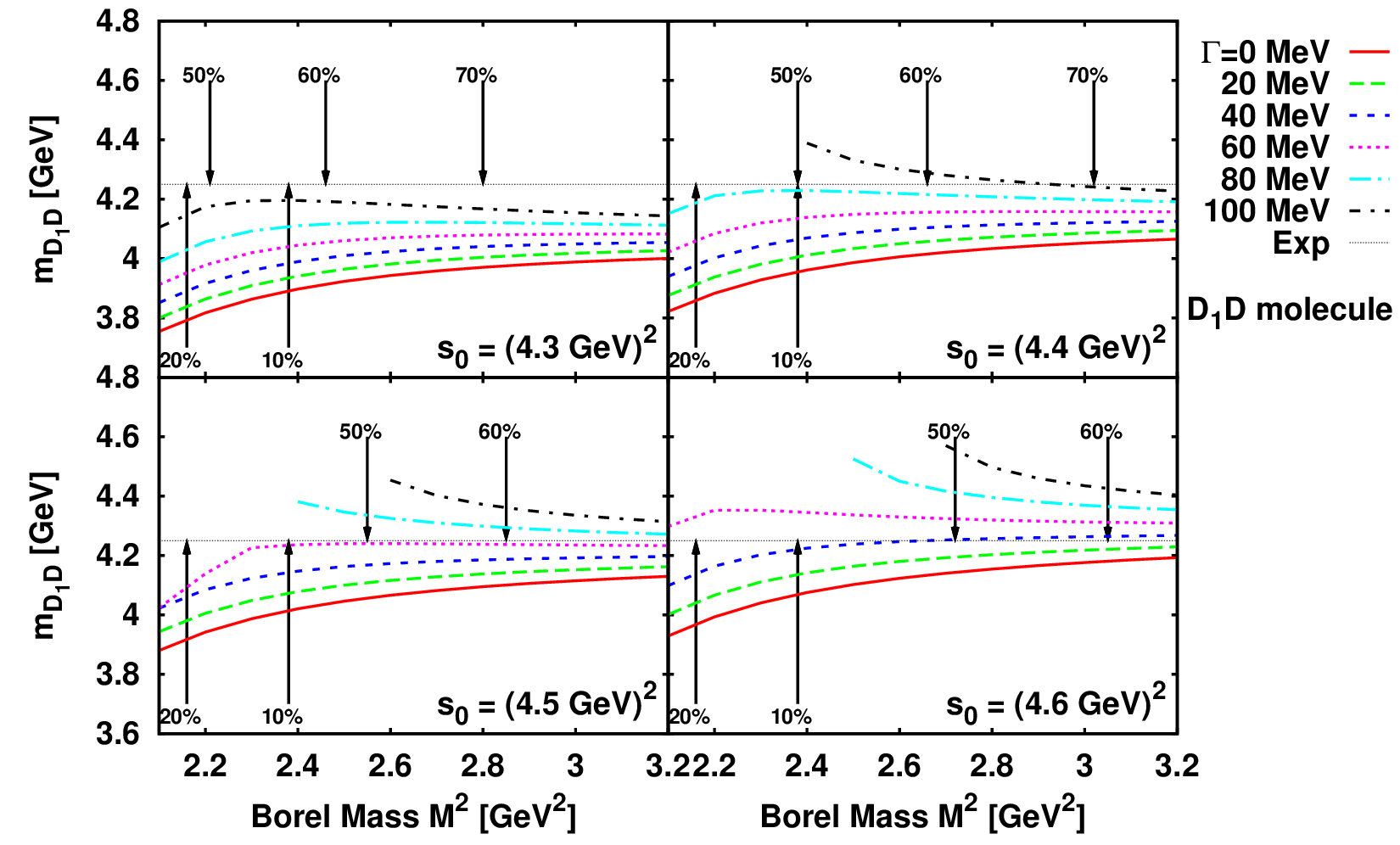}}
\caption{Results for the $D_1 D$ molecule from ref.~\cite{width}. Each panel 
shows a
  different choice of the continuum threshold. Upward and downward arrows
  indicate the region of the Borel window $M^2_{\text{min}}$ and
  $M^2_{\text{max}}$, respectively. Associated numbers in \% denote the
  dimension eight condensate  contribution for upward arrows and
  continuum contribution for downward ones.
\label{figz2}}
\end{figure}

In ref.~\cite{width}
it was found that the inclusion of the width, in the phenomenological side
of the sum rule, increases the obtained mass for molecular states. This means 
that the introduction of the width in the sum rule calculation, increases 
the mass of the states that couple to the $D^*\bar{D}^*$ and $D_1\bar{D}$ 
molecular currents.  As a result, using the current in Eq.~(\ref{curz2}),
it is possible to obtain a mass $m_{D_1D} = 4.25~\GeV$ with a width
$40\leq\Gamma\leq60~\MeV$, as can be seen in Fig.~\ref{figz2}.

On the other hand, the mass of the $D^*\bar{D}^*$ molecule will 
be far  from the $Z_1^+(4050)$ mass. Therefore, the authors of ref.~\cite{z12} 
conclude that it
is possible to describe the $Z_2^+(4250)$ resonance  structure as a
$D_1\bar{D}$ molecular state with $I^GJ^P=1^-1^-$ quantum numbers, and that
the $D^{*}\bar{D}^{*}$ state is probably a virtual state that is not
related with the $Z_1^+(4050)$ resonance-like structure. Considering the 
fact that the $D^*D^*$ threshold (4020) is so close to the $Z_1^+(4050)$ 
mass and that
the $\eta^{\prime\prime}_c(3^1S_0)$ mass is predicted to be around 4050 MeV
\cite{ss}, it is probable that the $Z_1^+(4050)$ is only a threshold
effect \cite{ss}.

\markboth{\sl Interpretation of the New Charmonium States }{\sl The $Y(3930)$ and
$Y(4140)$ states}
\section{\label{YY} The $Y(3930)$ and $Y(4140)$ states}

\subsection{Experiment versus theory}

\begin{figure}[h]
\scalebox{0.45}{\includegraphics[angle=0]{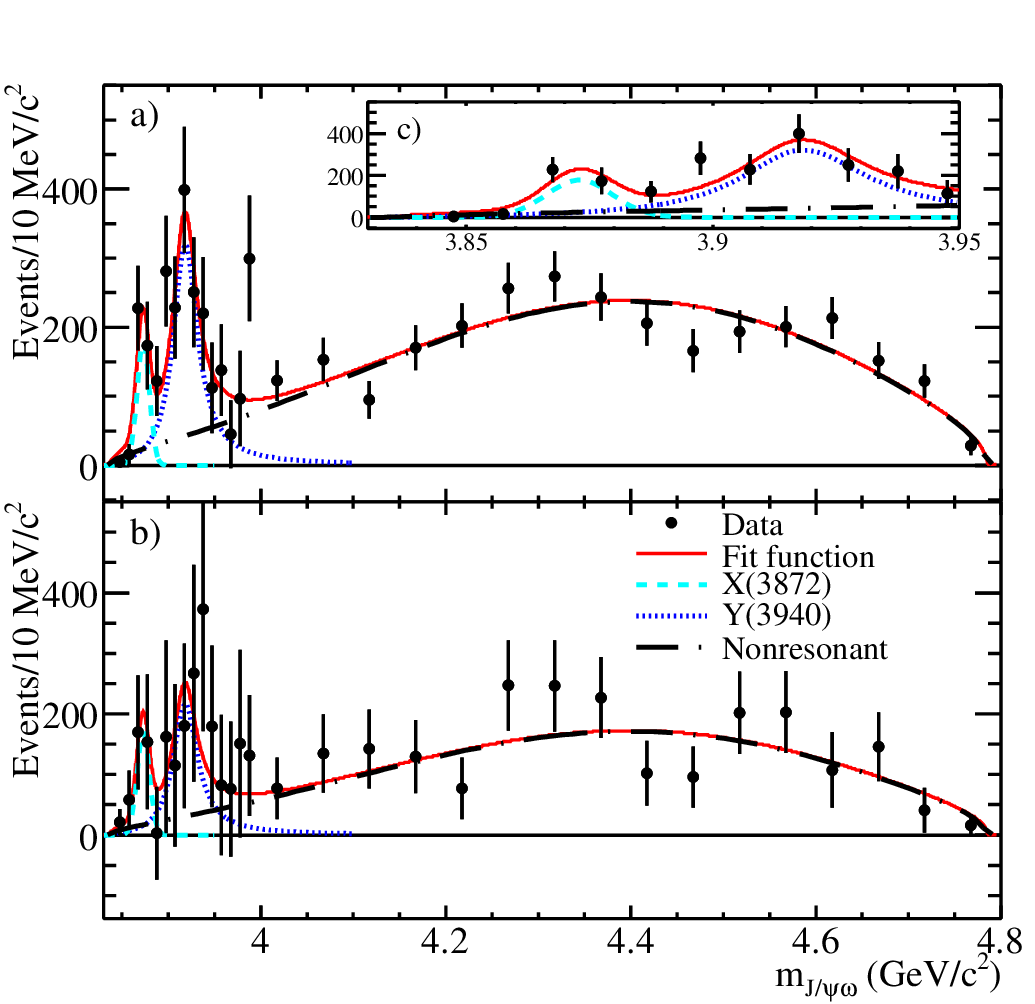}}
\caption{Reconstruction of the $X(3872)$ and $Y(3930)$ peaks from their decays
 into $J/\psi\omega$ mesons by BaBar (taken from ref.~\cite{babarate}).
\label{XY}}
\end{figure}

The $Y(3930)$ was first observed by the Belle Collaboration \cite{belley3}
in the decay $B\to KY(3930)\to K\omega J/\psi$. It was confirmed by BaBar
\cite{babary3,babarate} in two channels $B^+\to K^+\omega J/\psi$ and
$B^0\to K^0\omega J/\psi$.
The measured mass from these two Collaborations are:
$(3943\pm11)$ MeV from ref.~\cite{belley3} and $(3919.1^{+3.8}_{-3.4}\pm2)$ 
MeV from ref.~\cite{babarate}, which gives an average mass of 
$(3929\pm7)$ MeV. 
This state has positive $C$ and $G$ parities and the total width is 
$(31^{+10}_{-8}\pm5)$ MeV \cite{babarate}. The $m_{J/\psi\omega}$ mass
distribution observed by BaBar is shown in Fig.~\ref{XY}, from where we also 
see the $X(3872)$ observed in its decay into $J/\psi\omega$.

Since the decay 
$Y\to J/\psi\omega$ is OZI suppressed for a charmonium state \cite{egmr}, 
also for $Y(3930)$ it was
conjectured  that it could be a hybrid $c\bar{c}g$ state
\cite{closey}, a molecular state \cite{swan1,braaten3,llz,vol2,vol3,oset2},
or a tetraquark state \cite{maiani}.

\begin{figure}[h]
\vspace{-3.cm}
\scalebox{0.45}{\includegraphics[angle=0]{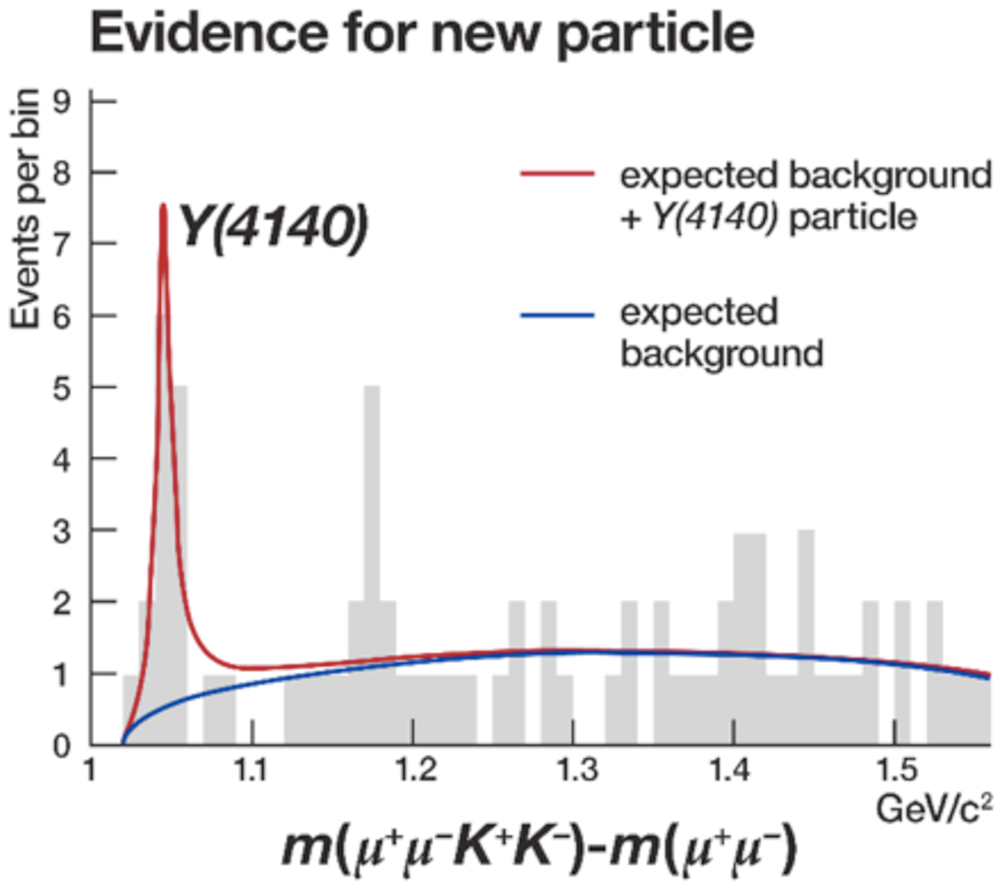}}
\vspace{-3.6cm}
\caption{Reconstruction of the $Y(4140)$ peak from its decay into muons and
$K$ mesons by CDF.
\label{peaky}}
\end{figure}

A recent acquisition to the list of
peculiar states is the narrow structure observed by the CDF Collaboration
in the decay $B^+\to Y(4140)K^+\to J/\psi\phi K^+$. The particle's signature 
peak can be seem in Fig.~\ref{peaky}. The mass and width
of this structure is $M=(4143\pm2.9\pm1.2)~\MeV$, $\Gamma=(11.7^{+8.3}_{-5.0}
\pm3.7)~\MeV$ \cite{cdfy}. Since the $Y(4140)$ decays into two 
$I^G(J^{PC})=0^-(1^{--})$ vector mesons, like the $Y(3930)$ it has positive 
$C$ and $G$ parities. The possible $J^P$ quantum numbers of a $S$-wave 
vector-vector system are $0^+,~1^+,~2^+$. However, since $C=(-1)^{L+S}$, the 
$J^P=1^+$ is forbidden for a state with $L=0$ and  $C=+1$. Therefore, the 
possible quantum numbers for $Y(3930)$ and $Y(4140)$ are $J^{PC}=0^{++},~
1^{-+}$ and $2^{++}$. At these quantum numbers, $1^{-+}$ is not consistent
with the constituent quark model and it is considered exotic.

There are already some theoretical interpretations of this structure. Its
interpretation as a conventional $c\bar{c}$ state is disfavored because,
as pointed out by the CDF Collaboration \cite{cdfy}, it lies well above the 
threshold for open charm decays and, therefore, a $c\bar{c}$ state with this 
mass would decay predominantly into an open charm pair with a large total 
width. It was also shown in ref.~\cite{liu} that the $Y(4140)$ is probably
not a $P$-wave charmonium state: $\chi_{cJ}^{''}~(J=0,
~1)$. If it were the case, the branching ratio of the hidden charm decay, 
$Y(4140)\to J/\psi\phi$, would be much smaller than
the experimental observation.

In ref.~\cite{zhuy}, the authors interpreted the $Y(4140)$ as the 
molecular partner of the charmonium-like state $Y(3930)$. They concluded 
that the $Y(4140)$ is probably a $D_s^{*}\bar{D}_s^{*}$ molecular state with 
$J^{PC}=0^{++}$ or $2^{++}$, while the $Y(3930)$ is its $D^*\bar{D}^*$ 
molecular partner. This idea is supported by the fact that the mass difference
between these two mesons is approximately the same as the mass difference 
between the $\phi$ and $\omega$ mesons: $m_{Y(4140)}-m_{Y(3930)}\sim m_\phi-
m_\omega\sim 210~ \MeV$. It is also interesting to notice that, if the 
$Y(4140)$ and the $Y(3930)$ mesons are $D_s^{*}\bar{D}_s^{*}$ and
$D^*\bar{D}^*$ molecular states, the binding energies of these states will be
approximately the same: $ m_{Y(4140)}-2m_{D_s^*}\sim m_{Y(3930)}-2m_{D^*}\sim
-90~\MeV$. However, with a meson exchange mechanism to bind the two charmed 
mesons, it seems natural to expect a more deeply bound system in the case that
pions can be exchanged between the two charmed mesons, as in the $D^*D^*$,
than when only $\eta$ and $\phi$ mesons can be exchanged, as in the $D_s^*
D_s^*$ system.

In ref.~\cite{maha} the author argue that  the $Y(4140)$ 
can be interpreted either as a $D_s^{*}\bar{D}_s^{*}$ molecular state or
as an exotic hybrid charmonium with $J^{PC}=1^{-+}$. A molecular $D_s^{*}
\bar{D}_s^{*}$ configuration was also considered in
refs.~\cite{wang,bgl,y4140,ding2,jrz}. Using QCD sum rules,
the authors of refs.~\cite{y4140,jrz} agree that it is possible to describe 
the $Y(4140)$ with a molecular $D_s^{*}\bar{D}_s^{*}$ current with 
$J^{PC}=0^{++}$.  Using one boson exchange model the author of 
ref.~\cite{ding2}
showed that the effective potencial of the $D_s^{*}\bar{D}_s^{*}$ system
supports the explanation of  $Y(4140)$ as a molecular state.
In ref.~\cite{stancu} the $Y(4140)$
was considered as a $J^{PC}=1^{++}$ $c\bar{c}s\bar{s}$ tetraquark state.
The $J^{PC}=1^{++}$ assignment reduces the coupling of the $Y(4140)$ with
the vector-vector channel and, therefore, a small decay width would be
possible in this case but not for a $J^{PC}=0^{++}$ tetraquark state, as
pointed out in ref.~\cite{zhuy}.
The authors of ref.~\cite{eef} argue that the $J/\psi\phi$ system has
quantum numbers $J^{PC}=1^{--}$ and that the enhancement observed by
CDF does not represent any kind of resonance. There is also a prediction
for the radiative open charm decay of the $Y(4140)$ that could test the
molecular assignment of this state \cite{liuke}.

\subsection{QCDSR calculation for $Y(3930)$ and $Y(4140)$}

Considering the $Y(3930)$ and $Y(4140)$ as $I^GJ^{PC}=0^+0^{++}$ states,
the possible currents that couple with a $D^{*}\bar{D}^{*}$ and $D_s^{*}
\bar{D}_s^{*}$ molecular  states are
\beq
j_q=(\bar{q}_a\gamma_\mu c_a)(\bar{c}_b\gamma^\mu q_b)
\;, \mbox{and }
j_s=(\bar{s}_a\gamma_\mu c_a)(\bar{c}_b\gamma^\mu s_b)
\;,
\label{fieldy}
\enq
respectively. These two currents were considered in a QCDSR study in
ref.~\cite{y4140}. Surprisingly the masses obtained were
\beq
m_{D_s^*D_s^*} = (4.14\pm0.09)~\GeV,
\enq
and
\beq
m_{D^*D^*} = (4.13\pm0.11)~\GeV.
\enq

\begin{figure}[h]
\centerline{\epsfig{figure=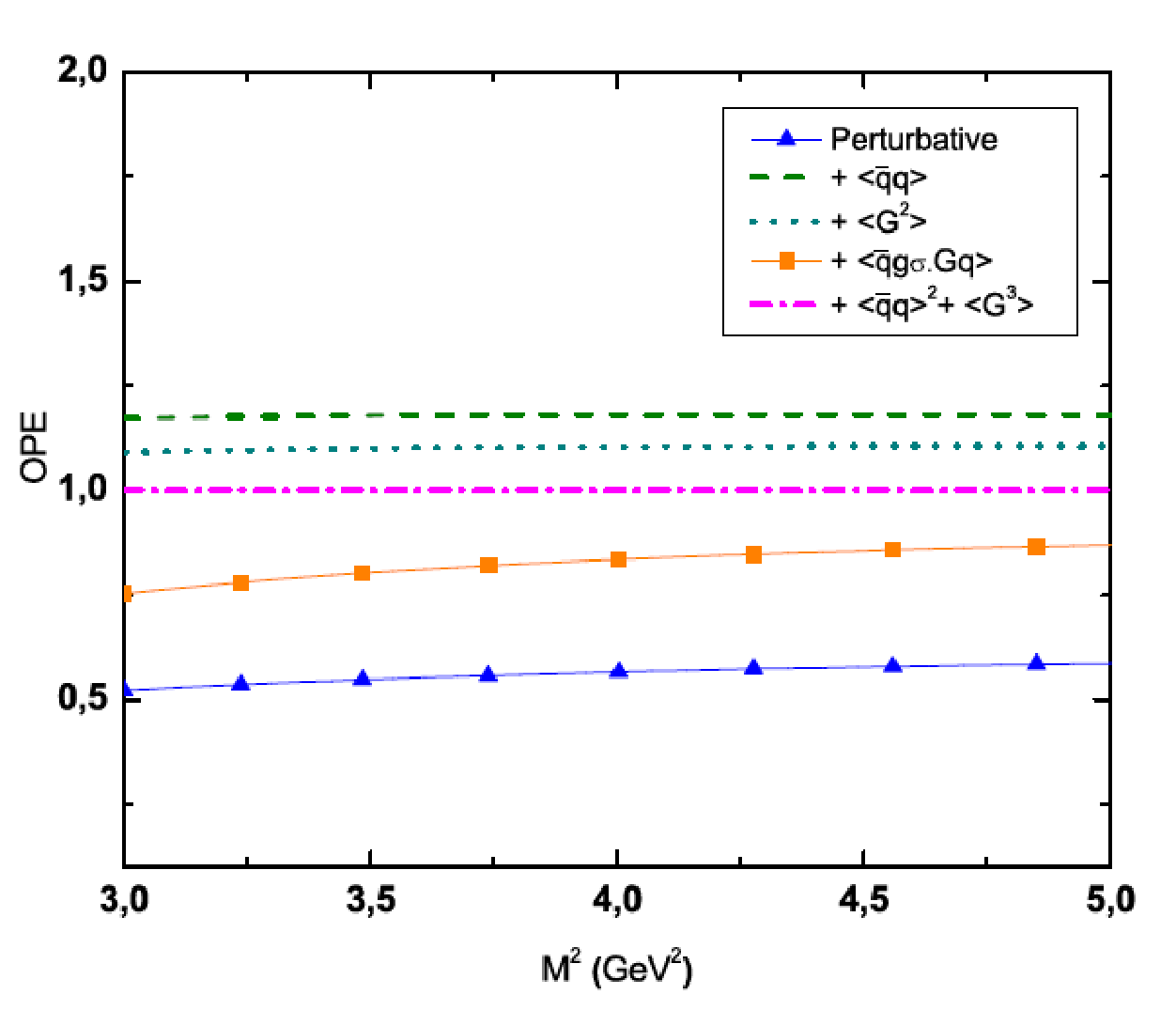,height=60mm}}
\vskip-.2cm
\caption{The OPE convergence for the $D^*D^*$ molecular current, as a function 
of the Borel mass, for $\sqrt{s_0} = 4.4$ GeV.  We plot the 
relative contributions starting with the perturbative one 
and each other line represents the total OPE
after adding of one extra condensate in the expansion.}
\label{conv}
\end{figure}

\begin{figure}[h]
\centerline{\epsfig{figure=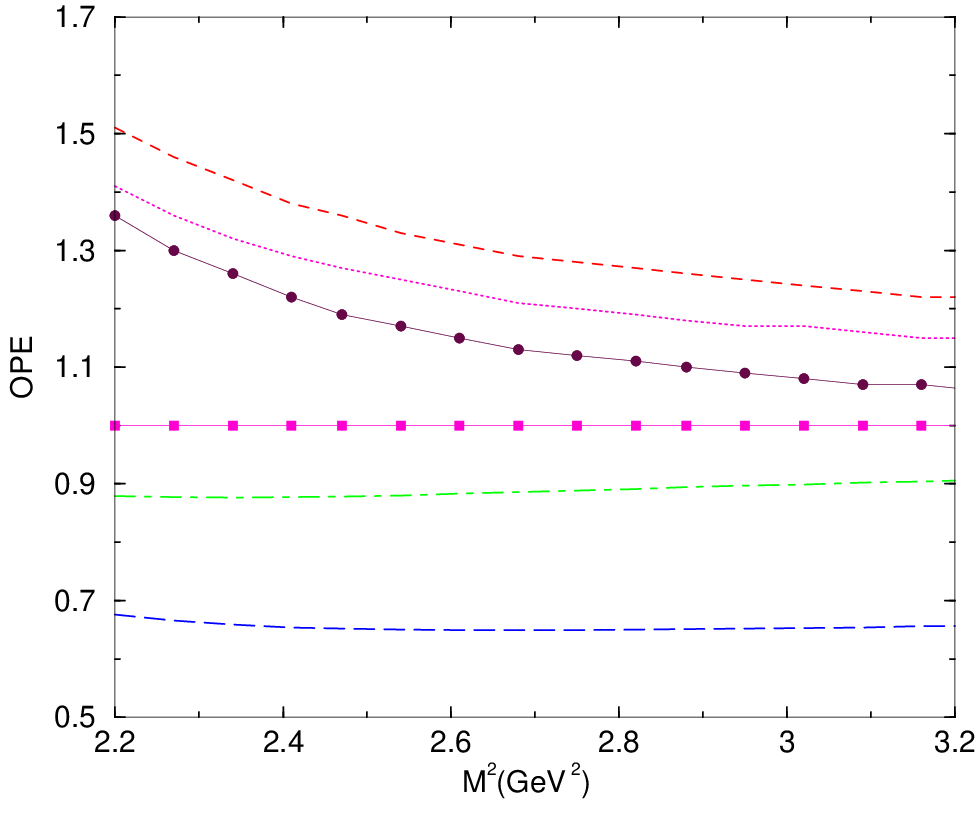,height=60mm}}
\caption{The OPE convergence for the $D^*D^*$ current for $\sqrt{s_0}=4.6~
\GeV$. We plot the 
relative contributions starting with the perturbative one
(long-dashed line), and each other line represents the total OPE 
after adding of one extra condensate in the expansion: + $\qqs$ 
(dashed line), 
+ $\langle g^2G^2\rangle$ (dotted line), + $m_0^2\qqs$ (dot-dashed line), 
+ $\qqs^2$ (line with circles), + $m_0^2\qqs^2$ (line with squares).
From ref.~\cite{y4140}.}
\label{opedd}
\end{figure}

There is another QCDSR calculation for the $D^*\bar{D}^*$ molecular
current with $J^{PC}=0^{++}$ \cite{jrong}, that finds a mass $m_{D^*D^*} = 
(3.91\pm0.11)~\GeV$, compatible with the $Y(3930)$ state. However, in 
ref.~\cite{jrong} the authors have considered only the condensates up to
dimension six. We show, in Fig.~\ref{conv}, the contribution of all the terms 
in the OPE side of the sum rule, up to dimension-six. From this figure we see 
that only for $M^2\geq 3.5$ GeV$^2$ the contribution of the dimension-six 
condensate is less than 20\% of the total contribution. Therefore, 
the lower value of $M^2$ in the sum rule window should be $M^2_{min}= 3.5$ 
GeV$^2$. The inclusion of the dimension-eight 
condensate improves the OPE convergence, and its contribution  
is less than 20\% of the total contribution for 
$M^2\geq 2.5~\GeV^2$, as it can be seen in Fig.~\ref{opedd}.

\begin{figure}[h]
\centerline{\epsfig{figure=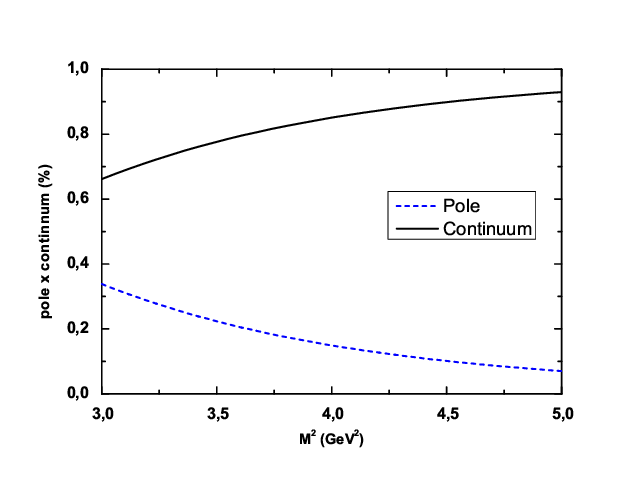,height=60mm}}
\caption{The pole and the continuum contributions
for the $D^*D^*$ current for $\sqrt{s_0}=4.4~\GeV$.}
\label{poledd}
\end{figure}

Besides, we observe that considering only condensates up to 
dimension-six, there is no pole dominance, in the Borel range considered
in ref.~\cite{jrong}, as it can be seen in Fig.~\ref{poledd} and, therefore,
no allowed Borel window can be found in this case.

In the 
case of the $D_s^*\bar{D}_s^*$ molecular current, the inclusion of the 
dimension-eight  condensate almost does not change the OPE convergence and the 
value of the mass. This is the reason why the results obtained with the 
$D_s^*\bar{D}_s^*$ molecular current are the same in refs.~\cite{y4140} and 
\cite{jrz}.

Therefore, from a QCD sum rule study, the results obtained in 
refs.~\cite{y4140} and \cite{jrz} indicate that the $Y(4140)$ narrow 
structure observed by  the CDF Collaboration  in the decay 
$B^+\to Y(4140)K^+\to J/\psi\phi K^+$ can be very well described by a 
scalar $D_s^*\bar{D}_s^*$ current.
The mass obtained with the $D^*\bar{D}^*$ scalar current, on the other hand,
depends on the dimension of the condensates considered in the OPE side,
showing that there is still no OPE convergence in the sum rule up 
to dimension-6 condensate. To test if the convergence is achieved up
to dimension-8, it is important to consider higher dimension condensates 
in the OPE side of the sum rule.

\markboth{\sl Interpretation of the New Charmonium States }{\sl The $X(3915)$ and
$X(4350)$ states}
\section{\label{XX} The $X(3915)$ and $X(4350)$ states}

\begin{figure}[h]
\scalebox{0.45}{\includegraphics[angle=0]{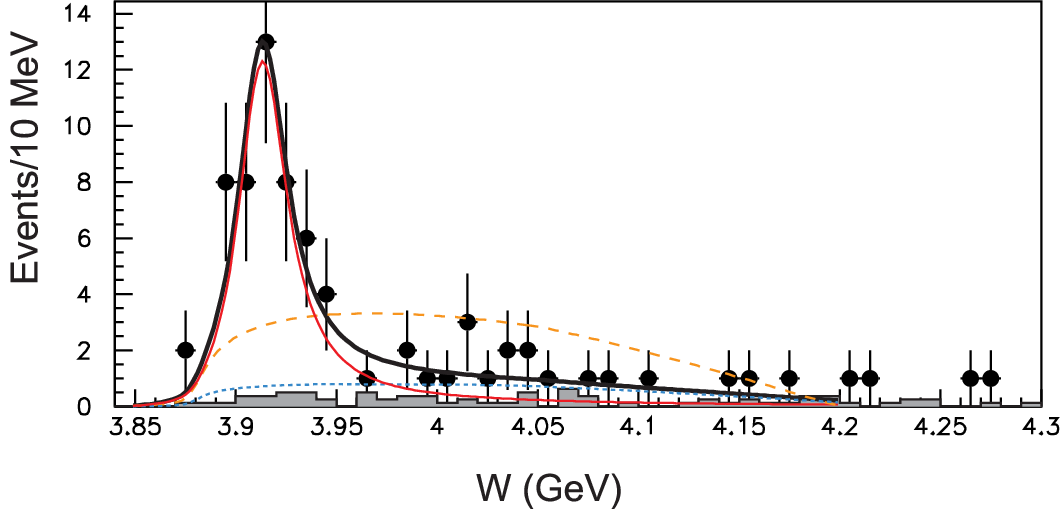}}
\caption{The $\omega J/\psi$ mass enhancement observed by Belle Coll.
in the $\gamma\gamma\to\omega J/\psi$ events. From ref.~\cite{czy}. 
\label{x3915}}
\end{figure}
\begin{figure}[h]
\scalebox{0.5}{\includegraphics[angle=0]{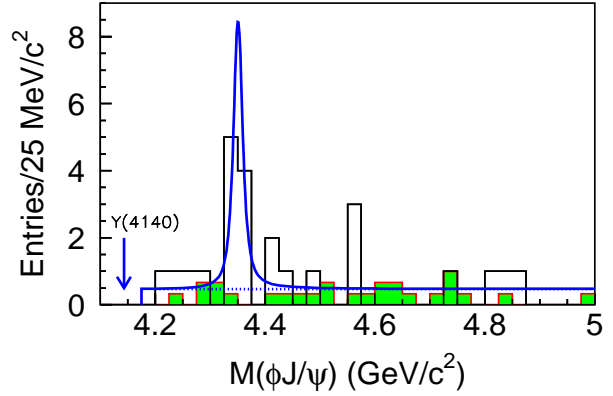}}
\caption{The $\phi J/\psi$ mass enhancement observed by Belle Coll.
in the $\gamma\gamma\to\phi J/\psi$ events. From ref.~\cite{czy}. 
\label{x4350}}
\end{figure}

In recent communications, the Belle Collaboration has reported the observation
of  two  narrow peaks in the two-photon processes   $\gamma\gamma\to\omega J/
\psi$ \cite{bellegg} and $\gamma\gamma\to\phi J/\psi$ \cite{belleggs}, as can 
be seen in Figs.~\ref{x3915} and \ref{x4350}.

These two states were  reported in the experimental reviews 
\cite{olsen,czy,zupa,godxx}. When fitted with  Breit-Wigner resonance 
amplitudes, the resonance parameters are:
\beqa
M&=&(3914\pm4\pm2)~\MeV,\nn\\
\Gamma&=&(28\pm12^{+2}_{-8})~\MeV,
\label{mx3915}
\enqa
for $X(3915)$, and
\beqa
M&=&(4350.6^{+4.6}_{-5.1}\pm0.7)~\MeV,\nn\\
\Gamma&=&(13.3^{+17.9}_{-9.1}\pm4.1)~\MeV.
\label{mx4350}
\enqa 
for $X(4350)$. 

Since these two states decay into two vector mesons, they have positive $C$ 
and $G$ parities, like the $Y(3930)$ and $Y(4140)$. As a matter of fact,
it is possible that the $X(3915)$ be the same state as the 
$Y(3930)$, observed by Belle \cite{belley3} and BaBar  \cite{babary3,babarate} 
in the 
decay channel $B\to KY(3930)\to K\omega J/\psi$, since the Babar mass and width
for this state are very close to the result in Eq.~(\ref{mx3915}): 
$M=(3919.1^{+3.8}_{-3.4}\pm2$ MeV and $\Gamma= (31^{+10}_{-8}\pm5)~\MeV$.
In any case, a charmonium assignment for this state is difficult \cite{olsen}.

In the case of the $X(4350)$ its mass is much higher than the $Y(4140)$ mass
and, as it can be seen in Fig.~\ref{x4350}, no  $Y(4140)$ signal is observed
in the two-photon process $\gamma\gamma\to\phi J/\psi$. This fact was 
interpreted, in ref.~\cite{czy}, as a point against the $D_s^{*+}D_s^{*-}$
molecular picture for the $Y(4140)$ states. As shown in ref.~\cite{bgl},
a $D_s^{*+}D_s^{*-}$ molecular state should be seen in the two-photon process.

The possible quantum numbers for a state decaying
into $J/\psi\phi$ are  $J^{PC}=0^{++},~1^{-+}$ and $2^{++}$. At these quantum 
numbers, $1^{-+}$ is not consistent with the constituent quark model and it is 
considered exotic. In ref.~\cite{belleggs} it was noted that
the mass of the $X(4350)$ is consistent with the prediction for a $cs\bar{c}
\bar{s}$ tetraquark state with $J^{PC}=2^{++}$ \cite{stancu} and a
$D_s^{*+}\bar{D}_{s0}^{*-}$ molecular state \cite{jrz2}. However, the
state considered in ref.~\cite{jrz2} has $J^P=1^-$ with no definite
charge conjugation. A molecular state with a vector and a scalar $D_s$ mesons
with negative charge conjugation was studied by the first time in 
ref.~\cite{rapha}, and the obtained mass was $(4.42\pm0.10)~\GeV$, also 
consistent with the $X(4350)$ mass, but with not consistent quantum numbers.
A molecular state with a vector and a scalar $D_s$ mesons
with positive charge conjugation can be constructed using the combination
$D_s^{*+}{D}_{s0}^{*-}-D_s^{*-}D_{s0}^{*+}$.

Some possible interpretations for this state are: a excited $P$-wave 
charmonium state $\Xi_{c2}^{\prime\prime}$ \cite{xliu}; a
mixed charmonium-$D_s^*D_s^*$ state \cite{wangx}.

\subsection{QCDSR calculation for $X(4350)$}

Following Belle Collaboration's suggestion \cite{belleggs}, 
in ref.~\cite{x4350}, a QCDSR calculation using a  $D_s^*{D}_{s0}^*$ current
 with $J^{PC}=1^{-+}$, was considered
to test if the new observed resonance structure, $X(4350)$, 
can be interpreted as such molecular state.

A current that couples with a $J^{PC}=1^{-+}$ $D_s^*{D}_{s0}^*$ molecular
state is given by:
\beq
j_\mu={1\over\sqrt{2}}\left[(\bar{s}_a\gamma_\mu c_a)(\bar{c}_bs_b)-
(\bar{c}_a\gamma_\mu s_a)(\bar{s}_bc_b)\right]
\;.
\label{fieldx4350}
\enq

\begin{figure}[h]
\centerline{\epsfig{figure=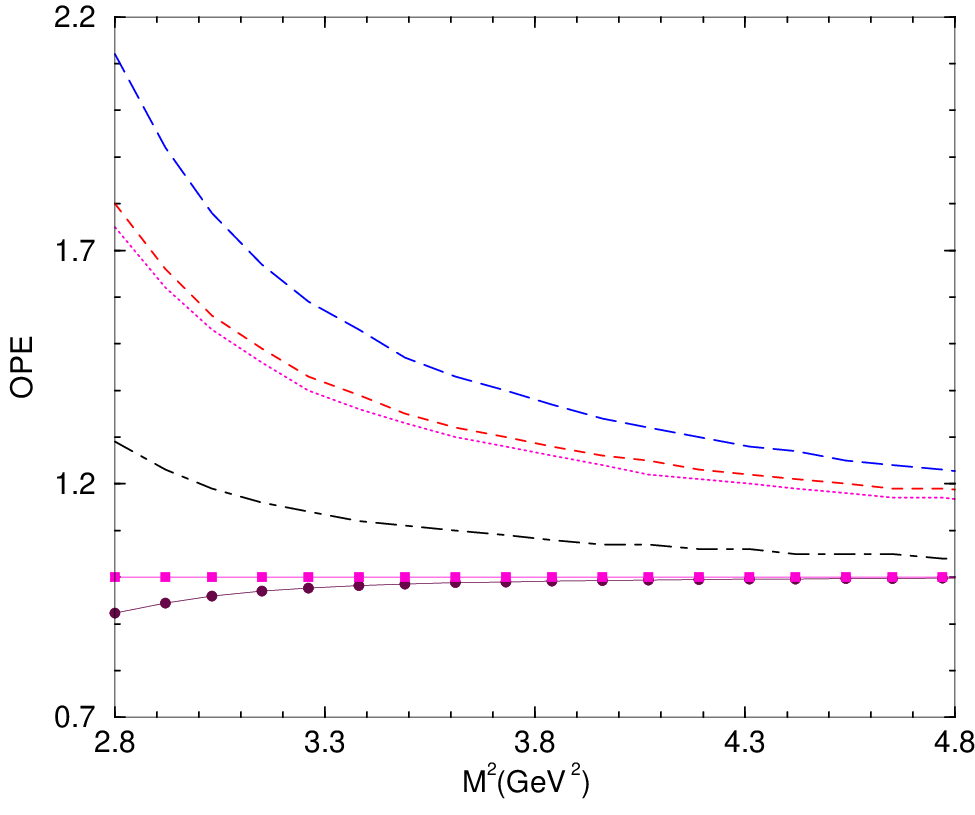,height=70mm}}
\caption{The OPE convergence for the $J^{PC}=1^{-+},~D_s^*D_{s0}^*$ 
molecule in the region
$2.8 \leq M^2 \leq4.8~\GeV^2$ for $\sqrt{s_0} = 5.3$ GeV.  We plot the 
relative contributions starting with the perturbative contribution plus de 
$m_s$ correction (long-dashed line), and each other line represents the 
relative contribution after adding of one extra condensate in the expansion: 
+ $\qqs+m_s\qqs$  (dashed line), 
+ $\langle g^2G^2\rangle$ (dotted line), + $\mixs+m_s\mixs$ 
(dot-dashed line), + $\qqs^2+m_s\qqs^2$ (line with circles), + $\qqs\mixs+
m_s\qqs\mixs$ (line with squares). Taken from ref.~\cite{x4350}.}
\label{conv4350}
\end{figure}

For this current the OPE convergence is very good and the dimension-8 
condensate (obtained using the factorization hypothesis) is almost negligible,
as can be seen by Fig.~\ref{conv4350}.

Considering condensates up to dimension-8 and keeping terms which are linear in
the strange quark mass $m_s$, the mass obtained in ref.~\cite{x4350}
was:
\beq
m_{D_s^*{D}_{s0}^*} = (5.05\pm0.19)~\GeV,
\label{mass4350}
\enq
where the uncertainty was obtained by considering the QCD parameters in the 
ranges: $m_c=(1.23\pm0.05)~\GeV$, $m_s=(0.13\pm 0.03)\,\GeV $,
$\lag\bar{q}q\rag=\,-(0.23\pm0.03)^3\,\GeV^3$, $m_0^2=(0.8\pm0.1)\GeV^2$, and 
also by allowing a violation of the factorization hypothesis by using
$\rho=2.1$.

The result in Eq.~(\ref{mass4350})
is much bigger than the mass of the narrow structure $X(4350)$
observed by Belle. Therefore, the authors of ref.~\cite{x4350} concluded that 
it is not possible to interpret the $X(4350)$ as a $D_s^*{D}_{s0}^*$ molecular
state with $J^{PC}=1^{-+}$.
It is also interesting to notice that the mass
obtained for a state described with a $1^{--},~ D_s^*D_{s0}^*$ molecular
current, given in Eq.~(\ref{massmols}), is much smaller than the result
obtained with the $1^{-+},~ D_s^*D_{s0}^*$ molecular current. This may be 
interpreted as an indication that it is easier to form molecular states with
not exotic quantum numbers.

\markboth{\sl Interpretation of the New Charmonium States }{\sl The $X(3940),
Z(3930),X(4160)$ and
$Y(4008)$ states}
\section{\label{XZXY} The $X(3940),Z(3930),X(4160)$ and
$Y(4008)$ states}


The $X(3940)$ \cite{bellecc} and the $X(4160)$ \cite{Abe:2007sya} were 
observed by the Belle Collaboration
in the analysis of the $M_{recoil}(J/\psi)$ recoil spectrum in $e^+e^-\to 
J/\psi(c \bar{c}) $. The $X(3940)$ mass and width are $(3942^{+7}_{-6} 
\pm 6)$
 MeV and $\Gamma=(37^{+26}_{-15} \pm 8)$ MeV, and was observed to decay into 
$D^*\bar{D}$.   The  $X(4160)$ was found to decay dominantly into $D^* 
\bar{D}^*$
 with  parameters given as $M=(4156^{+25}_{-20} \pm 15)$ MeV and $\Gamma=
(139^{+111}_{-61} \pm 21)$ MeV.  The only other known charmonium sate 
observed 
in $e^+e^-\to J/\psi X $ process has $J=0$.  This fact and the absence of 
these states in the $D\bar{D}$ decay suggests that these states favor $J^{PC}
=0^{-+}$. However, these states are either too low or too high to be the 
$\eta_c''$ or the $\eta_c'''$ states \cite{Olsen:2009gi}.   The X(4160) was 
identified as a $2^{++}$ state in two recent works:  generated from a 
relativistic four quark equations in ref.\cite{Gerasyuta:2008ps}, and from a 
coupled channel approach using a vector-vector interaction in 
ref.~\cite{oset2}.  In ref.~\cite{chao} two  assignments were found to be 
possible: $\eta_c(4S)$ and  the P-wave excited state $\chi_{c0}(3P)$.
So far, no other serious theoretical attempts were 
performed to investigate these states, and the nature of these states 
remains a puzzle.

The $Z(3930)$ was observed by Belle collaboration as an enhancement in the 
$\gamma \gamma \to D \bar{D}$ event.  The observed mass and width are 
$M=3929 
\pm5 \pm 2$ MeV and $\Gamma=29 \pm 10 \pm 2 $ MeV.  The measured properties 
are consistent with expectations for the previously unseen $\chi_{c2}'$ 
charmonium state\cite{Eichten:2005ga,bg}.

The $Y(4008)$ was seen by the Belle collaboration as a by-product while 
confirming the $Y(4260)$ in a two resonance fit to the $e^+e^- \to \pi^+ 
\pi^- J/\psi$ reaction \cite{yexp}.  There are theoretical speculations that 
this state might be the $\psi(3S)$ or is a $D^* \bar{D}^*$ molecular state 
\cite{Liu:2007ez,ding3}.  However, BaBar could not so far confirm
experimentally  its existence \cite{babary}.

\markboth{\sl Interpretation of the New Charmonium States }{\sl Other multiquark
states}
\section{\label{mul} Other multiquark states}

\subsection{A $D_s\bar{D}^*$ molecular state}

If the mesons $X(3872)$, $Z^+(4430)$, $Y(4260)$ and $Z_2^+(4250)$ are really
molecular states, then many other molecules should exist. A systematic study
of these molecular states and their experimental observation would confirm
its structure and provide a new testing ground for QCD within multiquark
configurations. In this context, a natural extension would be to probe the
strangeness sector. In particular, in analogy with the  meson $X(3872)$,
a $D_sD^*$ molecule with $J^{P}=1^{+}$ could be formed in the $B$ meson decay
$B\to\pi X_s\to\pi(J/\psi K\pi)$. Since it would decay into $J/\psi K^*\to J/
\psi K\pi$, it could be easily reconstructed.

In ref.~\cite{lnw} the QCD sum rules approach was used to predict the mass of
the $D_sD^*$  molecular state. Such prediction
is of particular importance for new upcoming experiments
which can investigate with much higher precision the charmonium energy regime,
like the PANDA experiment at the antiproton-proton facility at FAIR, or a
possible Super-B factory experiment. Especially PANDA can do a careful scan of
the various thresholds being present, in addition to going through
the exact form of the resonance curve.

The current used in ref.~\cite{lnw} is very  similar to the current in
Eq.~(\ref{cur-mol}) for the $X(3872)$:
\beq
j_\mu={1\over\sqrt{2}}\left[(\bar{s}_a\gamma_5 c_a)(\bar{c}_b\gamma_\mu
d_b)-(\bar{s}_a\gamma_\mu c_a)(\bar{c}_b\gamma_5 d_b)\right]\;.
\enq
The mass obtained for this state is \cite{lnw}:
\beq
M_{D_sD^*} = (3.96\pm0.10)~\GeV,
\label{dsdmass}
\enq
which is around 100 MeV bigger than the mass of the $X(3872)$ meson and
below the $D_sD^*$ threshold ($M(D_sD^*)\simeq 3980~\MeV$).

\begin{figure}[h]
\centerline{\epsfig{figure=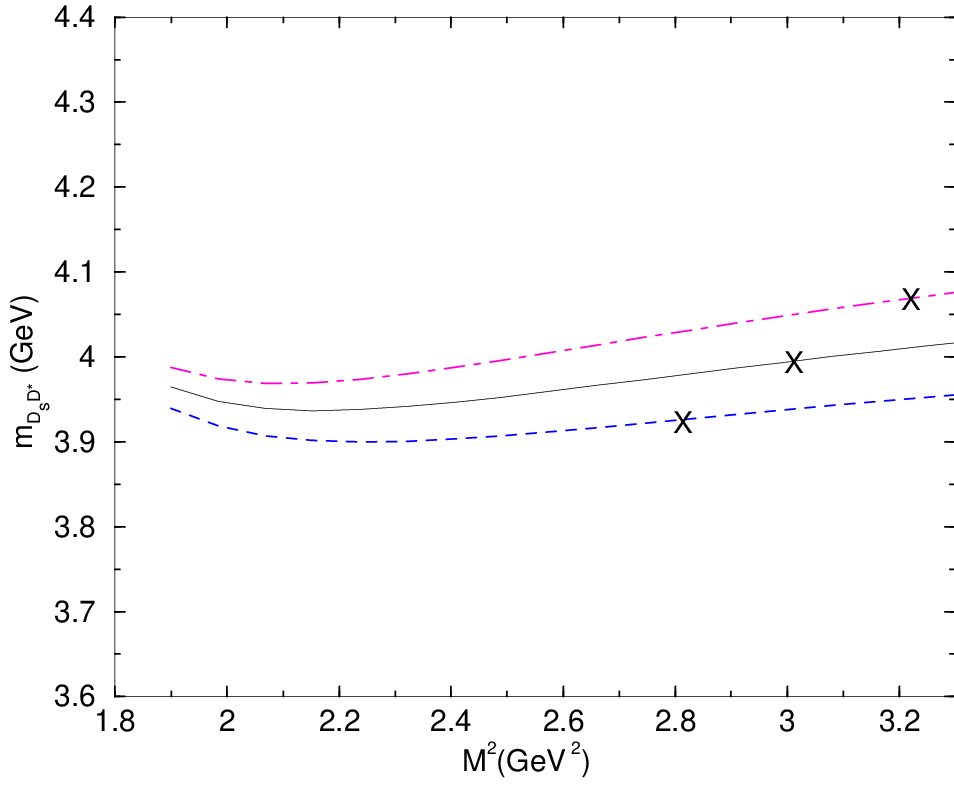,height=60mm}}
\caption{The $D_sD^*$ meson mass as a function of the sum rule parameter
($M^2$) for $\sqrt{s_0} =4.4$ GeV (dashed line), $\sqrt{s_0} =4.5$ GeV (solid
line) and $\sqrt{s_0} =4.6$ GeV (dot-dashed line). The crosses
indicate  the upper limit in the Borel region allowed
by the dominance of the QCD pole contribution (taken from ref.\cite{lnw}).}
\label{figmmol}
\end{figure}
The Borel curve for the mass of such state is quite stable, as can be seen
in Fig.~\ref{figmmol},
 and has a minimum within the
relevant Borel window.  Such a stable Borel curve strongly suggests
the possibility of the existence of a $D_sD^*$ molecular state with $J^P=1^+$.

Of course if such state exists, there will be no reason why its isospin
partners would not exist as well. Therefore,  charged states with the quark
content $c\bar{c}s\bar{u}$ and $c\bar{c}u\bar{s}$, and with a mass given
in Eq.~(\ref{dsdmass}), should also be observed
in the decay channels:  $J/\psi K^{*+}$ and $J/\psi K^{*-}$.

\subsection{A $[cc][\bar{u}\bar{d}]$ state}

Considering that  the double-charmonium production was already observed
in the reaction \cite{bellecc}
\beq
e^+e^-\to J/\psi+X(3940),
\label{prod-x}
\enq
it seems that it would be possible to form the tetraquark $[cc]
[\bar{u}\bar{d}]$.  Such state with quantum numbers $I=0, ~J=1$ and $P=+1$
which, following ref.\cite{ros},
we call $T_{cc}$, is especially interesting.  Although the process in 
Eq.~(\ref{prod-x}) will create two $c\bar{c}$ pairs, the production of 
$T_{cc}$ will further involve combining two light anti-quarks with $cc$ 
and might be suppressed.  On the other hand, heavy ion collisions at LHC 
might provide another opportunity for its non-trivial production\cite{Lee07}.
As already noted previously
\cite{ros,zsgr}, the $T_{cc}$ state cannot decay strongly
or electromagnetically into two $D$ mesons in the
$S$ wave due to angular momentum conservation nor in $P$ wave due to
parity conservation. If its mass is below the $DD^*$ threshold, this decay
is also forbidden, and this state would be very narrow.

Considering $T_{cc}$ as an axial diquark-antidiquark state, a possible
current describing it is given by:
\beqa
j_\mu&=&{i\over2}\epsilon_{abc}\epsilon_{dec}[(c_a^TC
\gamma_\mu c_b)(\bar{u}_d\gamma_5 C\bar{d}_e^T)\nn\\
&=&i[c_a^TC\gamma_\mu c_b][\bar{u}_a\gamma_5 C\bar{d}_b^T]\;.
\label{cur-tcc}
\enqa
This current represents well the most attractive
configuration expected with two heavy quarks.  This is so because
the most attractive light antidiquark is expected to be the in the
color triplet, flavor anti-symmetric and spin 0
channel \cite{Jaffe03,Shuryak03,Schafer93}.  This is also expected
quite naturally from the color magnetic interaction, which can be
phenomenologically parameterized as,
\begin{eqnarray}
V_{ij}=-\frac{C_H}{m_i m_j} \lambda_i \cdot \lambda_j \sigma_i \cdot
\sigma_j. \label{color-magnetic}
\end{eqnarray}
Here, $m,\lambda, \sigma$ are the mass, color and spin of the
constituent quark $i,j$.   Depending on the color state, the color factor 
$\lambda_i^a \lambda_j^a$ is $-\frac{8}{3},\frac{4}{3}, -\frac{16}{3}$ for 
quark quark in the color state $ \overline{3}, 6$ and for quark-antiquark in 
the singlet state respectively.  The phenomenological value of $C_H$ can be 
estimated from fits to the baryon and meson mass splitting, from which one 
finds that two constants $C_M$ and $C_B$ can fit all the mass splitting 
within the mesons and the baryons  respectively. Also, one finds that $C_M$ 
is about a factor of $3/2$ larger than $C_B$\cite{Lee07,Lee-Yasui09}.
Therefore, favorable multiquark configuration will inevitably involve diquark 
configuration with spin zero in color $\overline{3}$: this is a scalar 
diquark with anti symmetric flavor combination due to anti-symmetry of the 
total wave function involving color, spin and flavor.  The tetraquark picture
 for the light scalar nonet involves states composed of two scalar diquarks.  
However, here, one notes that the two pseudo scalar mesons will have a 
smaller mass and the tetraquark state not stable against strong decay.  
Possible stable configurations emerge if either the diquark or the 
anti-diquark in the tetraquark is composed of heavy quarks; this is so 
because the large attraction between quark-antiquark will be suppressed by 
the heavy quark mass while the strong attraction in either the diquark or 
the anti-diquark will remain.  Simple estimates within a constituent quark 
model predict stable configurations with spin zero, where a light diquark 
combines with a scalar $(cb)$ diquark, and with spin one, where a light 
diquark combines with
spin 1 heavy diquark \cite{Lee-Yasui09}. The $T_{cc}$ falls into the latter 
case where heavy anti-diquark is a color triplet $cc$ with  spin 1, as the 
pair should combine antisymmetrically. Although
the spin 1 configuration is repulsive, its strength is much
smaller than that of the light diquark due to the heavy charm
quark mass.  Therefore a constituent quark picture for $T_{cc}$
would be a light anti-diquark in color triplet, flavor  anti-symmetric and
spin 0 ($\epsilon_{dec}[\bar{u}_d\gamma_5 C\bar{d}_e^T]$) combined
with a heavy diquark of spin 1 ($\epsilon_{abc}[c_a^TC\gamma_\mu
c_b]$).  The simplest choice for the current which has a non zero
overlap with such a $T_{cc}$ configuration is given in
Eq.~(\ref{cur-tcc}).   While a similar configuration $T_{ss}$ is
also possible \cite{Morimatsu}, we believe that the repulsion in
the strange diquark with spin 1 will be larger and hence
energetically less favorable.

There are some predictions for the masses of the $T_{QQ}$ states.
In ref.~\cite{zhutcc} the authors use a color-magnetic interaction, with
flavor symmetry breaking corrections, to study heavy tetraquarks. They
assume that the Belle resonance, $X(3872)$, is a $cq\bar{c}\bar{q}$
tetraquark, and use its mass as input to determine the mass of other
tetraquark states. They get $M_{T_{cc}}=3966~\MeV$ and $M_{T_{bb}}=
10372~\MeV$. In ref.~\cite{ros}, the authors use  one-gluon exchange
potentials and two different spatial configurations to study the
mesons $T_{cc}$ and $T_{bb}$. They get $M_{T_{cc}}=3876 - 3905~\MeV$
and $M_{T_{bb}}=10519 - 10651~\MeV$. There are also calculations using
expansion in the harmonic oscillator basis \cite{sem}, and variational
method \cite{brst}.
In ref.~\cite{Lee-Yasui09}, the authors use Eq.~(\ref{color-magnetic}) together with a simple diquark picture to find stable heavy tetraquark configuration with spin zero $T_{cb}$ and spin one $T_{cc},T_{cb},T_{bb}$.

In ref.~\cite{tcc} a QCD sum rule calculation was done with the current in
Eq.~(\ref{cur-tcc}).  The authors found that the results are not very
sensitive to the value of the charm quark mass, neither to the value of the
condensates. The QCDSR predictions for the $T_{cc}$ mesons mass is:
\beq
M_{T_{cc}} = (4.0\pm0.2)~\GeV,
\enq
in a very good agreement with the predictions based on the one gluon exchange
potential model \cite{ros}, and color-magnetic model \cite{zhutcc}.

It is straightforward to extend the analysis done for the $T_{cc}$ to the
the bottom sector. The prediction for the $T_{bb}$ mass from ref.~\cite{tcc}
is:
\beq
\lb{massXb}
M_{T_{bb}}=  (10.2\pm0.3)~\GeV~,
\enq
also in a very good agreement with the results in refs.~\cite{ros},
\cite{zhutcc} and \cite{brst}.

The results from ref.~\cite{tcc} show that while the $T_{cc}$ mass is bigger
than the $D^*D$ threshold at about 3.875 GeV, the $T_{bb}$ mass is
appreciably below the $\bar{B}^*\bar{B}$ threshold at about
$10.6\;$GeV. Therefore, these results indicate that the  $T_{bb}$
meson should be stable with respect to strong interactions and
must decay weakly.  These results also confirm the naive expectation
that the exotic states with heavy quarks tend to be more stable
than the corresponding light states \cite{Lee05}.

\markboth{\sl Interpretation of the New Charmonium States }{\sl Summary}
\section{\label{sum} Summary} 

 We have presented a review, from the perspective of QCD sum rules, of the new
charmonium states recently observed by BaBar and Belle Collaborations. 
As it was seen case by case, this method has contributed a 
great deal to the understanding of the structure of these new states. When  a 
state 
is first observed and its  existence still needs confirmation, a QCDSR 
calculation can be very useful. It can provide evidence in favor or against
 the existence of the state. We have computed the masses of several $X$,$Y$ 
and $Z$ states  and they were supported by QCDSR. In some cases a tetraquark 
configuration 
was favored and in some other cases a molecular configuration was favored. 
QCDSR does 
not always corroborate previous indications. In Ref. \cite{mnnr07}, a 
comprehensive 
QCDSR analysis of light tetraquarks led to the conclusion that their existence 
(as tetraquarks states) is very unlikely. However, a more positive conclusion 
was found in the case of  tetraquarks with at least one heavy quark.

The limitations in statements made with QCDSR estimates come from 
uncertainties in the 
method. However these statements can be made progressively more precise as we 
know
more experimental information about the state in question. One good example 
is the
$X(3872)$, from which, besides the mass, several decay modes were measured. 
Combining all the available information and using QCDSR to calculate the 
observed decay widths, we were able to say that the $X(3872)$ is a mixed 
state, 
where the most important component is a $c \bar{c}$ pair, which is mixed 
with a 
small molecular component, out of which a large fraction is composed by 
neutral 
$D$ and $D^*$ mesons with  only a tiny fraction of  charged $D$ and $D^*$ 
mesons. 
This conclusion is very specific and precise and it is more elaborated than 
the other 
results presented addressing only the masses of the new charmonia. This 
improvement 
was a consequence of studying simultaneously the mass and the decay width. 
This type 
of combined calculation will eventually be extended to all states.

We close this review with some conclusions from the results presented in the
previous sections. They are contained in Tables ~\ref{tabfinal} and
\ref{tabbot}.

In Table~\ref{tabfinal} we present a summary of the most plausible 
interpretations for some of the states presented in Table~\ref{tabnew}, 
described in the previous sections.
\begin{table}[h]

  \begin{center}
    \caption{Structure and quantum numbers from  QCDSR studies.}
  \label{tabfinal}
    \begin{tabular}{|c|c|c|} \hline
      state    & structure& $J^{PC}$    \\ \hline
  $X(3872)$ & mixed $c\bar{c}-D\bar{D}^*$ & $1^{++}$    \\
  $Y(4140)$ & $D_s^*\bar{D}_s^*$ molecule  & $0^{++}$ \\
  $Z_2^+(4250)$ & $D\bar{D}_1$ molecule & $1^-$  \\
  $Y(4260)$ & $D_0\bar{D}^*$ or $D\bar{D}_1$ molecule & $1^{--}$  \\
  $Z^+(4430)$ & $D^*\bar{D}_1$ molecule & $0^-$  \\
  $Y(4660)$ & $[cs][\bar{c}\bar{s}]$ tetraquark & $1^{--}$ \\
\hline
     \end{tabular}
  \end{center}
\end{table}

In Table~\ref{tabbot} we present some predictions for bottomonium states, 
that may be observed at LHCb.
\begin{table}[h]

  \begin{center}
    \caption{Prediction for bottomonium states from  QCDSR studies.}
  \label{tabbot}
    \begin{tabular}{|c|c|c|c|} \hline
      state    & structure& $J^{PC}$ & mass (GeV)    \\ \hline
  $X_b$ &  $[bq][\bar{b}\bar{q}]$ tetraquark or & $1^{++}$& $10.27\pm0.23$  \\
        &   $B\bar{B}^*$ molecule & &  \\
  $Z_b$ & $[bq][\bar{b}\bar{q}]$ tetraquark or & $0^-$ & $10.74\pm0.12$  \\
   &    $B^*\bar{B}_1$ molecule &  &   \\
  $T_{bb}$ & $[bb][\bar{u}\bar{d}]$ tetraquark   & $1^{++}$ & $10.2\pm0.3$\\
        &   $B{B}^*$ molecule & &  \\
\hline
     \end{tabular}
  \end{center}
\end{table}

From the last Table we see that the mass predictions for the states $X_b$ 
and $T_{bb}$ are basically the same. These states, if they exist, may be 
produced at LHCb. Since the $T_{bb}$ is a charged state, it could be more
 easily identified. Moreover, since its mass prediction lies below the 
$B{B}^*$ threshold, it should be very narrow because it could not decay 
strongly.

Looking at Table~\ref{tabbot} we see that the predictions made with tetraquark 
and molecular currents are the same. Probably, in order to distinguish one
from the other, we will need to know the decay width of these states. On the
other hand, from the point of view of measurement, this convergence of
predictions makes them robust. In other words, even not knowing very precisely
their structure, we know that these states must be there.

Table~\ref{tabfinal} represents the final result of a comprehensive effort
and a careful analysis of several theoretical possibilities in the light
of existing data. It is an encouraging example of what QCDSR can do. This Table
contains a short summary of what we have learned about the new charmonium 
states in the recent past. Table~\ref{tabbot} addresses the near future and
shows some predictions.

From what was said above, hadron spectroscopy is and will be a very lively
field and QCDSR is a powerful tool to study it.
\vskip 2cm

{\bf Acknowledgements:} The authors would like to thank S. Narison, 
J.-M.~Richard, U. Wiedner,
 R.M. Albuquerque,  M.E. Bracco, R.D. Matheus, K. Morita, R.~Rodrigues da 
Silva and C.M.~Zanetti, with whom they have collaborated in one or more of 
the works described in this review.
The  authors  are indebted to the brazilian funding 
agencies FAPESP and CNPq, and
to the Korea Research Foundation KRF-2006-C00011. 

\markboth{\sl Interpretation of the New Charmonium States }{\sl Appendix}
\section{\label{app} Appendix: Fierz Transformation}

In general, the currents constructed from diquark antidiquark type of 
fields are related to those composed of meson type of fields.  However, 
the relation are suppressed by a typical color and Dirac factors so that 
we obtain a reliable sum rule only if we have chosen the current well to 
have a maximum overlap with the physical meson.  This is expected to be 
particularly true for multiquark configuration with special diquark or 
molecular structures.   Therefore, if we obtain a sum rule that reproduces 
the physical mass well, we can infer the  inner structure of the multiquark 
configuration.  For example, if the state is of molecular type, it will 
have a maximum overlap with a current that is constructed with the two 
corresponding meson current; the sum rule will then be able to reproduce 
the physical mass well.  On the other hand, if a current with a 
diquark-antidiquark type of current is used, the overlap to the physical 
meson will be small and the sum rule will not be able to reproduce the mass 
well.  The opposite will also be true.

To show the suppression, we discuss Fierz transformation of tetraquark 
current made of diquark fields into quark-antiquark fields.
The starting relation is given as follows.
\begin{eqnarray}
q_a^i \bar{q}_b^j & = &
-\frac{1}{8}\lambda^\alpha_{ab} \bigg[ \delta_{ij} (\bar{q} \lambda^\alpha q)+
\gamma_{ij}^\mu (\bar{q} \gamma_\mu \lambda^\alpha q)  \nonumber \\
&& - (\gamma^\mu \gamma^5)_{ij} (\bar{q} \gamma_\mu \gamma^5 \lambda^\alpha q)
+\gamma^5_{ij}(\bar{q} \gamma^5 \lambda^\alpha q) \nonumber \\
&& + \frac{1}{2} \sigma_{ij}^{\mu \nu} (\bar{q} \sigma_{\mu \nu} \lambda^\alpha q) \bigg] \nonumber \\
&& -\frac{1}{12}\delta_{ab} \bigg[ \delta_{ij} (\bar{q}  q)+
\gamma_{ij}^\mu (\bar{q} \gamma_\mu  q)  \nonumber \\
&& - (\gamma^\mu \gamma^5)_{ij} (\bar{q} \gamma_\mu \gamma^5 q)
+\gamma^5_{ij}(\bar{q} \gamma^5  q) \nonumber \\
&&
+ \frac{1}{2} \sigma_{ij}^{\mu \nu} (\bar{q} \sigma_{\mu \nu} q) \bigg].
\label{fiertz1}
\end{eqnarray}
Here, $a,b$ are the color and $i,j$ the Dirac spinor indices respectively.
The two color representations come from $3 \bigotimes \bar{3} = 8 
\bigoplus 1$.

In the formulas to follows, the two color terms can be worked out using the 
following formulas,
\begin{eqnarray}
\epsilon_{abc} \epsilon_{dec}   (q^T_a \Gamma_1 \lambda^\alpha_{bd}  \bar{c}^T_e)(\bar{q} \lambda^\alpha \Gamma_2 c) & =&
 -(\bar{c} \lambda^\alpha \Gamma_1^Tq ) (\bar{q} \lambda^\alpha \Gamma_2 c)
\nonumber \\
\epsilon_{abc} \epsilon_{dec}   (q^T_a \Gamma_1 \delta_{bd}  \bar{c}^T_e)(\bar{q} \lambda^\alpha \Gamma_2 c) & =&
2(\bar{c}  \Gamma_1^T q ) (\bar{q} \Gamma_2 c)
\end{eqnarray}

We start by applying the formula to the current composed of two scalar 
diquarks.  After Fierz transformation, we find,

\begin{eqnarray}
j^{di} &  = &  \epsilon_{abc} \epsilon_{dec} (q_a^T C \gamma_5 c_b)(\bar{q}_d \gamma_5 C \bar{c}^T_e) \nonumber \\
&=&
(-\frac{1}{8}) \bigg[ - (\bar{c} \lambda^\alpha  q ) (\bar{q} \lambda^\alpha  c) - (\bar{c} \lambda^\alpha \gamma_\mu q ) (\bar{q} \lambda^\alpha \gamma_\mu c) \nonumber \\ &&
- (\bar{c} \lambda^\alpha \gamma_\mu \gamma_5 q ) (\bar{q} \lambda^\alpha \gamma_\mu \gamma_5 c) \nonumber \\
&&
- (\bar{c} \lambda^\alpha \gamma_5 q ) (\bar{q} \lambda^\alpha \gamma_5 c)
+\frac{1}{2}(\bar{c} \lambda^\alpha \sigma_{\mu \nu}  q ) (\bar{q} \lambda^\alpha \sigma_{\mu \nu} c) \bigg] \nonumber \\
&& +(\frac{1}{6}) \bigg[ - (\bar{c} q ) (\bar{q} c) - (\bar{c} \gamma_\mu q ) (\bar{q} \gamma_\mu c) \nonumber \\&&
- (\bar{c} \gamma_\mu \gamma_5 q ) (\bar{q} \gamma_\mu \gamma_5 c) \nonumber \\
&& - (\bar{c} \gamma_5 q ) (\bar{q} \gamma_5 c)
+\frac{1}{2}(\bar{c} \sigma_{\mu \nu}  q ) (\bar{q} \sigma_{\mu \nu} c) \bigg].
\label{fiertz2}
\end{eqnarray}
This is a typical example of the expansion. The factor of $1/6$ multiplying 
the color singlet quark-antiquark types of currents, comes from the color 
and Dirac factors and is responsible for the small overlap to the various 
quark-antiquark  type of currents.  In Eq.~(\ref{fiertz2}), all meson type 
of currents contributes.  In general however, depending on the current 
type and charge conjugation, only certain types of meson-meson currents 
contributes.  Several examples for the expansion used in the text are given 
below.

\begin{enumerate}

\item For $X(3872)$ with $J^{PC}=1^{++}$,  we find,
\begin{eqnarray}
j_\mu^{(q,di)} &  = & \frac{i \epsilon_{abc} \epsilon_{dec}}{\sqrt{2}} 
[(q_a^T C \gamma_5 c_b)(\bar{q}_d \gamma_\mu C \bar{c}^T_e) \nonumber \\
&& +  (q_a^T C \gamma_\mu c_b)(\bar{q}_d \gamma_5 C \bar{c}^T_e)] 
\nonumber \\
&=& \frac{i}{\sqrt{2}}(-\frac{1}{8})(-1)^2  \bigg[
\sum_{\Gamma} \bigg(\bar{c} \lambda^\alpha ( \gamma_\mu C\Gamma^T C 
\gamma_5 \nonumber \\
&&+ \gamma_5 C \Gamma^T C \gamma_\mu ) q \bigg)(\bar{q} 
\lambda^\alpha \Gamma c) \bigg]\nonumber \\
&& 
 + \frac{i}{\sqrt{2}}(-\frac{1}{12})(2)(-1)\bigg[
\sum_{\Gamma} \bigg(\bar{c} ( \gamma_\mu C\Gamma^T C \gamma_5 \nonumber \\
&&+ \gamma_5 
C \Gamma^T C \gamma_\mu ) q \bigg) (\bar{q} \Gamma c) \bigg] \nonumber
\enqa
\beqa
j_\mu^{(q,di)} &=&
\frac{i}{\sqrt{2}}(-\frac{1}{8}) \bigg[ 2 (\bar{c} \lambda^\alpha \gamma_5 
q ) (\bar{q} \lambda^\alpha \gamma_\mu c) \nonumber \\
&&-  2 (\bar{c} \lambda^\alpha 
\gamma_\mu q )(\bar{q} \lambda^\alpha \gamma_5 c)
-2i (\bar{c} \lambda^\alpha \sigma_{\mu \nu} q ) \nonumber \\
&&\times(\bar{q} \lambda^\alpha 
\gamma_\nu \gamma_5 c) 
+  2i (\bar{c} \lambda^\alpha \gamma_\nu \gamma_5 q ) (\bar{q} 
\lambda^\alpha \sigma_{\mu \nu} c)
\bigg] \nonumber \\
&& +\frac{i}{\sqrt{2}}(\frac{1}{6}) \bigg[ 2 (\bar{c}  \gamma_5 q ) 
(\bar{q}  \gamma_\mu c) \nonumber \\
&&-  2 (\bar{c}  \gamma_\mu q ) (\bar{q}  \gamma_5 c) 
-2i (\bar{c}  \sigma_{\mu \nu} q ) (\bar{q}  \gamma_\nu \gamma_5 c)
\nonumber \\
&&+  
2i (\bar{c}  \gamma_\nu \gamma_5 q ) (\bar{q} \sigma_{\mu \nu} c)
\bigg] .
\label{fiertz3}
\end{eqnarray}
As can be seen from the above equation, the molecular component as given 
in  $j_\mu^{(q,mol)}=\frac{i}{\sqrt{2}} \bigg[ (\bar{c}  \gamma_5 q ) 
(\bar{q}  \gamma_\mu c) - (\bar{c}  \gamma_\mu q ) (\bar{q}  \gamma_5 c) 
\bigg]$ comprises only a small part of the wave function given in 
Eq.~(\ref{fiertz3}).  The overlap to the molecular type is suppressed by 
$\frac{1}{3}$ so that the overall contribution to the correlation function 
is suppressed by $\frac{1}{9}$.

\item For $Z^+(4430)$

\begin{itemize}
\item  with $J^P=0^-$,  we find the following current works well.

\begin{eqnarray}
j_{0-}^{(di)} &  = & \frac{i \epsilon_{abc} \epsilon_{dec}}{\sqrt{2}} 
[(u_a^T C \gamma_5 c_b)(\bar{d}_d C \bar{c}^T_e)\nonumber \\
&& -  (u_a^T C c_b)(\bar{d}_d \gamma_5 C \bar{c}^T_e)] \nonumber \\
&=&
\frac{i}{\sqrt{2}}(-\frac{1}{8}) \bigg[ 2 (\bar{c} \lambda^\alpha 
\gamma_\mu \gamma_5 u ) (\bar{d} \lambda^\alpha \gamma_\mu c)\nonumber \\
&& 
+  2 (\bar{c} \lambda^\alpha \gamma_\mu u ) (\bar{d} \lambda^\alpha 
\gamma_\mu \gamma_5 c)
\bigg] \nonumber \\
&& +\frac{i}{\sqrt{2}}(\frac{1}{6}) \bigg[ 2 (\bar{c} \gamma_\mu 
\gamma_5 u ) (\bar{d}  \gamma_\mu c) \nonumber \\
&&+  2 (\bar{c}  \gamma_\mu u ) (\bar{d} \gamma_\mu  \gamma_5 c)
\bigg] .
\label{fiertz4}
\end{eqnarray}
As can be seen from the above equation, the molecular component as given 
in  Eq.~(\ref{zmol}), contributes with a  suppression factor of 1/3.  
It is useful to compare this factor to the ratio of the overlaps found in 
QCD sum rule analysis and given in Eqs.~(\ref{ladi}) and (\ref{lamo}); 
numerically, 
the ratio $\lambda_{Z_{0-}}/\lambda_{D^*D_1}= 0.66$.  On the other hand, 
assuming that the $Z^+(4430)$ has only a molecular component that couples 
dominantly to the molecular current, the ratio should be 1/3, as can be 
seen from Eq.~(\ref{fiertz4}).  The fact that it is larger than 1/3 suggest 
that  while the $Z^+(4430)$ is dominantly of molecular type, there still 
remains some coupling to other type of non-molecular type of current 
component, such as the diquark-antidiquark type.

\item with $J^P=1^-$, we find the following current does not fit any of the 
observed $Z$ states.

\begin{eqnarray}
j_\mu  &  = & \frac{i \epsilon_{abc} \epsilon_{dec}}{\sqrt{2}} [(u_a^T C 
\gamma_5 c_b)(\bar{d}_d \gamma_\mu \gamma_5 C \bar{c}^T_e)\nonumber \\
&&
 +  (u_a^T C \gamma_\mu \gamma_5 c_b)(\bar{d}_d \gamma_5 C \bar{c}^T_e)] 
\nonumber \\
&=&
\frac{i}{\sqrt{2}}(-\frac{1}{8})
 \bigg[ -2 (\bar{c} \lambda^\alpha \gamma_5 u) (\bar{d} \lambda^\alpha 
\gamma_\mu \gamma_5 c) \nonumber \\ && -  2 (\bar{c} \lambda^\alpha 
\gamma_\mu \gamma_5 u ) (\bar{d} \lambda^\alpha \gamma_5 c) \nonumber \\
&& -2i (\bar{c} \lambda^\alpha \sigma_{\mu \nu} u ) (\bar{d} \lambda^\alpha 
\gamma_\nu  c) -  2i (\bar{c} \lambda^\alpha \gamma_\nu  u ) (\bar{d} 
\lambda^\alpha \sigma_{\mu \nu} c)
\bigg] \nonumber \\
&& +\frac{i}{\sqrt{2}}(\frac{1}{6}) \bigg[  -2 (\bar{c}  \gamma_5 u) 
(\bar{d}  \gamma_\mu \gamma_5 c) -  2 (\bar{c}  \gamma_\mu \gamma_5 u ) 
(\bar{d}  \gamma_5 c) \nonumber \\
&& -2i (\bar{c} \sigma_{\mu \nu} u ) (\bar{d}  \gamma_\nu  c) -  2i 
(\bar{c}  \gamma_\nu  u ) (\bar{d}  \sigma_{\mu \nu} c)
\bigg] .
\label{fiertz5}
\end{eqnarray}

\end{itemize}

\end{enumerate}

\markboth{\sl Interpretation of the New Charmonium States } {\sl Bibliography  }

\bibliographystyle{apsrmp}
\bibliography{ref-intro,ref-hq,ref-light,ref-mod}

\end{document}